\documentclass[11pt,a4paper]{article}
\pdfoutput=1

\usepackage{jcappub, slashed}
      
\usepackage{bm,graphicx,amssymb}

\usepackage[italicdiff]{physics}
\usepackage{ulem}

\begin{document}

\title{Axion-Gauge Field Dynamics with Backreaction}
\author[a]{Koji Ishiwata,}
\author[b,c]{Eiichiro Komatsu,}
\author[b]{Ippei Obata}

\affiliation[a]{Institute for Theoretical Physics, Kanazawa
  University, Kanazawa 920-1192, Japan}
\affiliation[b]{Max-Planck-Institute for Astrophysics,
  Karl-Schwarzschild-Str. 1, 85741 Garching, Germany}
\affiliation[c]{Kavli Institute for the Physics and Mathematics of the
  Universe (Kavli IPMU, WPI), UTIAS, The University of Tokyo, Chiba,
  277-8583, Japan}

\emailAdd{ishiwata@hep.s.kanazawa-u.ac.jp}
\emailAdd{komatsu@MPA-Garching.MPG.DE}
\emailAdd{obata@mpa-garching.mpg.de}

\abstract{ Phenomenological success of inflation models with axion and
  SU(2) gauge fields relies crucially on control of backreaction from
  particle production. Most of the previous study only demanded the
  backreaction terms in equations of motion for axion and gauge fields
  be small on the basis of order-of-magnitude estimation. In this
  paper, we solve the equations of motion with backreaction for a wide
  range of parameters of the spectator axion-SU(2) model. First, we
  find a new slow-roll solution of the axion-SU(2) system in the
  absence of backreaction. Next, we obtain accurate conditions for
  stable slow-roll solutions in the presence of backreaction. Finally,
  we show that the amplitude of primordial gravitational waves sourced
  by the gauge fields can exceed that of quantum vacuum fluctuations
  in spacetime by a large factor, without backreaction spoiling
  slow-roll dynamics. Imposing additional constraints on the power
  spectra of scalar and tensor modes measured at CMB scales, we find
  that the sourced contribution can be more than ten times the vacuum
  one. Imposing further a constraint of scalar modes non-linearly
  sourced by tensor modes, the two contributions can still be
  comparable.  }

\maketitle
 
\section{Introduction}
\label{sec:intro}    
\setcounter{equation}{0} 

Detection of a stochastic background of primordial gravitational waves
(tensor modes)~\cite{Grishchuk:1974ny,Starobinsky:1979ty} provides a
strong evidence for cosmic
inflation~\cite{Guth:1980zm,Sato:1980yn,Linde:1981mu,Albrecht:1982wi}. Statistical
properties of the density fluctuations (scalar modes) measured by
cosmic microwave background (CMB) experiments agree well with the
basic
predictions~\cite{Mukhanov:1981xt,Hawking:1982cz,Starobinsky:1982ee,Guth:1982ec,Bardeen:1983qw}
of the simplest models of inflation based on a single scalar field
(``inflaton'' field) rolling down on its potential
slowly~\cite{Komatsu:2014ioa,Planck:2018jri}.

Polarization data of CMB experiments place the strongest constraint on
the amplitude of primordial tensor
modes~\cite{Tristram:2020wbi,SPIDER:2021ncy,BICEP:2021xfz}. The
current upper bound on the ratio of squared amplitudes of tensor and
scalar modes (tensor-to-scalar ratio) is
$r<0.036$~(95\%~C.L.)~\cite{BICEP:2021xfz}.  Future experiments aim at
discovering $r$ at the level of $r=10^{-3}$ or
below~\cite{SimonsObservatory:2018koc,CMB-S4:2016ple,Hazumi:2019lys,NASAPICO:2019thw}. However,
even when $r$ is detected, interpretation of the result regarding the
origin of tensor modes may not be unique.

What generated primordial tensor modes? The leading idea is the
quantum vacuum fluctuation in
spacetime~\cite{Grishchuk:1974ny,Starobinsky:1979ty}, which relates
$r$ to the Hubble expansion rate during inflation, $H$. The current
bound on $r$ gives $H<1.9\times 10^{-5}M_{pl}\simeq 4.7\times
10^{13}$~GeV ($M_{pl}$ is the reduced Planck mass). Future detection
of $r$ has profound implications for the fundamental physics behind
inflation at such a high energy~\cite{Lyth:1996im,Lyth:1998xn}. This
relation, however, does not hold when tensor modes are sourced by
matter fields during inflation. In this paper, we study tensor modes
sourced by non-Abelian (SU(2)) gauge fields coupled to an axion field,
which evade the above relationship
\cite{Adshead:2013qp,Dimastrogiovanni:2012ew}.

Unlike scalar- and tensor-mode perturbations described above, massless
free gauge fields are conformally coupled to gravitation and cannot be
excited during inflation.  Conformal invariance can be broken if a
non-Abelian gauge field has a coupling
$(F^a_{\mu\nu}\tilde{F}^{a\,\mu\nu})^2$, where $F^a_{\mu\nu}$ and
$\tilde{F}^{a\,\mu\nu}$ are the field strength tensor of the gauge
field with the color index $a$ and its dual field, respectively. Then
an isotropic and homogeneous gauge configuration is established and
leads to a slow-roll inflationary dynamics called
``Gauge-flation''~\cite{Maleknejad:2011sq,Maleknejad:2011jw}. One can
also break conformal invariance by coupling a pseudo-scalar axion
field, $\chi$, to a non-Abelian gauge field via the Chern-Simons
interaction, $\chi F^a_{\mu\nu}\tilde F^{a\mu\nu}$. This set up, in
which $\chi$ acts as the inflaton field with a cosine
potential~\cite{Freese:1990rb} is called ``Chromo-natural
inflation’’~\cite{Adshead:2012kp}.  We can obtain the
$(F_{\mu\nu}^a\tilde{F}^{a\,\mu\nu})^2$ term of Gauge-flation by
integrating out the inflaton field of Chromo-natural inflation near
the minimum of the
potential~\cite{Adshead:2012qe,Sheikh-Jabbari:2012tom}.  Both models
share similar cosmological features~\cite{Maleknejad:2012fw} and reach
the isotropic gauge field configuration as an attractor, even if the
initial configuration is highly
anisotropic~\cite{Maleknejad:2013npa,Adshead:2018emn,Wolfson:2020fqz,Wolfson:2021fya}.

These models have unique cosmological consequences compared to single
real-scalar inflaton models.  During the slow-roll regime, a copious
amount of spin-2 particles are produced, which source gravitational
waves. Such gravitational waves are strongly scale-dependent and
chiral
\cite{Adshead:2013qp,Dimastrogiovanni:2012ew,Maleknejad:2012fw}, and
highly
non-Gaussian~\cite{Agrawal:2017awz,Agrawal:2018mrg,Dimastrogiovanni:2018xnn,Fujita:2018vmv,Fujita:2021flu}. Since
the production of gravitational waves is very efficient, the original
Gauge-flation and Chromo-natural inflation models are excluded by the
lack of detection of $r$~\cite{Adshead:2013nka,Namba:2013kia}.
Several models have been proposed to avoid this constraint;
modification to the inflaton
potential~\cite{Maleknejad:2016qjz,Caldwell:2017chz} or the kinetic
term~\cite{Watanabe:2020ctz}; introduction of another scalar field as
the inflaton field~\cite{Dimastrogiovanni:2016fuu,Iarygina:2021bxq};
mass of the gauge field via spontaneous symmetry
breaking~\cite{Nieto:2016gnp,Adshead:2016omu,Adshead:2017hnc}; and
delaying the period of Chromo-natural inflation to be out of the
observable scale of CMB
observation~\cite{Obata:2014loa,Obata:2016tmo,Domcke:2018rvv}.  The
scale-dependent and chiral gravitational wave spectrum is testable not
only with CMB experiments but also with direct detection experiments
(such as laser interferometers) across a wide range of
frequencies~\cite{Thorne:2017jft,Campeti:2020xwn}.

Such particle production causes backreaction on dynamics of the
axion-gauge field system during inflation; the energy and momentum of
the produced particles affect the equations of motion for axion and
gauge fields~\cite{Dimastrogiovanni:2016fuu,Fujita:2017jwq}.
Ref.\,\cite{Maleknejad:2018nxz} derives analytical formulae of the
backreaction terms. Dynamics of the axion-gauge system can be altered
significantly for a large value of the background gauge field.  Most
of the previous study
\cite{Dimastrogiovanni:2016fuu,Maleknejad:2018nxz,Papageorgiou:2019ecb}
discuss impacts of the backreaction terms by comparing them to the
other terms in the equations of motion on the basis of
order-of-magnitude estimation. Ref.~\cite{Fujita:2017jwq} solves the
equations of motion numerically including backreaction, but the
analysis is limited to one particular choice of the model parameters.

In this paper, we study dynamics of the spectator axion-SU(2) model of
Ref.~\cite{Dimastrogiovanni:2016fuu} with backreaction in the
comprehensive parameter space.  We solve the equations of motion for
the background axion and gauge fields to find slow-roll solutions. We
find two solutions; one is a known solution that has already been
studied in the literature, while the other is new. We then find that
the dynamics becomes unstable if the backreaction terms dominate in
the equations of motion. We provide accurate conditions for avoiding
this instability.

This paper is organized as follows. In Sec.\,\ref{sec:model}, the
Lagrangian of the model is given, leading to the equations of motion
for the background axion and gauge fields. In
Sec.\,\ref{sec:backreaction}, we present two slow-roll solutions,
conditions for the parameters to give stable solutions, and an example
of the time evolution of the axion-gauge system.  In
Sec.\,\ref{sec:gw}, we calculate the tensor power spectra sourced by
the gauge fields and compare them to those of the vacuum
fluctuation. We conclude in Sec.\,\ref{sec:conclusion}.

For the rest of the paper we take $M_{pl}=1$ and use a convention of
the metric tensor, $g_{\mu\nu}={\rm diag}(-1,1,1,1)$, in the flat
limit.

\section{Spectator axion-SU(2) model}
\label{sec:model}    
\setcounter{equation}{0} 

\subsection{Field equations without backreaction}

The action of the model \cite{Dimastrogiovanni:2016fuu} is given by 
$S=\int d^4x \sqrt{-g}({\cal L}_{\chi \mathchar`-{\rm gauge}}
  + {\cal L}_{\rm inf})$,
where $g$ is the determinant of the metric tensor and 
\begin{align}
  {\cal L}_{\chi \mathchar`-{\rm gauge}} &= 
  \frac{1}{2}R
  -\frac{1}{2}(\partial_\mu \chi)^2-V_\chi(\chi)
    -\frac{1}{4}F^a_{\mu\nu}F^{a\,\mu\nu}
    -\frac{\lambda }{4f}\chi \tilde{F}^{a\,\mu\nu}F^a_{\mu\nu}
    \,,
  \\
  {\cal L}_{\rm \inf} &=-\frac{1}{2}(\partial_\mu \phi)^2-V_\phi(\phi)\,.
\end{align}
Here, $\phi$ is the inflaton field with a potential $V_{\phi}(\phi)$,
$\chi$ is the pseudo scalar (axion) field with a potential
$V_\chi(\chi)$, $F_{\mu \nu}^a$ ($a=1,2,3$) is the field strength
tensor of the SU(2) gauge field, and we define
$\tilde{F}^{a\,\mu\nu}=\epsilon^{\mu\nu\rho\sigma}F^a_{\rho\sigma}/(2\sqrt{-g})$
with $\epsilon^{0123}=1$. The axion decay constant and the coupling
constant are given by $f$ and $\lambda$, respectively. When computing
the Ricci scalar, $R$, we use the
Friedmann-Lema\^{i}tre-Robertson-Walker metric for $g_{\mu\nu}$,
$ds^2=g_{\mu\nu}dx^\mu dx^\nu=-dt^2+a^2(t)\delta_{ij}dx^idx^j$, with
$a(t)$ being the scale factor.

The SU(2) gauge field $A^a_{\mu}$ acquires an isotropic and
homogeneous background solution $\bar{A}^a_{\mu}$ given by
$\bar{A}_0^a(t)=0$ and $\bar{A}_i^a(t)=a(t)\psi(t)\delta^a_i$ for
$i=1,2,3$~\cite{Maleknejad:2011sq,Maleknejad:2011jw}.  We have
introduced a field $\psi(t)$ that describes the background gauge field
configuration.  Here we assume that the axion and inflaton fields are
also homogeneous. Then, in terms of $\psi$, the Lagrangian densities
are written as
\begin{align}
  {\cal L}_{\chi \mathchar`-{\rm gauge}} &= 
  \frac{1}{2}R
  +\frac{1}{2}\dot{\chi}^2-V_{\chi}(\chi)
    +\frac{3}{2}
    \left[\frac{1}{a^2}\left(\pdv{(a\psi)}{t}\right)^2-g_A^2\psi^4\right]
    -3g_A\frac{\lambda}{f}\chi \frac{\psi^2}{a}\pdv{(a\psi)}{t}\,,
    \\
    {\cal L}_{\rm \inf} &=\frac{1}{2}\dot{\phi}^2-V_\phi(\phi)\,,
\end{align}
where $g_A$ is the gauge coupling constant of the SU(2) field and the dot denotes the time derivative. 

The equations of motion for $\chi$ and $\psi$ derived from the action are given by
\begin{align}
  &\ddot{\psi}+3H\dot{\psi}
  +(\dot{H}+2H^2)\psi +2g_A^2\psi^3
  =\frac{g_A\lambda}{f}\dot{\chi}\psi^2\,,
  \label{eq:EoM_psi} \\
  &\ddot{\chi}+3H\dot{\chi}+V_{\chi,\chi}(\chi)
  =-\frac{3g_A\lambda}{f}\psi^2(\dot{\psi}+H\psi)\,,
 \label{eq:EoM_chi}
\end{align}
where  $V_{\chi,\chi}(\chi)=dV_\chi(\chi)/d\chi$ and $H$ is the Hubble expansion rate given by
\begin{align}
  3H^2
  =\frac{1}{2}\dot{\phi}^2+V_\phi(\phi)
  +\frac{1}{2}\dot{\chi}^2+V_\chi(\chi)
  +\frac{3}{2}\left[(\dot{\psi}+H\psi_A)^2+g_A^2\psi_A^4\right]\,.
  \label{eq:H}
\end{align} 
The equation of motion for the inflaton field,
  $\ddot{\phi}+3H\dot{\phi}+V_{\phi,\phi}=0$,
with $V_{\phi,\phi}(\phi)=dV_\phi(\phi)/d\phi$ is decoupled from the other
field equations. 

Instead of using $\psi$, we use the following dimensionless parameter,
\begin{align}
  \xi_A \equiv \frac{g_A\psi}{H}\,,
  \label{eq:xi_A}
\end{align}
when studying dynamics of the background gauge field.

\subsection{Field equations with backreaction}

We now include the fluctuation, $\delta A_i^a$, around the homogeneous and isotropic background solution for the gauge field. We write $A_i^a=a(t)\psi(t)\delta_i^a+\delta A_i^a$,
where $\delta A_i^a$ contains scalar, vector, and tensor (spin-2) perturbations~\cite{Maleknejad:2011sq,Maleknejad:2011jw}. While $\delta A_i^a$ does not affect the background equations of motion for $\psi$ and $\chi$ at first order by construction, it does affect them at second order~\cite{Dimastrogiovanni:2016fuu,Fujita:2017jwq}.

Writing the spin-2 perturbation as $\delta A_i^a=\delta^{aj}B_{ij}$ and expanding $B_{ij}$ into $\sigma=\pm 2$ helicity states in Fourier space, the corrections to the equations of motion are given by~\cite{Maleknejad:2018nxz}
\begin{align}
    {\cal J}_A&=\frac{g_A}{3a^3}\sum_{\sigma=\pm 2}\int d^3k\left(
    -\frac{\sigma}{2}k+a\dot{\bar\alpha}\right)\left|B_\sigma\right|^2\,,\\
    {\cal P}_\chi&=\frac{\lambda}{2a^3f}\sum_{\sigma=\pm 2}\frac{d}{dt}\int d^3k\left(
    -\frac{\sigma}{2}k+aH\xi_A\right)\left|B_\sigma\right|^2\,,
\end{align}
for $\psi$ and $\chi$, respectively. Here, $B_\sigma$ is the mode function of the spin-2 perturbation and $\dot{\bar\alpha}$ is a model-dependent parameter which will be specified below.
This integral is divergent and requires regularization. 

Eqs.\,\eqref{eq:EoM_psi} and \eqref{eq:EoM_chi} are modified to\footnote{The sign of ${\cal J}_A$ term is different from Eq.\,(6.9) of Ref.\,\cite{Maleknejad:2018nxz}. This is a typo and their conclusions are not affected. We thank A. Maleknejad for the confirmation.}
\begin{align}
  &\ddot{\psi}+3H\dot{\psi}
  +(\dot{H}+2H^2)\psi +2g_A^2\psi^3
  =\frac{g_A\lambda}{f}\dot{\chi}\psi^2
  -{\cal J}_A\,,
  \label{eq:EoM_psi_BR} \\
  &\ddot{\chi}+3H\dot{\chi}+V_{\chi,\chi}(\chi)
  =-\frac{3g_A\lambda}{f}\psi^2(\dot{\psi}+H\psi)
  +{\cal P}_\chi\,.
  \label{eq:EoM_chi_BR}
\end{align}
where ${\cal J}_A$ and ${\cal P}_\chi$ are 
\begin{align}
  {\cal J}_A&=
  \frac{g_A H^3}{6\pi^2}
       {\cal K}_{\rm reg}[\dot{\bar{\alpha}}/H;\,\xi,\xi_A]\,,
  \label{eq:JA} \\
        {\cal P}_\chi&=
        \frac{3\lambda H^4}{4\pi^2 f}
             {\cal K}_{\rm reg}[\xi_A;\,\xi,\xi_A]\,,
        \label{eq:Pchi}
\end{align}
with ${\cal K}_{\rm reg}[X;\,\xi,\xi_A]$ 
computed by adiabatic subtraction of the diverging terms.
The expressions for ${\cal K}_\mathrm{reg}$ given in Appendix D of Ref.\,\cite{Maleknejad:2018nxz} are valid in the slow-roll regime, which is suitable for our purpose. We aim at finding a set of stable, slow-roll solutions in the presence of backreaction. If we had non-slow-roll $\psi$ and $\chi$, the backreaction would still lead to non-slow-roll solutions, which we are not interested in. Rather, we ask whether slow-roll solutions can be spoiled by the backreaction. To this end, we use the slow-roll expressions for ${\cal K}_\mathrm{reg}$.

The slow-roll assumption may not be valid near the end of inflation, when $\xi_A$ can also become large. Our approach breaks down in that regime, but one may use a new gradient expansion formalism of Refs.~\cite{Gorbar:2021rlt,Gorbar:2021zlr} to deal with such a case.

For the spectator axion-SU(2) model, $\dot{\bar{\alpha}}$ and $\xi$ are given by
\begin{align}
  \frac{\dot{\bar{\alpha}}}{H}=\xi\equiv\frac{\lambda \dot{\chi}}{2fH}\,.
\label{eq:xi}
\end{align}
If we ignore $\ddot{\psi}$, $\dot{\psi}$, $\dot{H}$ and the
backreaction term in Eq.\,\eqref{eq:EoM_psi_BR}, we obtain the slow-roll result $\dot{\bar{\alpha}}/H\simeq \xi_A +\xi_A^{-1}$\,. However, we find that this approximation (especially neglecting $\dot{\psi}$) is not
always valid in describing the dynamics of the axion-gauge system.  In the later analysis,
we solve the differential equations without using this slow-roll result.

So far, the discussion is general and applicable to any $V_\phi(\phi)$ and $V_\chi(\chi)$. In this paper, we do not specify the inflaton sector, hence $V_\phi$, but simply assume that it provides (quasi) de Sitter background with a given value of $H$~\cite{Dimastrogiovanni:2016fuu}. Specifically, the energy density $\rho$ of the universe is dominated by that of the slowly-rolling inflaton field, i.e., $\rho
\simeq \rho_\phi\simeq V_\phi(\phi)\simeq {\rm constant}$, which implies
\begin{align}
 \dot{H}\simeq 0\,,~~~~~V_\chi(\chi)\ll H^2\,. 
\end{align}
Then, using Eq.\,\eqref{eq:xi_A} and $\tilde{\chi}\equiv \chi/f$,
Eqs.\,\eqref{eq:EoM_psi_BR} and \eqref{eq:EoM_chi_BR} are rewritten as
\begin{align}
  &\xi_A''+3\xi'_A+2\xi_A(1+\xi^2_A)=
  \lambda \xi_A^2 \tilde{\chi}'-\tilde{{\cal J}}_A\,,
  \label{eq:EoM_xi_A_BR} \\
  &\tilde{\chi}''+3\tilde{\chi}'+\frac{V_{\chi,\chi}(\chi)}{H^2f}=
  -\frac{3}{\lambda\kappa}\xi_A^2(\xi_A+\xi_A')+\tilde{{\cal P}}_\chi\,,
  \label{eq:EoM_chi_til_BR}
\end{align}
where the primes denote derivatives with respect to $x\equiv Ht$ and 
\begin{align}
  \tilde{{\cal J}}_A&\equiv
  \frac{g_A^2}{6\pi^2}
  {\cal K}_{\rm reg}[\dot{\bar{\alpha}}/H;\,\xi,\xi_A]\,,
  \label{eq:JA_til} \\
  \tilde{{\cal P}}_\chi&\equiv
  \frac{3g_A^2}{4\pi^2 \kappa \lambda}
  {\cal K}_{\rm reg}[\xi_A;\,\xi,\xi_A]\,.
  \label{eq:Pchi_til}
\end{align}
Here we have introduced a dimensionless parameter $\kappa$,
\begin{align}
  \kappa \equiv \left(\frac{g_Af}{\lambda H}\right)^2\,.
\end{align}

We still keep our discussion general for $V_\chi(\chi)$, but will eventually
use the usual cosine potential~\cite{Freese:1990rb}, $V_\chi(\chi)=\mu^4[1+\cos(\chi/f)]$, when presenting numerical solutions.

\section{Backreaction}
\label{sec:backreaction}

In this section, we solve the equations of motion given in
Eqs.\,\eqref{eq:EoM_xi_A_BR} and \eqref{eq:EoM_chi_til_BR}.  The background axion and gauge fields may not be rolling slowly, depending
on the model parameters. We study a generic picture of
the axion-gauge dynamics and discuss the impact of the backreaction
effects.

\subsection{Stationary points}
\label{sec:stationary_points}

Although the slow-roll conditions are not always satisfied for
axion and gauge fields, they may be satisfied around possible
stationary point(s), if any. Therefore, it is legitimate to assume slow-roll motions for both fields for the moment to find stationary
points. Neglecting $\xi''_A$ and $\chi''$,
Eqs.\,\eqref{eq:EoM_xi_A_BR} and \eqref{eq:EoM_chi_til_BR} are
\begin{align}
  &3\xi'_A+2\xi_A(1+\xi_A^2)\simeq
  \lambda \xi_A^2 \tilde{\chi}'-\tilde{{\cal J}}_A\,,  
  \label{eq:EoM_xi_A_BR_slowroll} \\
  &3\tilde{\chi}'+\frac{V_{\chi,\chi}(\chi)}{H^2f}\simeq
  -\frac{3}{\lambda\kappa}\xi_A^2(\xi_A+\xi_A')+\tilde{{\cal P}}_\chi\,.
  \label{eq:EoM_chi_BR_slowroll}
\end{align}
They are easily diagonalized as
\begin{align}
  \xi'_A&=F_A(\xi_A,\chi)\,,
  \label{eq:dxiA}\\
  \tilde{\chi}'&=F_\chi(\xi_A,\chi)\,,
  \label{eq:dchi}
\end{align}
where
\begin{align}
  &F_A(\xi_A,\chi)=
  \frac{
    -\kappa \lambda \xi_A^2 V_{\chi,\chi}(\chi)/(H^2 f)-3\xi_A^5
    -6\kappa \xi_A(1+\xi_A^2)
    +\kappa \lambda \xi_A^2\tilde{\cal P}_\chi-3\kappa\tilde{J}_A}
       {3(3\kappa + \xi_A^4)}\,,
       \label{eq:FA}
       \\
       &F_\chi(\xi_A,\chi)=
       \frac{-\lambda \kappa V_{\chi,\chi}(\chi)/(H^2 f)-\xi_A^3+2\xi_A^5
         +\kappa \lambda \tilde{\cal P}_\chi+\xi_A^2 \tilde{J}_A}
            {\lambda(3\kappa + \xi_A^4)}\,.
            \label{eq:Fchi}
\end{align}
Then the stationary points for $\xi_A$ are given by $F_A(\xi_A,\chi)=0$. 

Let us first consider the case in which the backreaction terms are negligible. We find
stationary points for $\xi_A$ by solving
\begin{align}
  \kappa \beta \xi_A^2 -3\xi_A^5
  -6\kappa \xi_A(1+\xi_A^2)=0\,,
  \label{eq:findstationary_xiA}
\end{align}
where
\begin{align}
  \beta (\chi)\equiv -\frac{\lambda V_{\chi,\chi}(\chi)}{H^2 f} (>0)\,.
\end{align}
The solutions can be separated into two cases:
\begin{itemize}
\item[(a)] $\xi_{A0}^4\ge 2\kappa (1+\xi_{A0}^2)$
\item[(b)] $\xi_{A0}^4\le 2\kappa (1+\xi_{A0}^2)$
\end{itemize}
where $\xi_{A0}$ is the solution of
Eq.\,\eqref{eq:findstationary_xiA}.  For $\xi_{A0}\sim
\order{1}$, we expect that $\kappa\ll 1$ and $\kappa\gg 1$
roughly correspond to the case (a) and (b), respectively, which we will check later. If we further take a limit $\xi_{A0}^4\gg 2\kappa
(1+\xi_{A0}^2)$ in the case (a), the solution is obtained
analytically as
\begin{align}
  \xi_{A0}(\chi) \simeq
  \left[\frac{\kappa \beta(\chi)}{3}\right]^{1/3}\,.
  \label{eq:xiA0_1}
\end{align}
This is the known solution and is equal to $g_A\psi_0/H$ with $\psi_0=[\mu^4
  \sin(\chi/f)/(3g\lambda H)]^{1/3}$ given in
Ref.\,\cite{Adshead:2012kp} for $V_\chi(\chi)=\mu^4[1+\cos(\chi/f)]$.
In the other limit, $\xi_{A0}^4\ll 2\kappa (1+\xi_{A0}^2)$, in the case (b), the root
of Eq.\,\eqref{eq:findstationary_xiA} is given by\footnote{There is
another solution; however, it is unstable.}
\begin{align}
  \xi_{A0} (\chi)\simeq \frac{1}{12}
  \left[\beta(\chi)+ \sqrt{\beta(\chi)^2-144}\right]\,.
  \label{eq:xiA0_2}
\end{align}
This is a new solution. In summary, there are two types of stationary points for $\xi_A$,
which are given in Eqs.\,\eqref{eq:xiA0_1} and \eqref{eq:xiA0_2}, respectively,
depending on the parameters. We note that there are solutions that satisfy
Eq.\,\eqref{eq:findstationary_xiA} but are not given by Eqs.\,\eqref{eq:xiA0_1} and \eqref{eq:xiA0_2} which are valid only in the limits $\xi_{A0}^4\gg 2\kappa (1+\xi_{A0}^2)$ and 
$\xi_{A0}^4\ll 2\kappa (1+\xi_{A0}^2)$, respectively. We take into account all solutions
by solving Eq.\,\eqref{eq:findstationary_xiA} numerically and classify them
by $\xi_{A0}^4\gtrless 2\kappa (1+\xi_{A0}^2)$. Then the time evolution of $\xi_A$
is expected to be stable and given by
\begin{align}
  \xi_A(t)\simeq \xi_{A0}(\chi(t))\,.
  \label{eq:stationary_xiA}
\end{align}
However, the backreaction terms may change the trajectory of $\xi_A(t)$
or break the slow-roll dynamics.

\begin{figure}
 \begin{center}
   \includegraphics[width=7cm]{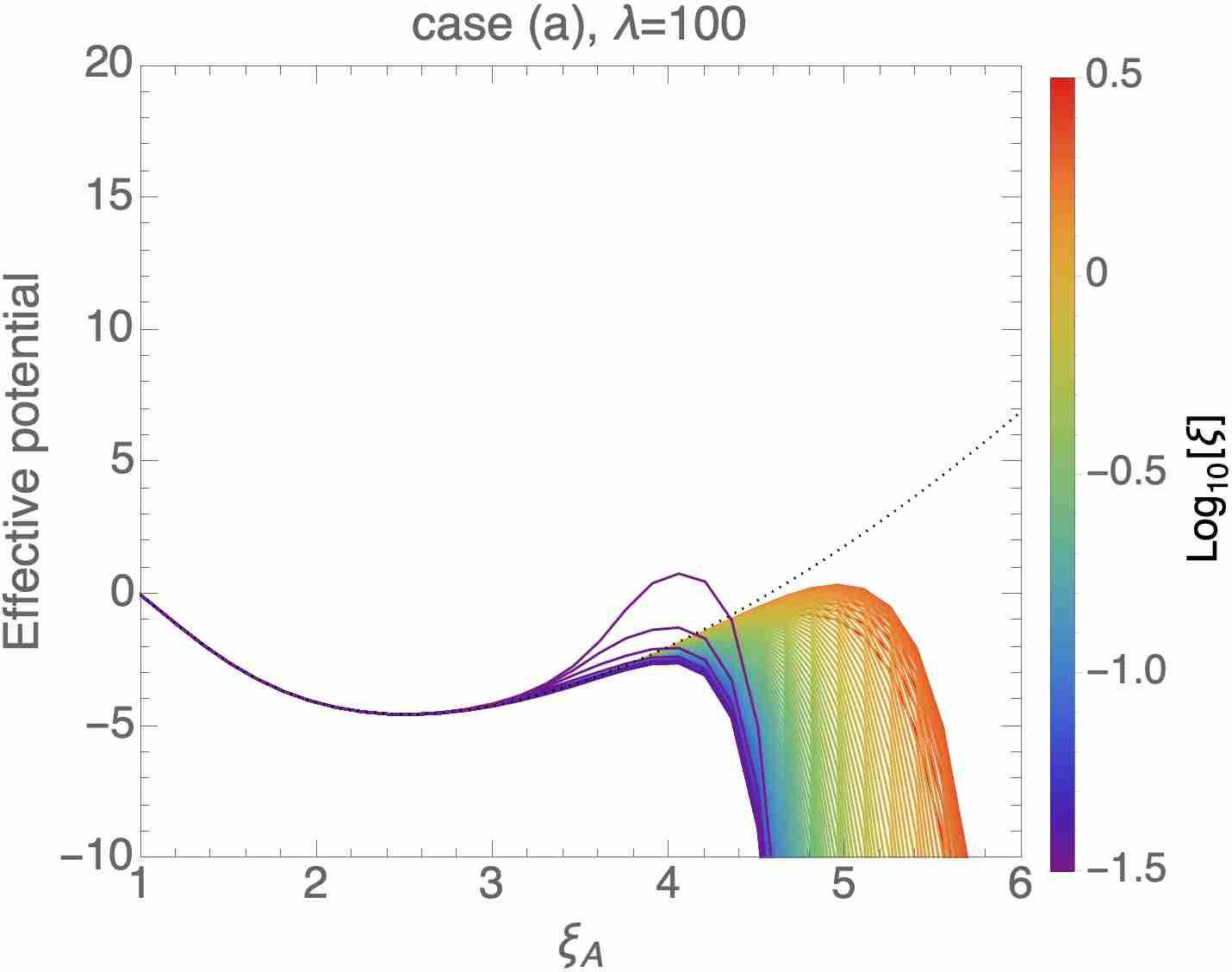}
   \includegraphics[width=7cm]{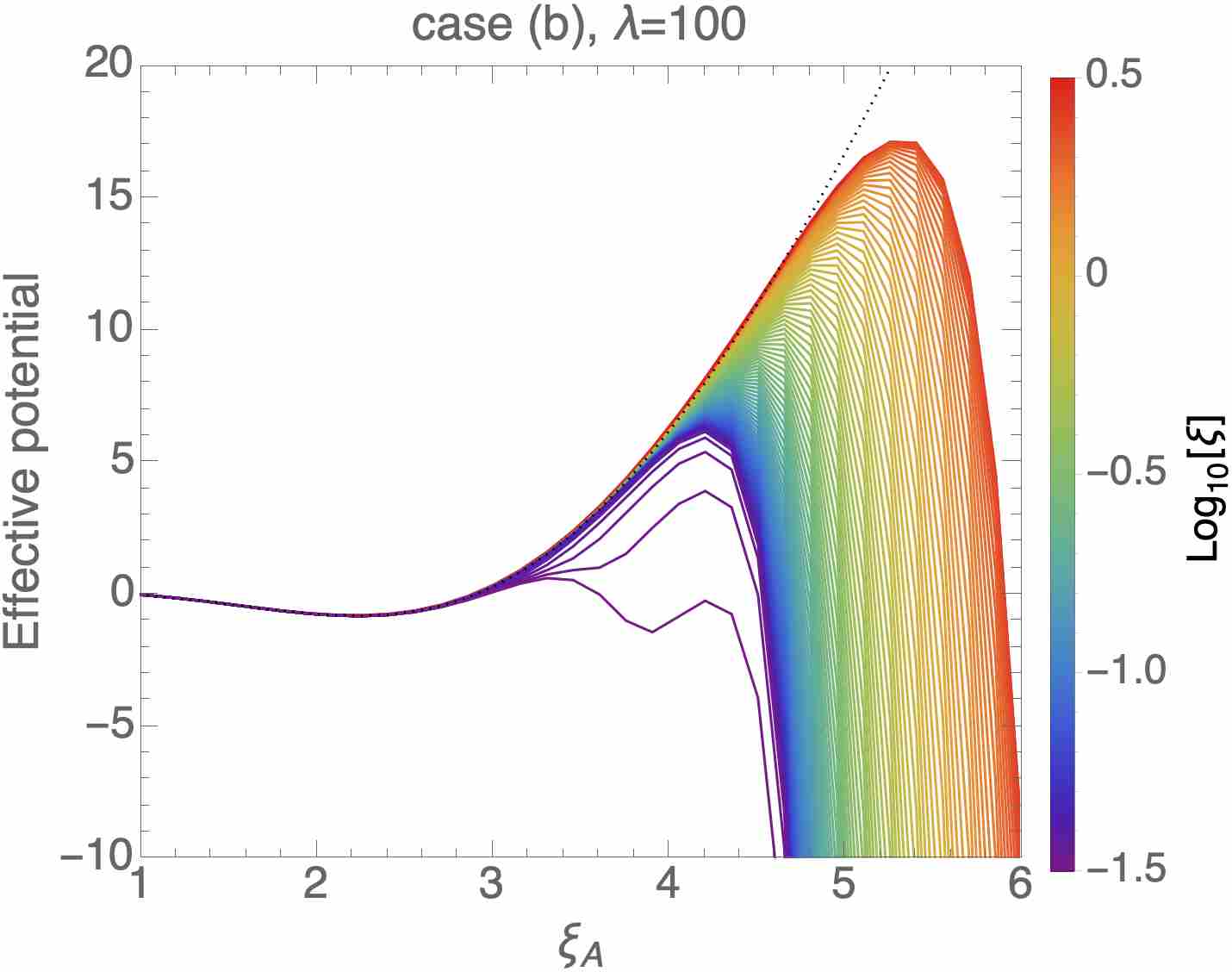}
   \caption{Arbitrarily normalized effective potential as a function of
     $\xi_A$.  Color of each line shows the value of
     $\xi\,(=\dot{\alpha}/H)$.  For comparison, the potential without the
     backreaction terms is shown in the dotted line.
     We use $\lambda=10^2$ and
     $\chi/(f\pi)=0.3$. The other parameters are 
     $g=1.1\times 10^{-2}$, $f=6.8\times 10^{-2}$, $H=9.6\times
     10^{-6}$ and $\mu= 8.3\times 10^{-4}$ (left panel; case (a)), and  $g=1.3\times
     10^{-2}$, $f=4.4\times 10^{-1}$, $H=7.9\times 10^{-6}$ and $\mu=
     1.3\times 10^{-3}$ (right panel; case (b)). }
  \label{fig:potential_eg}
 \end{center}
\end{figure}

We now include the backreaction terms in the calculation. To understand dynamics of the two fields, it is
useful to define an effective potential for $\psi$,
\begin{align}
  V^{\rm eff}_A(\xi_A,\chi) =
  -\int^{\xi_A}dy ~F_A(y,\chi)\,.
\end{align}
In Fig.\,\ref{fig:potential_eg}, we show two examples of the (arbitrarily
normalized) effective potentials of case (a) (left panel) and case (b) (right panel) as a function of $\xi_A$. The values of the other parameters, $\lambda$, $g$, $H$,
$f$ and $\mu$, are given in the figure caption. As expected, $\kappa$
is smaller (larger) than unity in the left (right) panel; we find $\kappa=6.6\times 10^{-1}$ and $5.4\times
10^{1}$ for the left and right panels, respectively. 
We find that the backreaction terms deform the effective potential, depending on $\xi\,(=\dot{\bar{\alpha}}/H=\lambda
\tilde{\chi}'/2)$. 
We show various values of $\xi$ in the range of $-1.5 \le \log_{10} \xi \le 0.5$ in color. 

Which value of $\xi$ is relevant in the figure? 
We can estimate $\xi$ from Eqs.\,\eqref{eq:xi} and \eqref{eq:dchi} and find that it is of order unity for the parameters used in the figure. This is expected from $\xi\simeq \xi_A+\xi_A^{-1}$ when the slow-roll 
approximation is valid.
Therefore, if the initial value of $\xi_A$ is around the
stationary point and the initial velocity of $\chi$ is within the 
slow-roll approximation, the effect of the backreaction is expected to be subdominant or negligible. We obtained similar results for smaller and larger values of $\lambda$.

Although we will solve the differential equations without the
slow-roll approximation, the effective potential for $\xi_A$ is useful for understanding the impact of the backreaction. First, the
backreaction terms modify the effective potential significantly for $\xi_A\gtrsim \order{10}$, and the modification is highly sensitive to $\xi$. 
Therefore, even if we find a stationary solution for $\xi_A\gtrsim
\order{10}$, the stationary solution might be merely temporal and lead
to an unstable behavior for the two fields. The conditions to avoid large contributions of the backreaction terms are roughly
estimated as
\begin{align}
  \frac{\{\kappa \lambda \xi_A^2\tilde{{\cal P}}_\chi,\,
    3\kappa \tilde{J}_A \}}
       {3\xi_A^5}
       \sim
       \frac{g_A^2{\cal K}_{\rm reg}[\xi_A;\xi,\xi_A]}
            {4\pi^2\xi_A^3}\,\{1,\,\kappa/\xi_A^2\}
       \ll 1\,,
\end{align}
for case (a) and
\begin{align}
  \frac{\{\kappa \lambda \xi_A^2\tilde{{\cal P}}_\chi,\,
    3\kappa \tilde{J}_A \}}
       {6\kappa\xi_A^3}
       \sim
       \frac{g_A^2{\cal K}_{\rm reg}[\xi;\xi,\xi_A]}{12\pi^2\xi_A^3}
       \{\xi_A^2/\kappa,\,1\}
       \ll 1\,,
\end{align}
for case (b) by assuming $\xi_A\gtrsim 1$. Since $\kappa$ is expected
to be small (large) for case (a) ((b)), the above two conditions
are satisfied when
\begin{align}
  \frac{g_A^2{\cal K}_{\rm reg}}{2\pi^2\xi_A^3} \ll 1\,,
  \label{eq:condition_for_stability}
\end{align}
where we have suppressed the argument of ${\cal K}_{\rm reg}$.  This leads to the following conditions:
\begin{itemize}
  \item Stationary solution with $\xi_A \gtrsim \order{1}$ needs a small $g_A$
  \item Stationary solution with $\xi_A \sim \order{1}$ allows for $g_A\sim
    \order{1}$
\end{itemize}
To make the criterion \eqref{eq:condition_for_stability} more
quantitative, we define a parameter ${\cal I}$,
\begin{align}
  {\cal I}\equiv
  \frac{g_A^2}{4\pi^2\xi_A^3}\times
  \left\{ \begin{array}{ll}
 {\cal K}_{\rm reg}[\xi_A;\xi,\xi_A] &~~~~~ {\rm case~(a)}
 \\[2mm]
 \frac{1}{3}{\cal K}_{\rm reg}[\xi;\xi,\xi_A] &~~~~~ {\rm case~(b)}
  \end{array} \right.\,,
  \label{eq:R}
\end{align}
and we will check whether ${\cal I}\ll 1$ is satisfied or not to get the 
stable solutions.  Note that only the case (a) was considered in the literature. Refs.\,\cite{Dimastrogiovanni:2016fuu,Fujita:2017jwq,Maleknejad:2018nxz,Papageorgiou:2019ecb} discussed the conditions for the backreaction terms to be subdominant for the case (a).

\subsection{Parameters of stable solutions}

We perform a parameter search to find stable slow-roll solutions for $\xi_A$ and $\chi$. Here, by ``stable slow-roll'' we mean that both $\xi_A$ and $\chi$ relax to the global minima ($\xi_A\to 0$ and $\chi\to f\pi$) at $x\to \infty$ satisfying Eq.\,\eqref{eq:findstationary_xiA}.

We check the stability by solving the differential equations \eqref{eq:EoM_xi_A_BR}
and \eqref{eq:EoM_chi_til_BR}. Here is the concrete procedure of the
analysis:
\begin{itemize}
\item[1.] Find stationary point $\xi_{A0}(\chi)$ from the effective
  potential $V_A^{\rm eff}(\xi_A,\chi)$ for $\chi_{0.3}\equiv 0.3\pi
  f$ and $\chi_{0.5}\equiv 0.5\pi f$. Here we take $\xi=1$.
  
\item[2.] For given initial values of $\xi_A=\xi_{A,{\rm ini}}$ and
  $\chi=\chi_{\rm ini}$ at $x=0$, we compute $\xi_A'$ and $\chi'$ at
  $x=0$ from Eqs.\,\eqref{eq:dxiA} and \eqref{eq:dchi} without backreaction. Inclusion of the backreaction terms at this stage is not possible because Eqs.~\eqref{eq:FA} and \eqref{eq:Fchi} are functions of $\chi'$. We take $(\chi_{\rm ini},\xi_{A,\rm ini})=(\chi_{0.3},1.1\xi_{A0}(\chi_{0.3}))$ and $(\chi_{0.5},1.1\xi_{A0}(\chi_{0.5}))$.
  
\item[3.] Using them as the initial conditions, we solve
  Eqs.\,\eqref{eq:EoM_xi_A_BR} and \eqref{eq:EoM_chi_til_BR} and find
  the solutions that satisfy the slow-roll conditions.
\end{itemize}
 We have checked that dynamics of
$\chi(t)$ and $\xi_{A}(t)$ is not affected significantly by taking
different initial conditions, unless we take very large values for $\xi_{A,{\rm ini}}$ for which the backreaction terms become extremely enhanced and inevitably give rise to unstable
solutions. For reasonable initial conditions, $\xi_A(t)$ soon reaches the
local minimum $\xi_{A0}(\chi(t))$ and finally relaxes to the global minimum.

The ranges of the parameters used in this analysis are
\begin{align}
  g_A:&~10^{-10}\, {\rm \mathchar`-\mathchar`-}\, 1
  \label{eq:range_gA}
  \\
  f:&~10^{-10}\, {\rm \mathchar`-\mathchar`-}\, 1
  \\
  H/f:&~10^{-10}\, {\rm \mathchar`-\mathchar`-}\, 10^{-1}
  \\
  \mu^4/H^2:&~10^{-10}\, {\rm \mathchar`-\mathchar`-}\, 10^{-1}
\end{align}
We explore three values of $\lambda=50$, 100 and 500.

\begin{figure}
  \begin{center}
    \includegraphics[width=7cm]{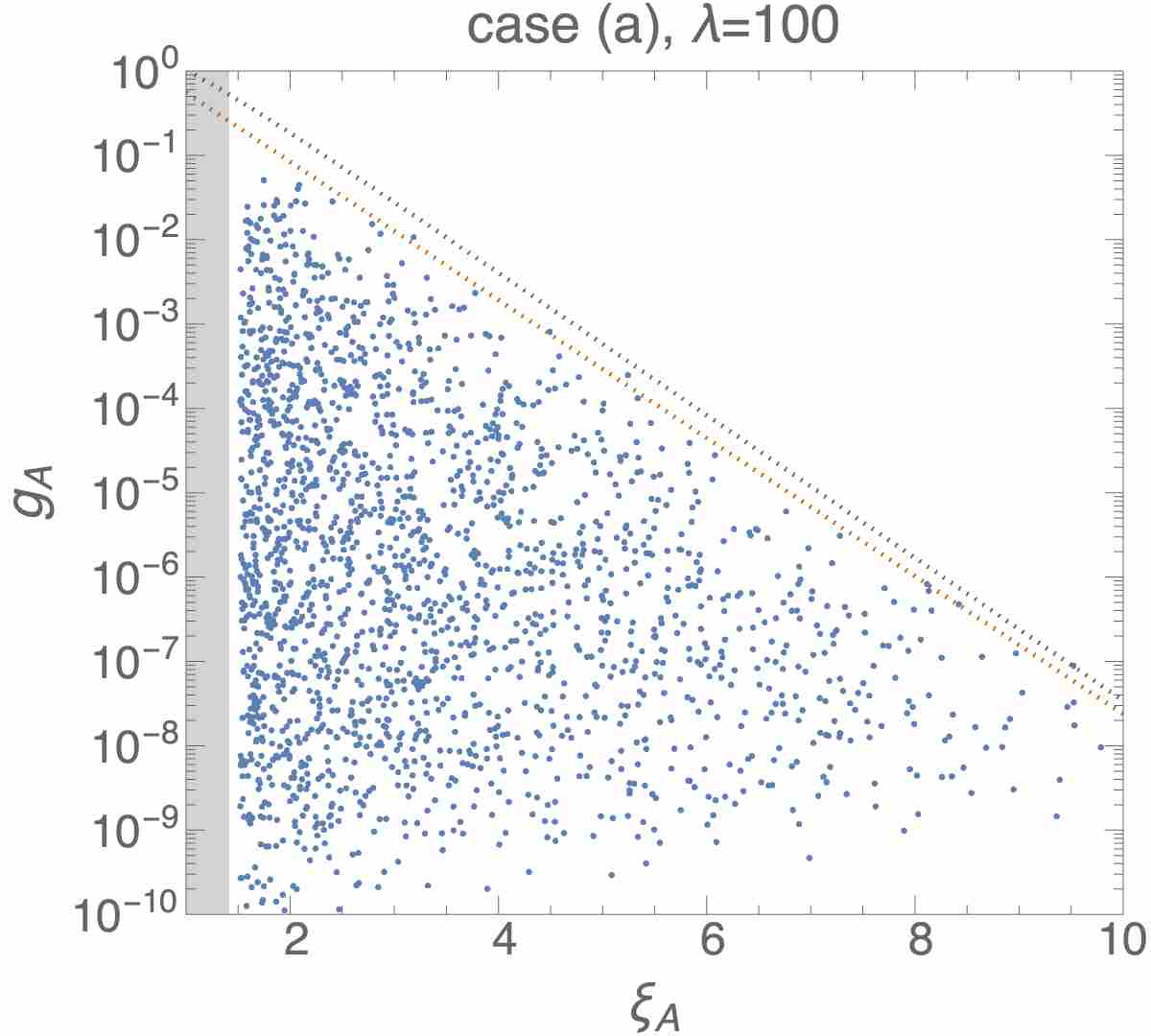}
    \includegraphics[width=7cm]{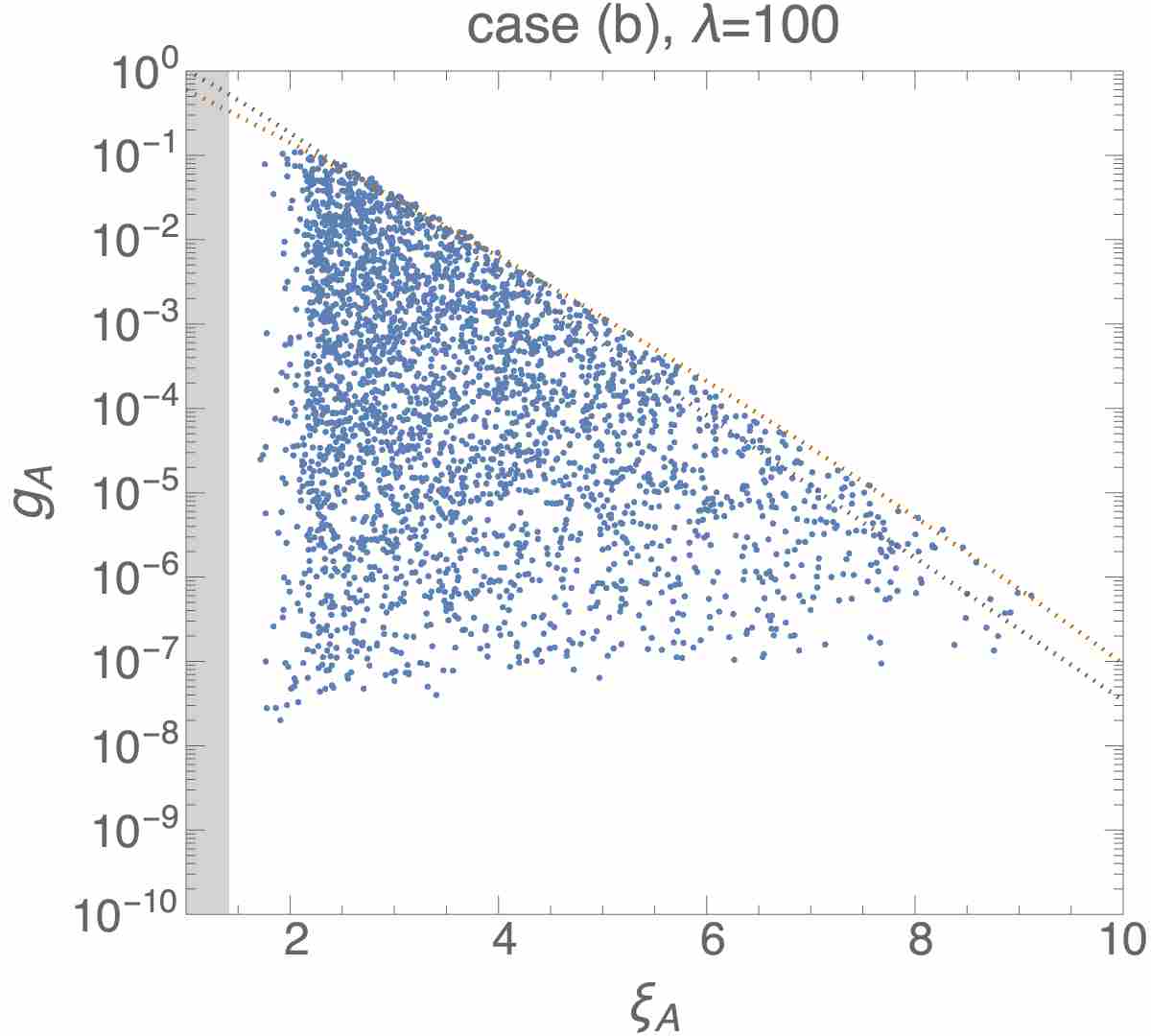}
    \caption{Scatter plots of the parameters of stable slow-roll solutions on
      $(\xi_{A0}(\chi_{0.5}),g_A)$ plane for case (a) (left) and (b) (right). We use $\lambda=100$. We show upper bounds on $g_A$ (orange, dashed), given in Eq.\,\eqref{eq:gmax}
      and Table~\ref{tab:gmax}. The shaded region shows
      $\xi_A<\sqrt{2}$ which we discard in the current study. For comparison,
      we show Eq.\,(5.5) in Ref.\,\cite{Papageorgiou:2019ecb}
      (black, dashed). }
  \label{fig:lam_100_g}
 \end{center}
\end{figure}

Fig.\,\ref{fig:lam_100_g} shows the parameters of stable slow-roll solutions on
$(\xi_{A0}(\chi_{0.5}),g_A)$ plane for $\lambda=100$. In Appendix~\ref{app:figs} the results for $\lambda=50$ and 500 are shown in Figs.\,\ref{fig:lam_050_gHf} and \ref{fig:lam_500_gHf}, respectively. Here we have
chosen $\xi_{A0}(\chi_{0.5})$ to give the maximum value for
$\xi_{A0}$. As expected from Eq.~\eqref{eq:condition_for_stability}, upper bounds on $g_A$ as a function of
$\xi_{A0}$ are obtained.  They are approximately given by an analytic
function
\begin{align}
  g_A^{\rm max} = e^{a \xi_{A0}^2+b \xi_{A0}+c}\,,
  \label{eq:gmax}
\end{align}
with $a$, $b$ and $c$ given in Table~\ref{tab:gmax}. 
In the calculation we use data points
in the range of $\xi_{A}>\sqrt{2}$, as the scalar modes in the model
would have a negative frequency-squared that may cause instability for
$\xi_A<\sqrt{2}$~\cite{Dimastrogiovanni:2012ew,Adshead:2013nka}. For both cases, we find lower bounds on $g_A$ as well as upper bounds. This is because $g_A$ can not be arbitrary small due to the lower limit on $H/f$ in the current parameter search to obtain  $\xi_{A0}\gtrsim \order{1}$. In addition, for case (b), 
the lower bound is more stringent since a too
small $g_A$ can not give a large $\kappa$.

\begin{table}[t]
  \centering
  \begin{tabular}{lccc}
    \hline
    \hline
    & $a$ & $b$ & $c$  \\
    \hline
    (a) $\lambda=50 $ & $-0.0133$ & $-1.71$ & 0.831  \\
    (a) $\lambda=100 $ & $0.000351$ & $-1.89$ & 1.29  \\
    (a) $\lambda=500 $ & $-0.0101$ & $-1.81$ & 1.13 \\
    \hline
    (b) $\lambda=50 $ & $-0.0594$ & $-0.981$ & $-0.232$  \\
    (b) $\lambda=100 $ & $-0.0378$ & $-1.32$ & 0.846  \\
    (b) $\lambda=500 $ & $-0.0652$ & $-1.13$ & 0.718  \\
    \hline
    \hline
  \end{tabular}
  \caption{Parameters of the fitting function for the upper bound on
    $g_A$, given in Eq.\,\eqref{eq:gmax}, for $\lambda=50$, 100, and 500.}
  \label{tab:gmax}
\end{table}

\begin{figure}
  \begin{center}
    \includegraphics[width=7cm]{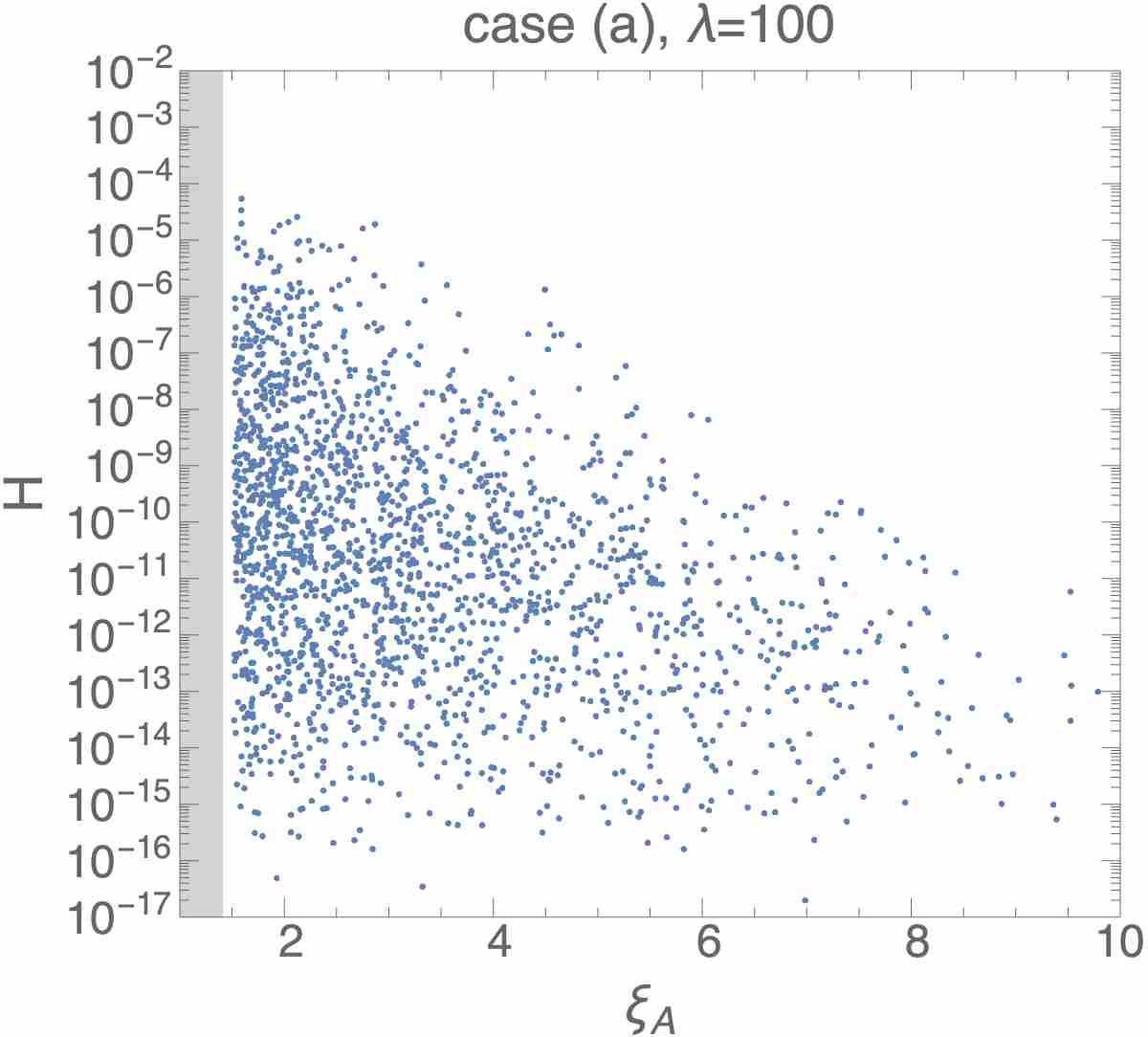}
    \includegraphics[width=7cm]{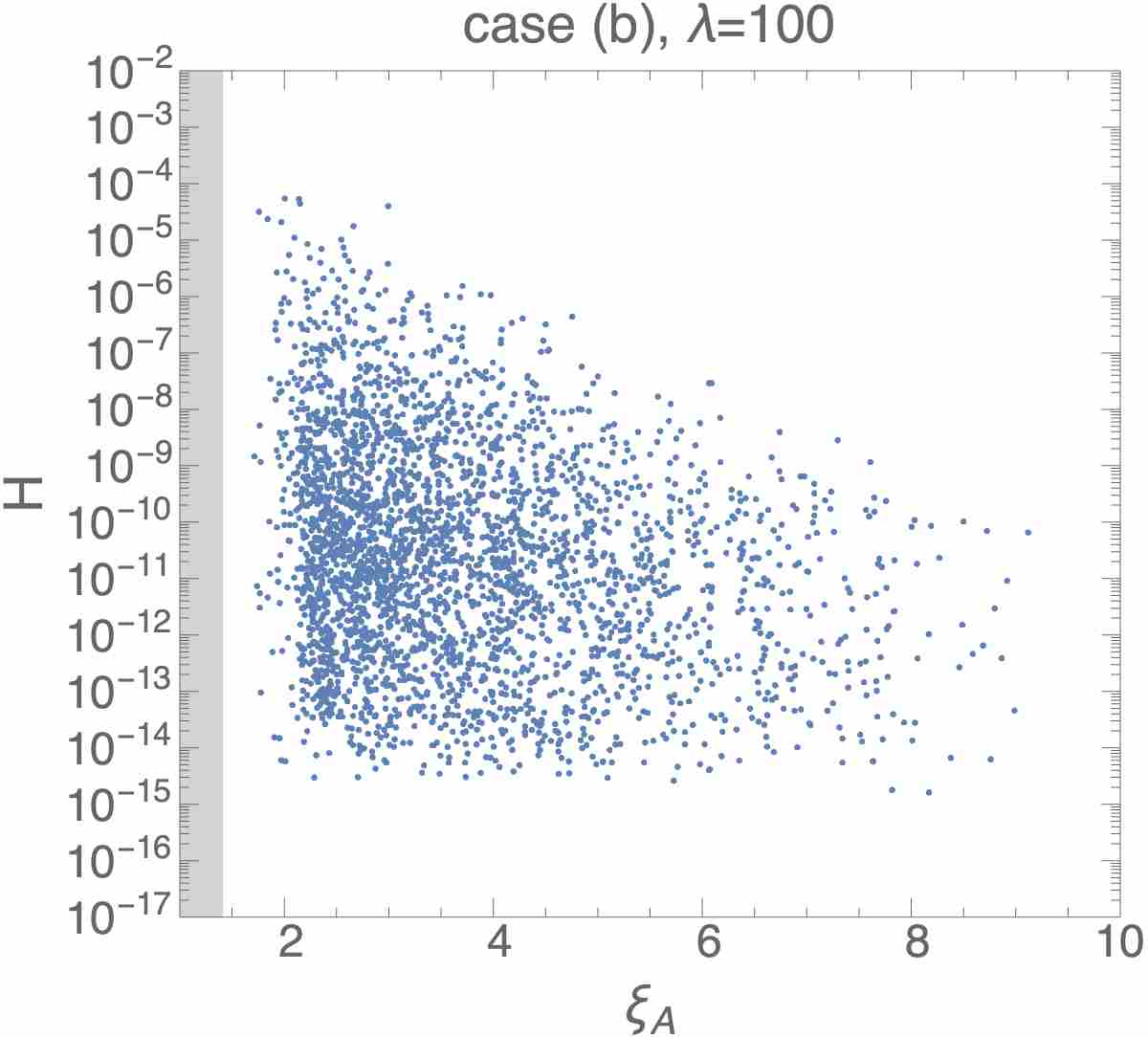}
    \includegraphics[width=7cm]{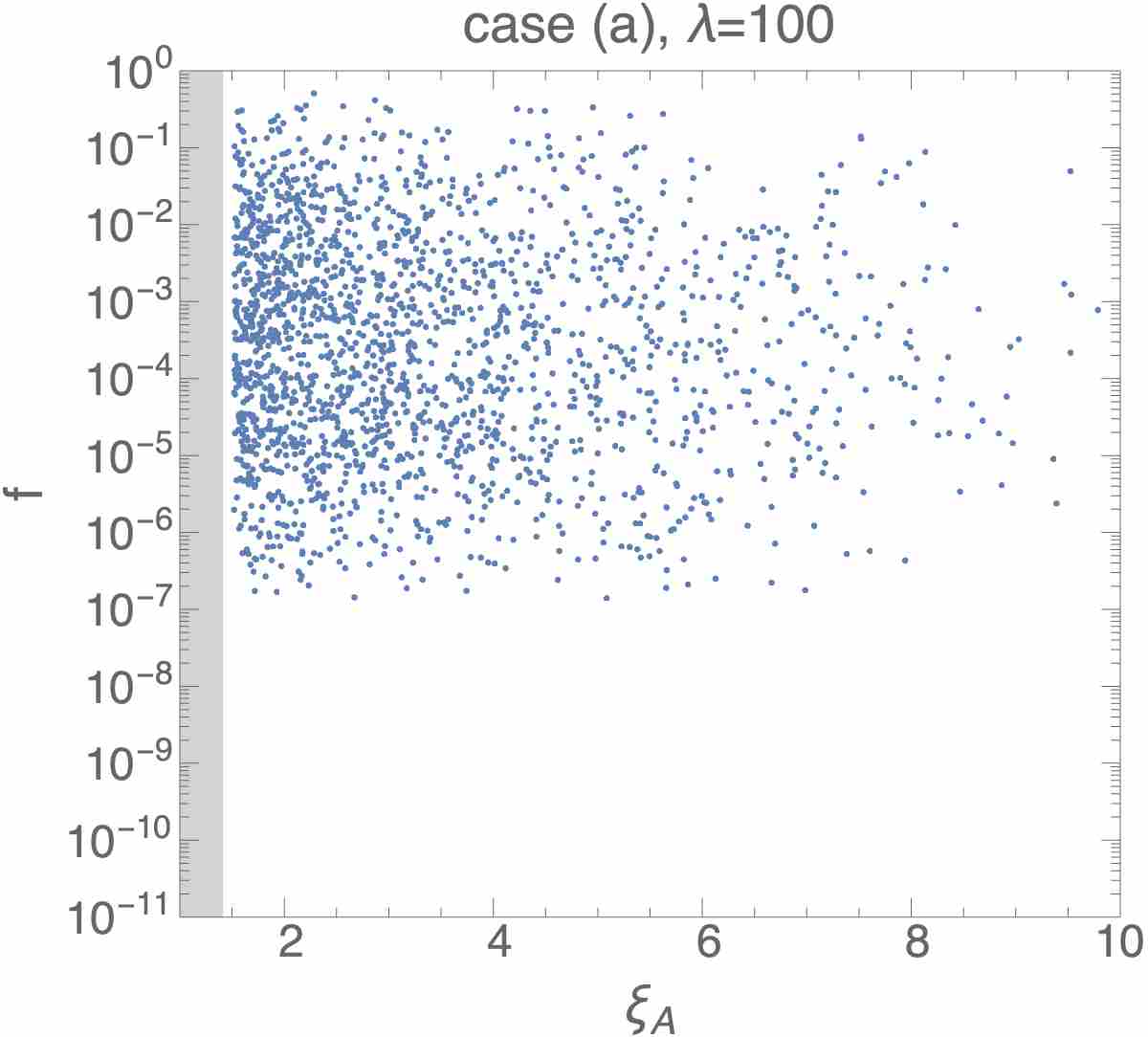}  
    \includegraphics[width=7cm]{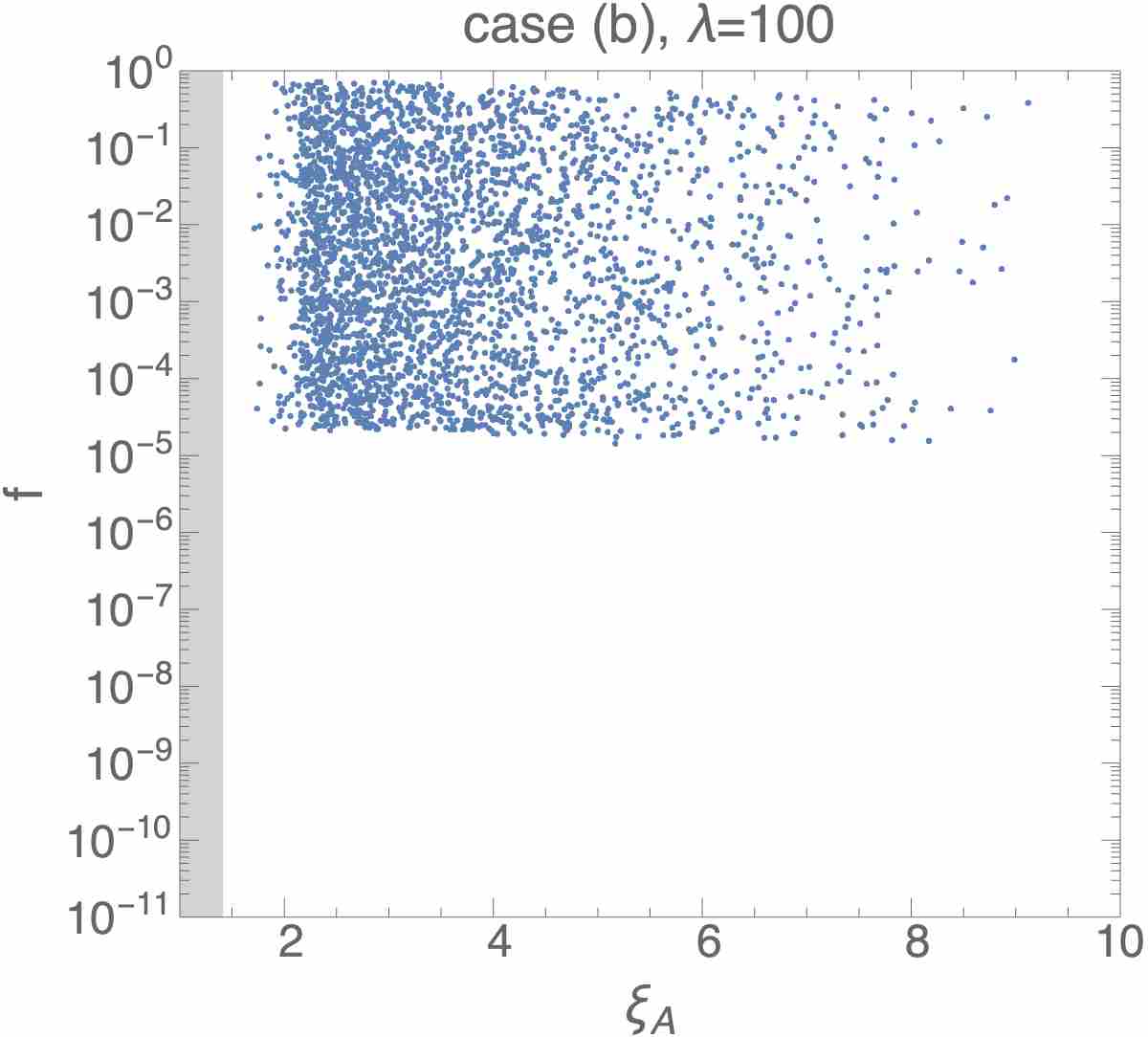}
    \caption{Scatter plots of the parameters of stable slow-roll solutions on
      $(\xi_{A0}(\chi_{0.5}),H)$ and $(\xi_{A0}(\chi_{0.5}),f)$ planes.
      We use $\lambda=100$.  }
  \label{fig:lam_100_Hf}
 \end{center}
\end{figure}

We find ${\cal I}\lesssim 0.1$ and
$\kappa/\xi_{A0}^2< 1$ for case (a) and ${\cal I}\lesssim 0.2$ and
$\kappa/\xi_{A0}^2\gtrsim 1$ for case (b). While the discussion given at
the end of Sec.\,\ref{sec:stationary_points} is qualitatively correct,
we find that the condition ${\cal I}\ll 1$ is too stringent; ${\cal I}<0.1$
is sufficient for obtaining stable slow-roll  solutions. In summary, we have
quantitatively confirmed that dynamics of the axion-gauge field system
becomes unstable when the backreaction terms become larger than the other
terms in the equations of motion, as argued in Refs.\,\cite{Dimastrogiovanni:2016fuu,Fujita:2017jwq,Maleknejad:2018nxz,Papageorgiou:2019ecb} for case (a), and sharpened the required condition, ${\cal I}<0.1$. The results for case (b) are new.

Fig.\,\ref{fig:lam_100_Hf} shows $H$ and $f$ of stable slow-roll solutions  (see Figs.\,\ref{fig:lam_050_gHf} and \ref{fig:lam_500_gHf} in Appendix~\ref{app:figs} for $\lambda=50$ and 500, respectively). We find upper bounds on $H$, which become more restrictive as $\xi_{A0}$ increases. On the other hand, $f$ can be as large as $\order{1}$. For case (b), more stringent lower bounds on $f$ are obtained. This is because too small $f$ can not give a large $\kappa$ required for case (b).

\subsection{Dynamics of the axion and gauge fields}

\begin{figure}
  \begin{center}
    \includegraphics[width=6.5cm]{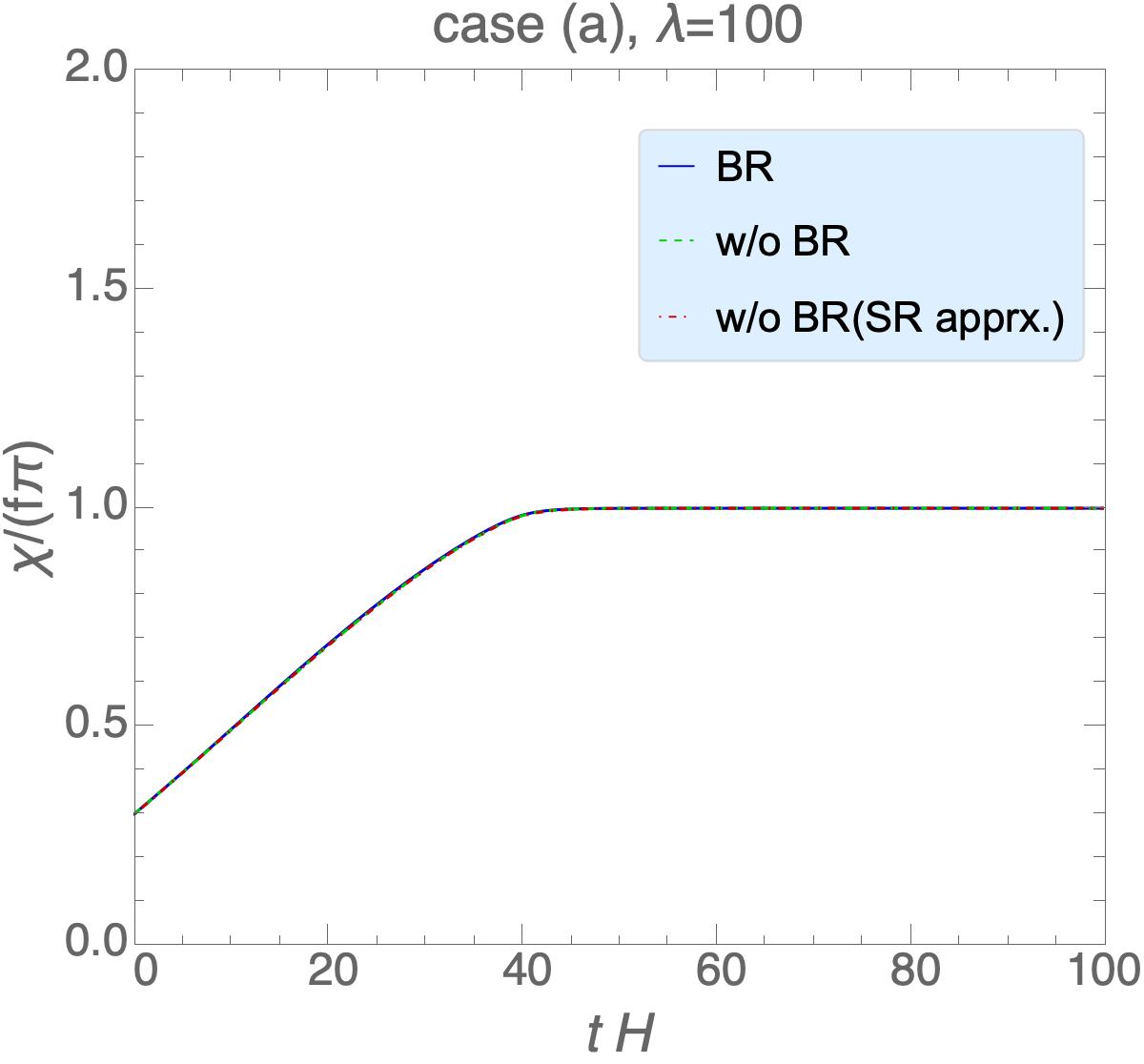}
    \includegraphics[width=6.5cm]{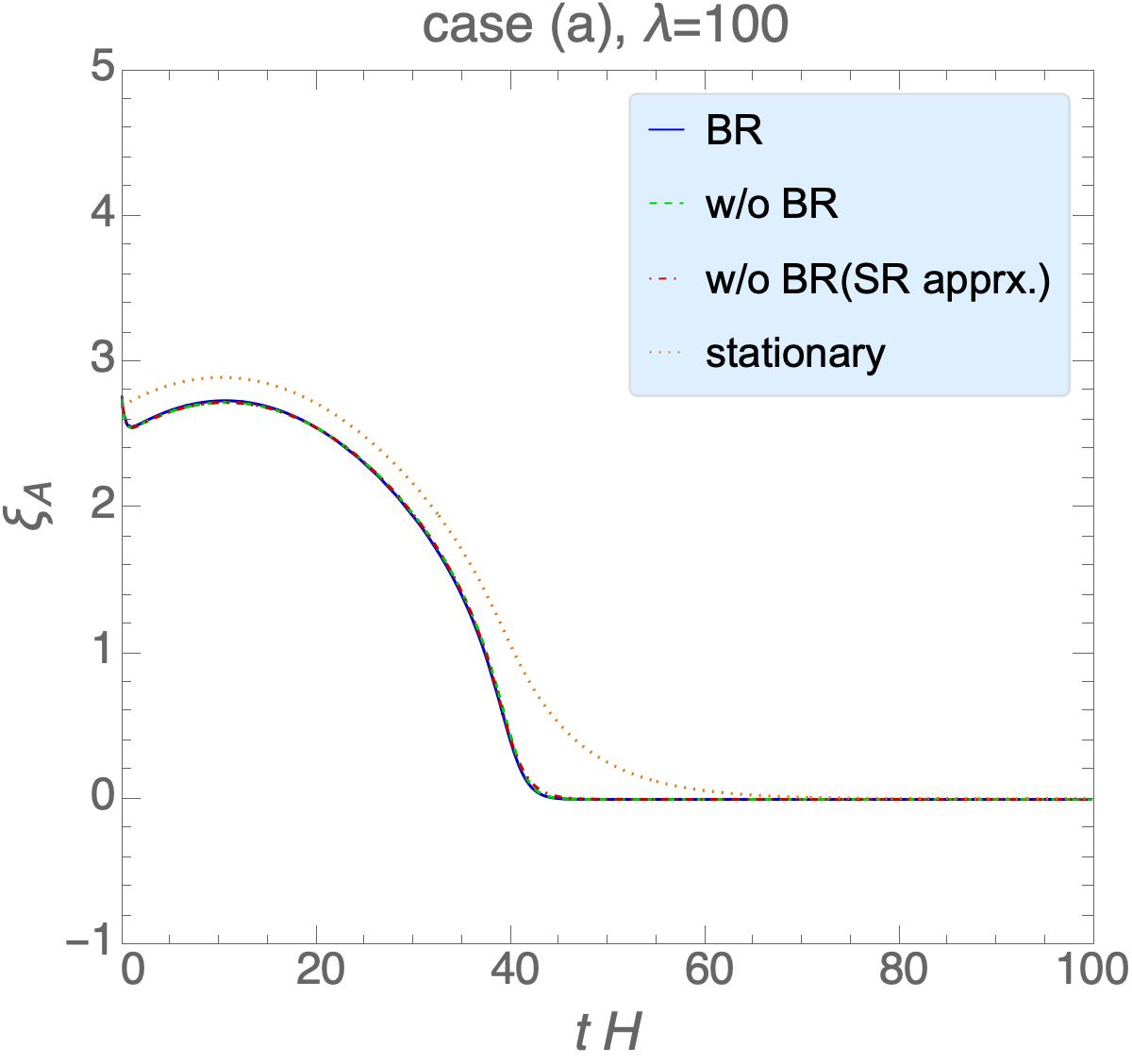}
     \caption{Time evolution of the axion ($\chi/(f\pi)$, left panel) and gauge  ($\xi_A=g_A\psi/H$, right panel) fields for case (a).
       The initial conditions are $\chi_{\rm
         ini}=\chi_{0.3}=0.3\pi f$ and $\xi_{A,\rm
         ini}=1.1\xi_{A0}(\chi_{0.3})$. The other parameters are the
       same as in the left panel of Fig.\,\ref{fig:potential_eg}. In
       both panels, we show numerical solutions of Eqs.\,\eqref{eq:EoM_xi_A_BR} and \eqref{eq:EoM_chi_til_BR} with backreaction (``BR'', blue solid); those without backreaction (``w/o BR'', green dashed); and
       Eqs.\,\eqref{eq:EoM_xi_A_BR_slowroll} and
       \eqref{eq:EoM_chi_BR_slowroll} without  backreaction (``w/o BR (SR apprx.)'', red dot-dashed). All curves lie on top of each other. In the right panel, we also show the trajectory given by Eq.\,\eqref{eq:xiA0_1} (``stationary'', orange dotted). }
  \label{fig:dy_a_lam100}
 \end{center}
\end{figure}

\begin{figure}
  \begin{center}
    \includegraphics[width=6.5cm]{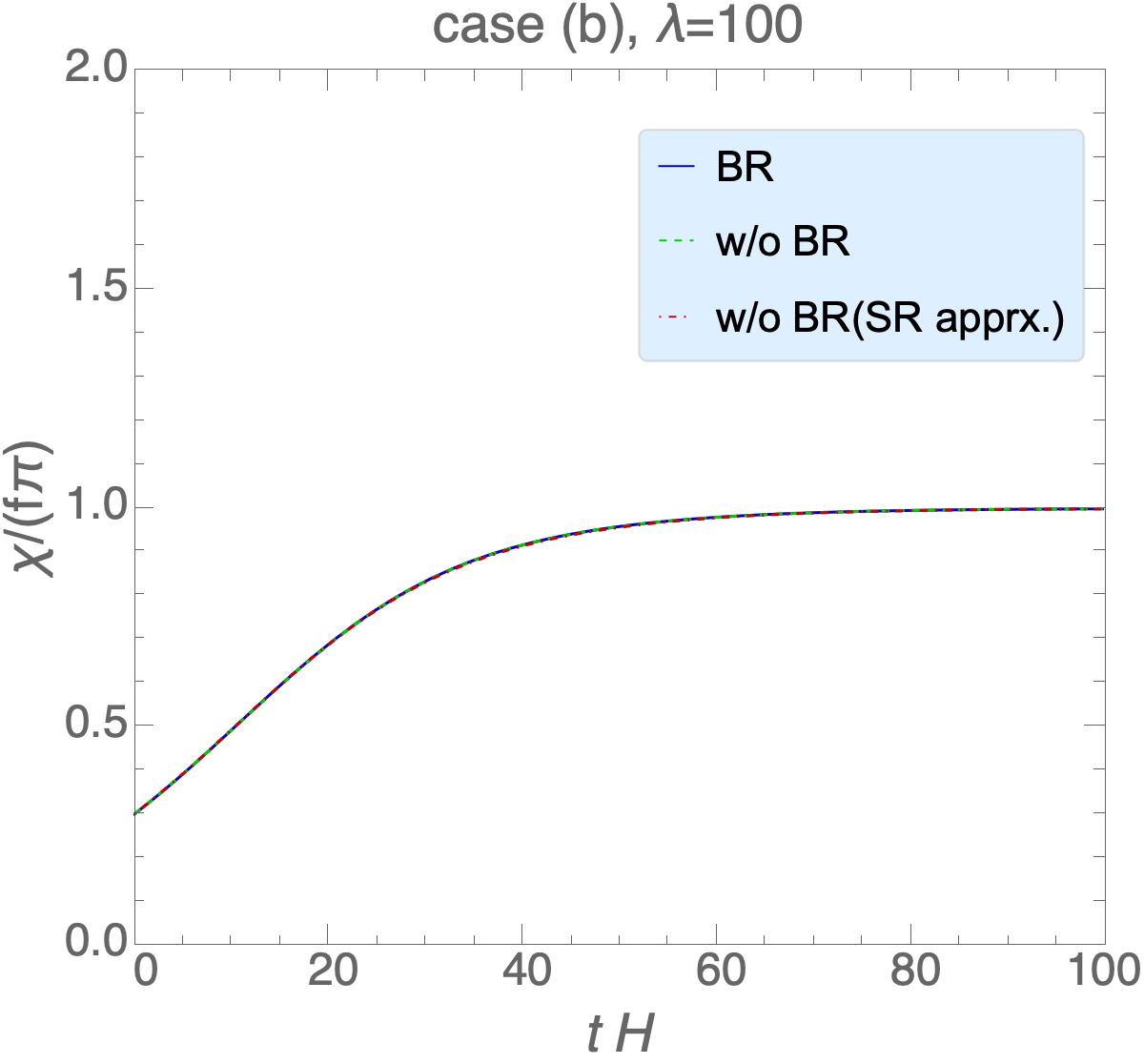}
    \includegraphics[width=6.5cm]{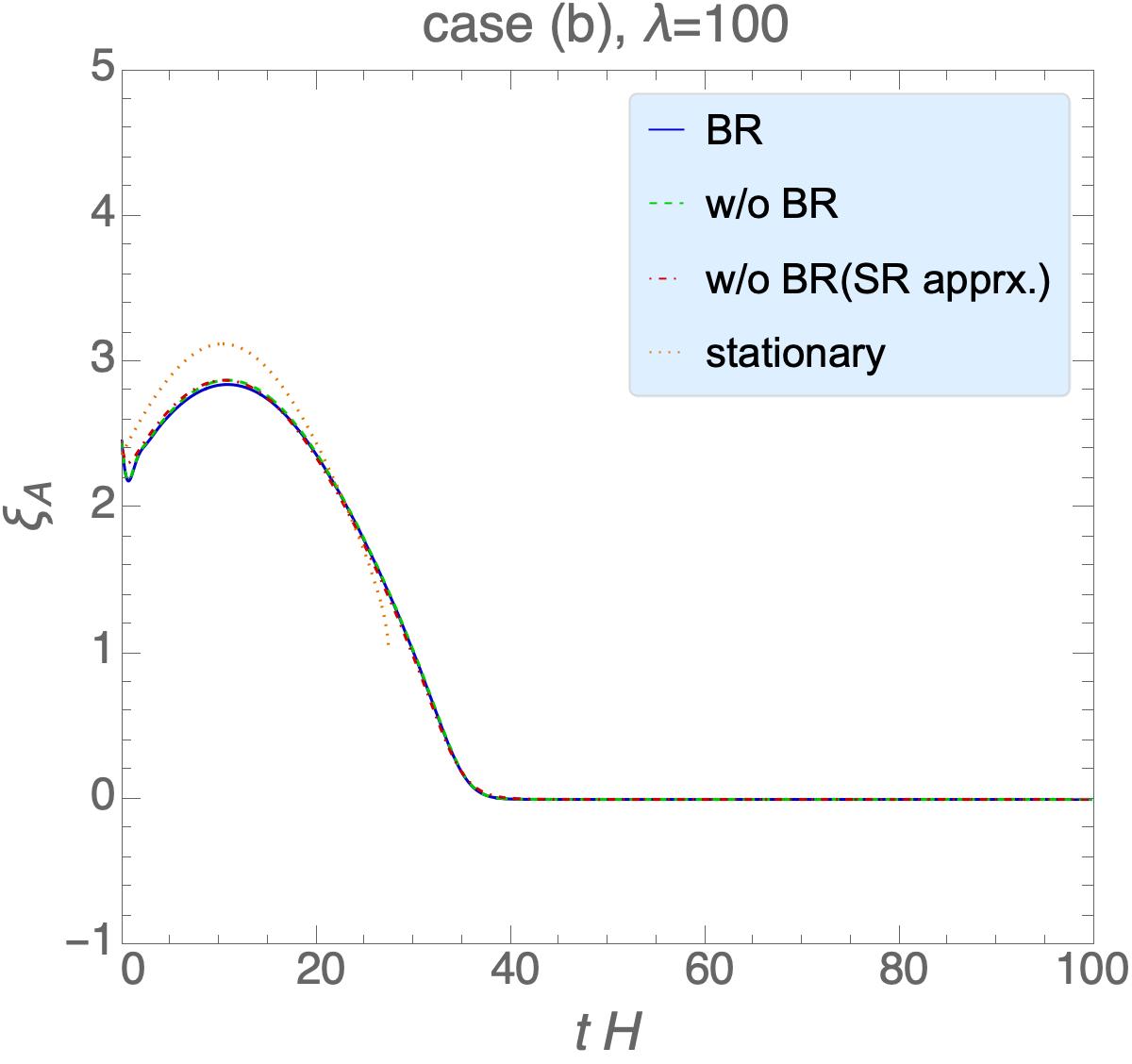}
     \caption{Same as Fig.\,\ref{fig:dy_a_lam100} but for case (b).}
  \label{fig:dy_b_lam100}
 \end{center}
\end{figure}

We now discuss dynamics of the fields. Figs.\,\ref{fig:dy_a_lam100} and \ref{fig:dy_b_lam100} show the time evolution of the axion-gauge system for case (a) and (b), respectively, for $\lambda=100$
(see Figs.\,\ref{fig:dy_a_lam050_500} and \ref{fig:dy_b_lam050_500} in Appendix~\ref{app:figs} for $\lambda=50$ and 500, respectively). The parameters are the same as those in
Fig.\,\ref{fig:potential_eg} and we take an initial condition
$\chi_{\rm ini}=\chi_{0.3}$ and $\xi_{A,\rm
  ini}=1.1\xi_{A0}(\chi_{0.3})$. For case (a), we find that the
time evolution of $\chi$ and $\xi_A$ coincides precisely with that without backreaction (see ``BR'' and ``w/o BR'' in the figure). This can
be understood from the values of ${\cal I}$ and $\kappa/\xi_{A0}^2$ in the
present choice of parameters: Since $({\cal I},\,\kappa/\xi_{A0}^2)\simeq
(0.014,0.10)$ and $(0.014,11)$ for case (a)
and (b), respectively, the backreaction terms are subdominant.  
Here we have taken $\chi=\chi_{0.3}$.
We also find that the solution
agrees with that in the slow-roll approximation.

Let us turn our attention to the new solution, i.e., case (b).
As $\xi_A$ relaxes to the stationary point, which is
approximately given in Eq.\,\eqref{eq:xiA0_2}, $\chi$ slowly rolls down on
the potential. We find that the qualitative behaviors of the fields are
similar to those for case (a), while they give quantitatively different
observational consequences, e.g., the spectrum of the gravitational waves, which
we discuss in the next section.

We find that the slow-roll approximation tends to be unreliable for smaller values of $\lambda$.  Even in the case where the
slow-roll approximation is valid, both fields relax too quickly to the global
minimum after $\order{1}$ Hubble time. This is true for both case (a)
and (b).  Recalling that the effective potential and the allowed
parameter space are less dependent on the value of $\lambda$, one may
wonder why dynamics of $\chi$ and
$\xi_A$ depends on $\lambda$. This can be understood from a factor of
$1/\lambda$ in $F_\chi(\xi_A,\chi)$ (Eq.~\eqref{eq:Fchi}).
We have also numerically checked that
$F_\chi(\xi_A,\chi)$ is smaller for larger $\lambda$, for which $\chi$
tends to slowly roll to the minimum and the slow-roll behavior of
$\xi_A$ is obtained. 

Such large values of $\lambda$ are difficult to realize in a more fundamental theory~\cite{Agrawal:2018mkd}. This issue is specific to the cosine potential and the canonical kinetic term for $\chi$, and reasonable values of $\lambda\simeq \order{1}$ are possible for axion monodromy potentials~\cite{Maleknejad:2016qjz} or non-canonical kinetic terms~\cite{Watanabe:2020ctz}.

\section{Gravitational waves}
\label{sec:gw}

\begin{figure}
  \begin{center}
    \includegraphics[width=6.5cm]{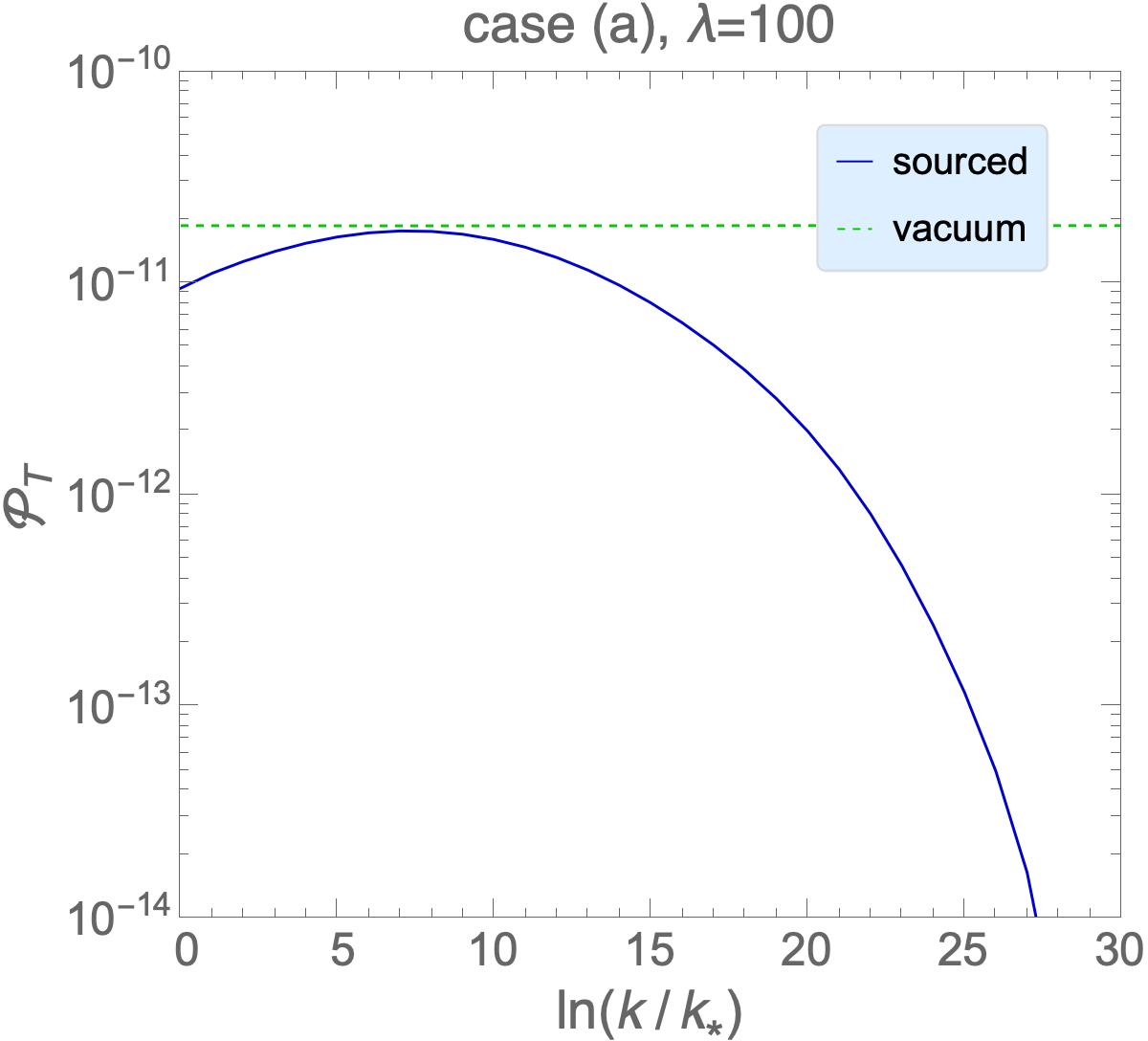}
    \includegraphics[width=6.5cm]{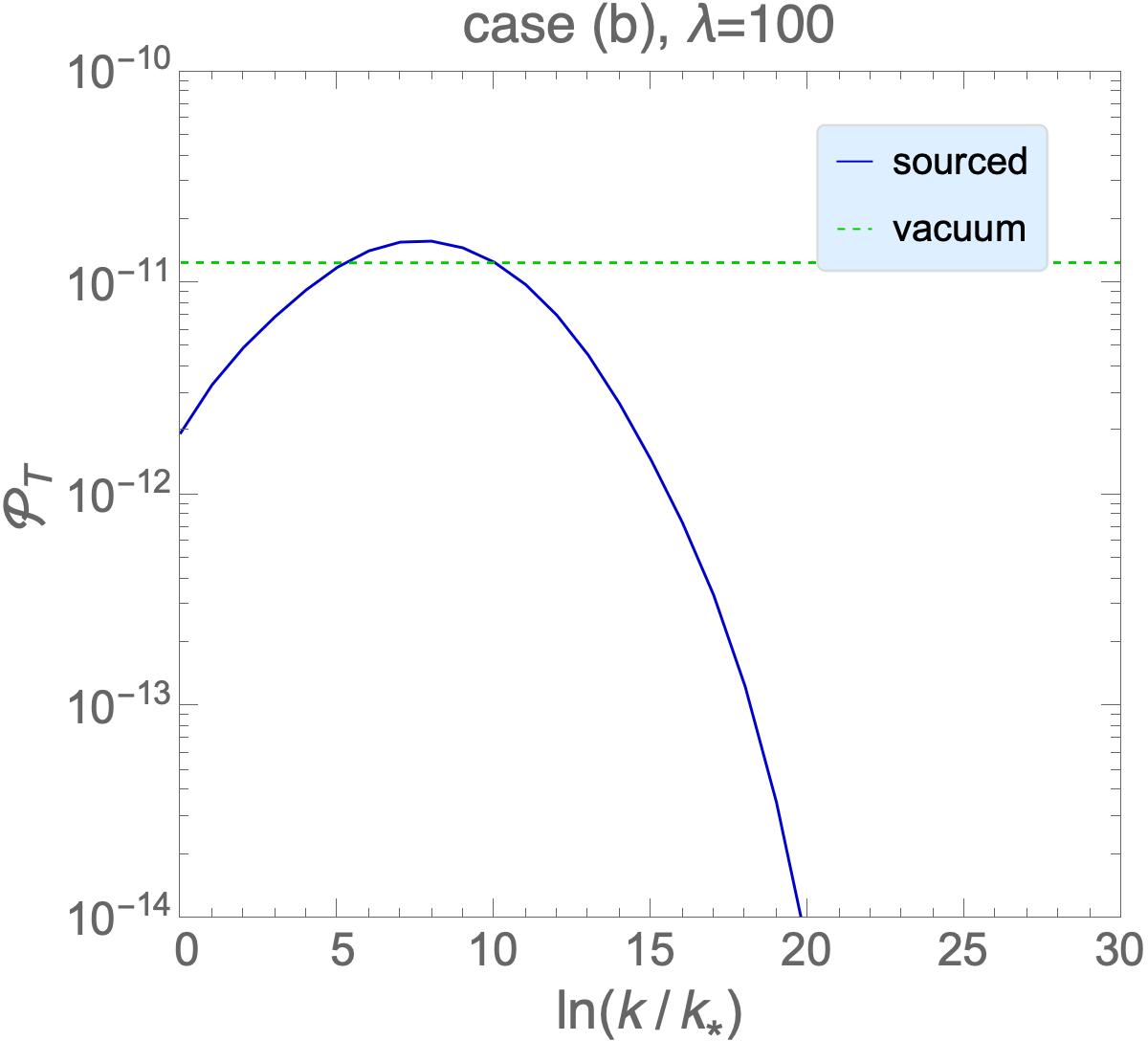}
    \caption{Power spectra of the gravitational
      waves sourced by the gauge field, ${\cal P}_T^{\rm (s)}$. The parameters are the same as those in Fig.\,\ref{fig:dy_a_lam100} (left) and
      Fig.\,\ref{fig:dy_b_lam100} (right). $k_*$ is some reference scale.}
  \label{fig:PT_lam_100}
 \end{center}
\end{figure}

The power spectrum of primordial gravitational waves is given by the sum of two contributions: the quantum vacuum fluctuation in spacetime, ${\cal P}_T^{\rm (v)}$~\cite{Grishchuk:1974ny,Starobinsky:1979ty}, and the sourced contribution, ${\cal P}_T^{\rm (s)}$~\cite{Adshead:2013qp,Dimastrogiovanni:2012ew,Maleknejad:2012fw}:
\begin{align}
  {\cal P}_T= {\cal P}_T^{\rm (v)} +   {\cal P}_T^{\rm (s)}\,,
\end{align}
where \cite{Lyth:1998xn}
\begin{align}
{\cal P}_T^{\rm (v)}=\frac{2H^2}{\pi^2}\,.
\end{align}
In Refs.\,\cite{Maleknejad:2018nxz,Fujita:2018ndp} ${\cal P}_T^{\rm (s)}$ is computed by both analytical and numerical methods. For instance, the maximum amplitude for the case (a) is estimated as
\begin{align}
  {\cal P}_{T\,{\rm est}}^{\rm (s)\,max}\approx
  \frac{H^4\xi_{A}^4e^{p\xi_{A}}}{\pi^2g_A^2}\,,
  \label{eq:PTes}
\end{align}
where $p\approx 3.6$\,--\,4 and $\xi_A$ is to be the maximum value
in the time evolution.

In Fig.~\ref{fig:PT_lam_100}, we show ${\cal P}_T^{\rm (s)}$
computed by numerically solving the equations of the tensor
perturbation given in Appendix of Ref.\,\cite{Fujita:2018ndp}. 
We find that the
the peak value obtained from the numerical calculation can be different
from the estimate given in Eq.\,\eqref{eq:PTes} as $\xi_A$ becomes large. 
Although it does not change the
order of magnitude, the estimate is not accurate enough for quantitative study. 
In Appendix~\ref{app:figs}, we show the power spectra for $\lambda=50$ and 500 in Fig.\,\ref{fig:PT_lam_050_500}.

\begin{figure}
  \begin{center}
    \includegraphics[width=7cm]{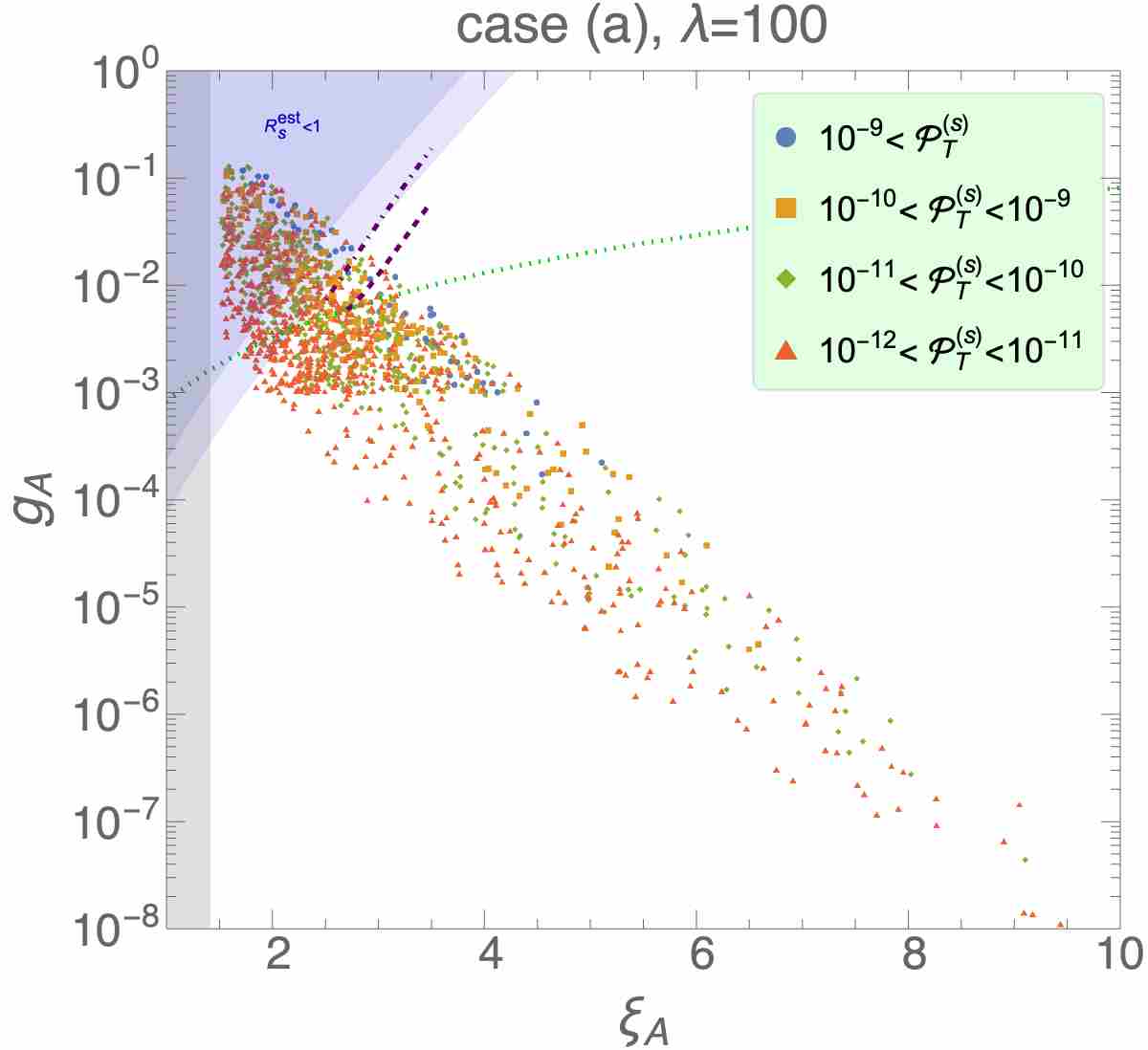}
    \includegraphics[width=7cm]{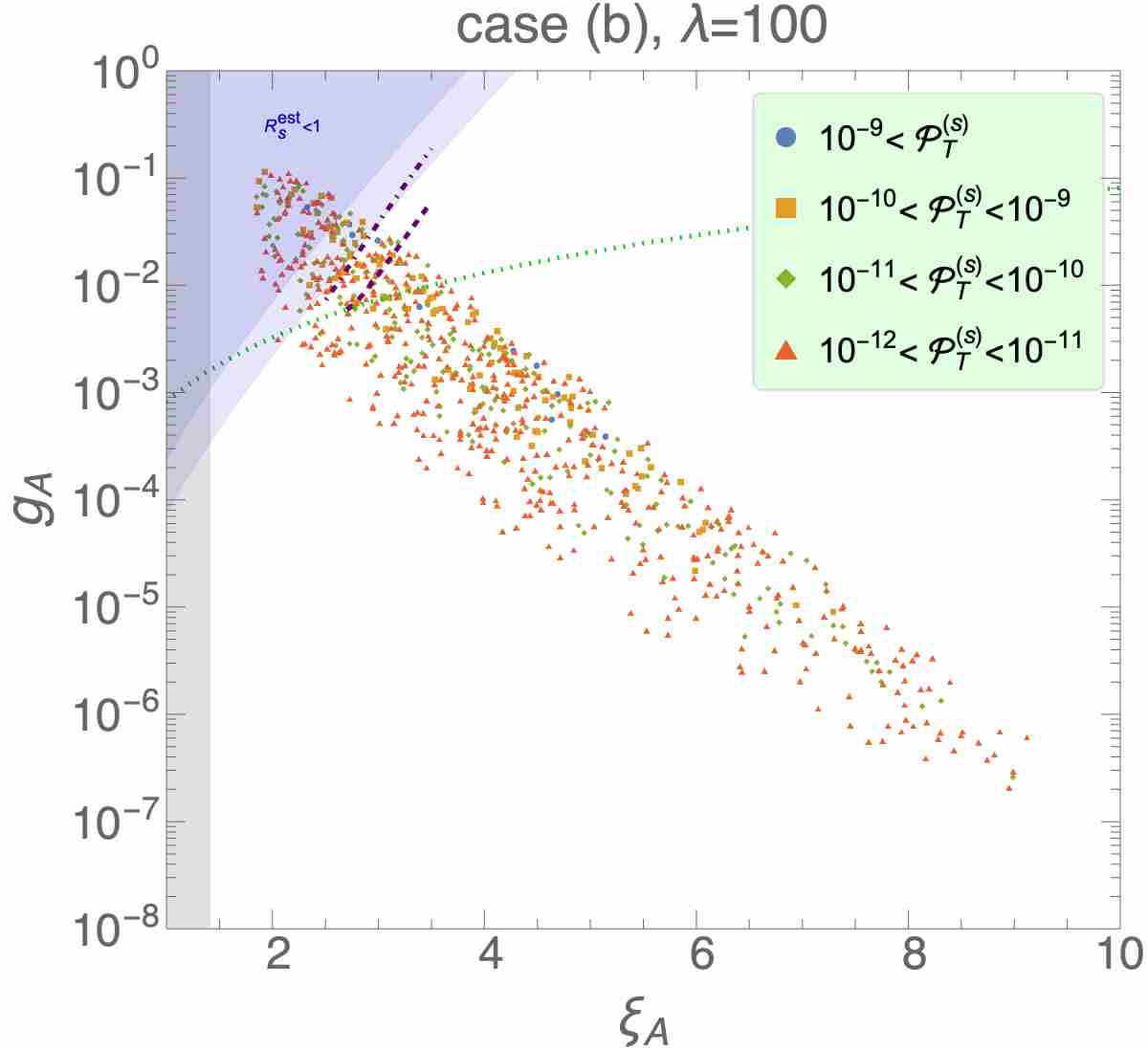}
    \caption{Values of ${\cal P}_T^{\rm (s)\,max}$ on $(\xi_{A0}(\chi_{0.5}),g_A)$
      plane.   The shaded region shows
      $\xi_A<\sqrt{2}$ which we discard in the current study. 
      The dark and light blue shaded regions show $R_s^{\rm est}<1$ for the vacuum tensor-to-scalar ratios of $r_{\rm v}={\cal
  P}_T^{\rm (v)}/{\cal P}_\zeta^{\rm obs}=0.1$ and $0.01$, respectively. 
      The dotted green line shows $g_A>g_A^{\rm min}$ required from the observed scalar power spectrum on CMB scales, ${\cal
  P}_\zeta^{\rm obs}\simeq 2.1\times
10^{-9}$.
      The dot-dashed and dashed purple lines show lower bounds on $g_A$ 
      from ${\cal R}_{\delta \phi}<0.1$ for $r_{\rm v}=0.1$ and 0.01, respectively. We take $\epsilon_B>\epsilon_\phi$ and $N_k=10$ when computing ${\cal R}_{\delta \phi}$.}
  \label{fig:GW_lam_100_g}
 \end{center}
\end{figure}

In our analysis, we solve the equations for the tensor perturbation and calculate the maximum tensor spectrum sourced by the gauge field. After computing the tensor spectrum, we select the data points that give ${\cal P}_T^{\rm (s)\,max}\ge 10^{-12}$, which is a reference value for the possibility to detect in future experiments~\cite{SimonsObservatory:2018koc,CMB-S4:2016ple,Hazumi:2019lys,NASAPICO:2019thw}. In Fig.\,\ref{fig:GW_lam_100_g}, we show the resultant scatter plot on
$(\xi_{A0}(\chi_{0.5}),g_A)$ plane. We find that ${\cal P}_T^{\rm (s)\,max}\ge 10^{-12}$ is possible in wide parameter space for both cases (a) and (b).\footnote{For case (a) the data points are denser in the region where $g_A\gtrsim 10^{-3}$, as we performed an additional parameter search to obtain more points in this region.}

\begin{figure}
  \begin{center}
    \includegraphics[width=7cm]{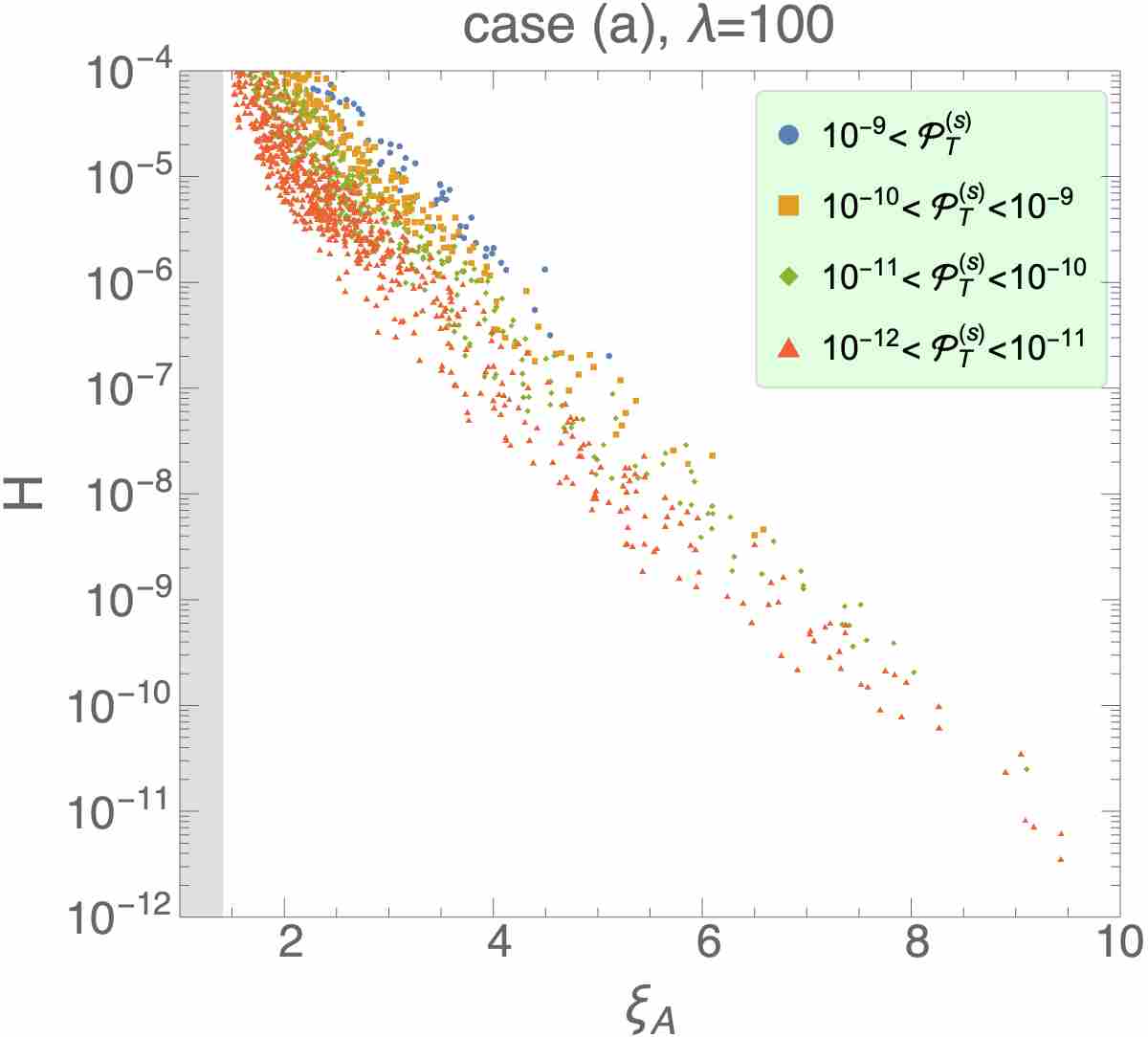}
    \includegraphics[width=7cm]{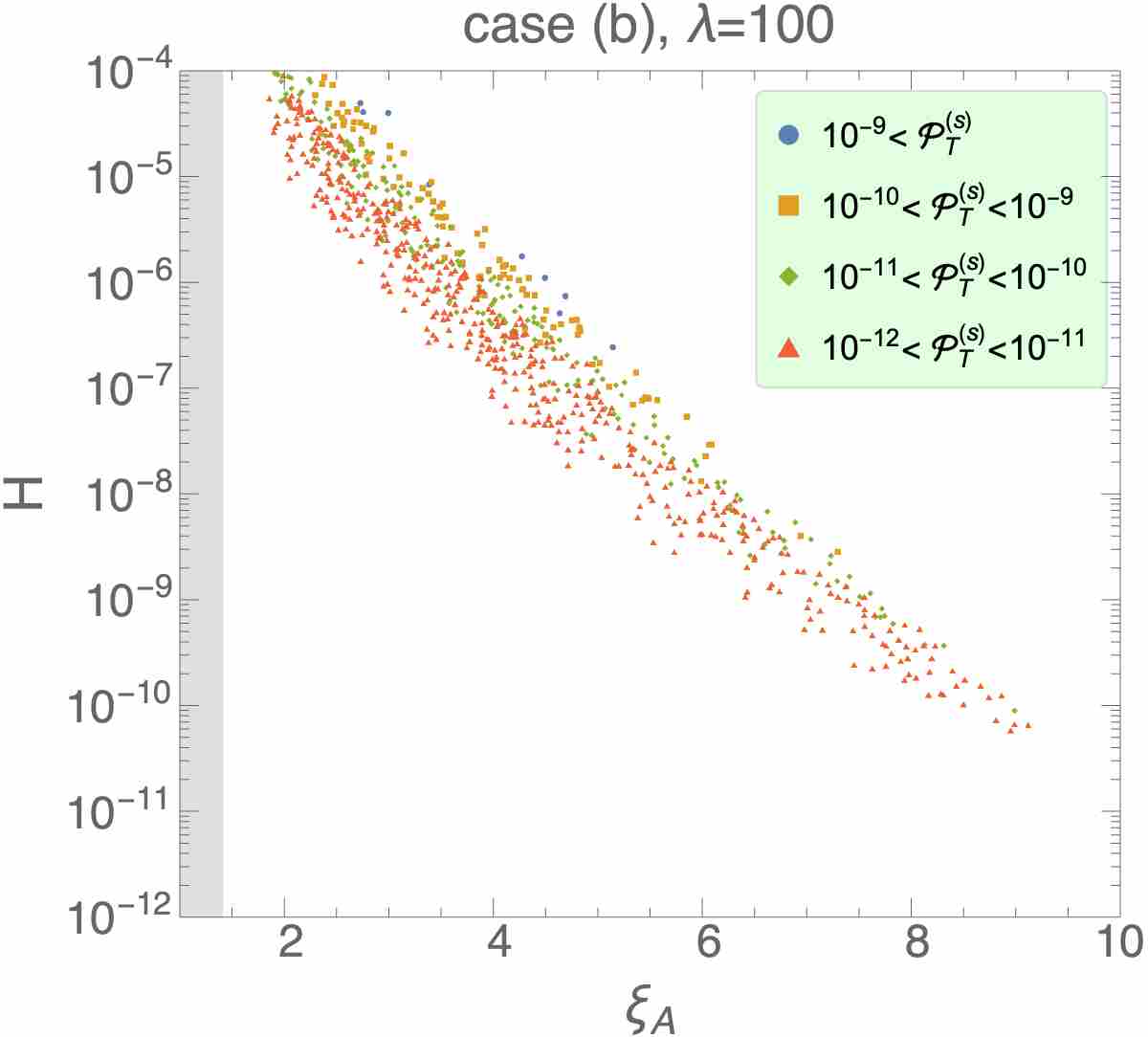}
    \includegraphics[width=7cm]{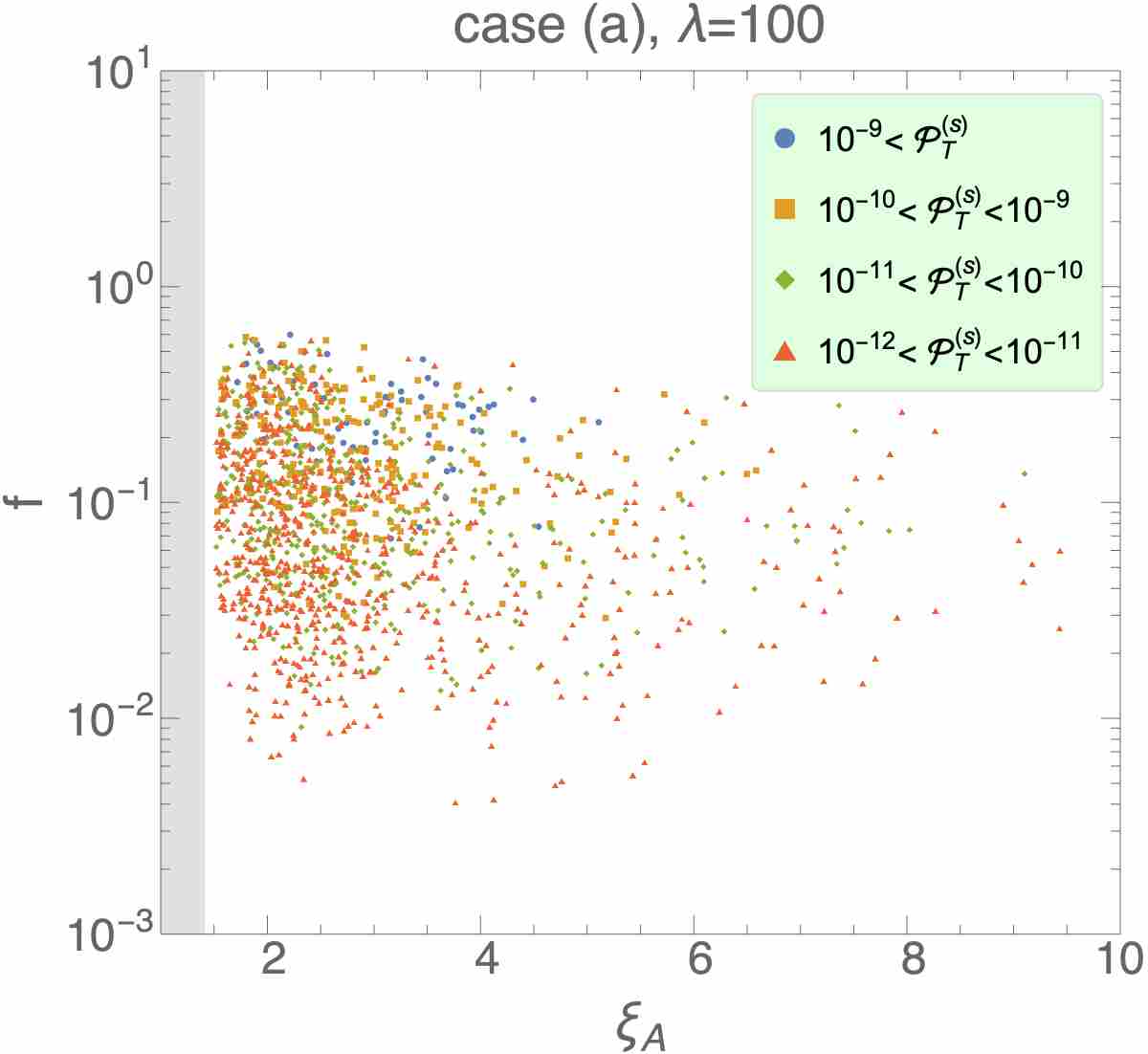}   
    \includegraphics[width=7cm]{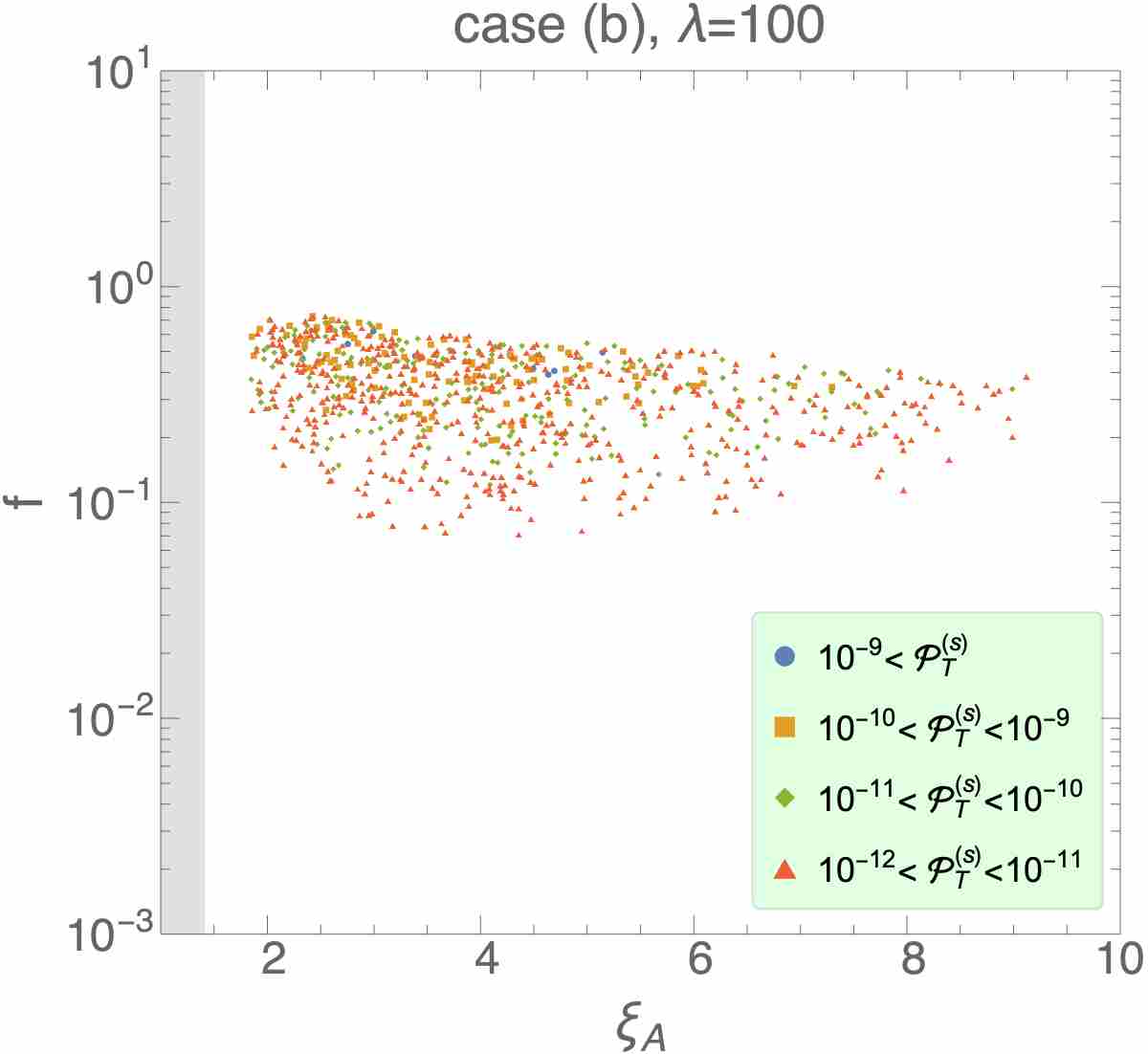}
    \includegraphics[width=7cm]{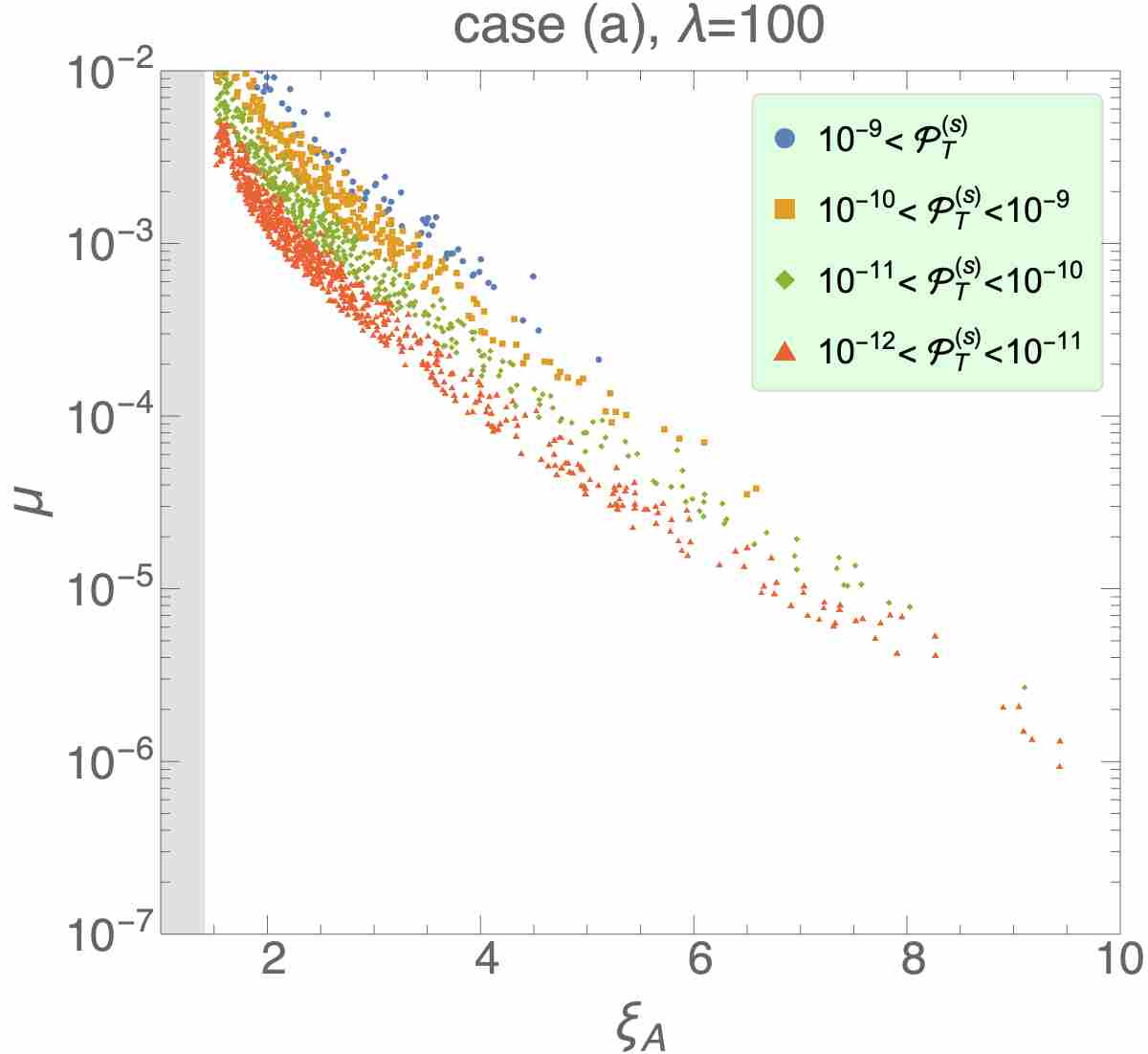}   
    \includegraphics[width=7cm]{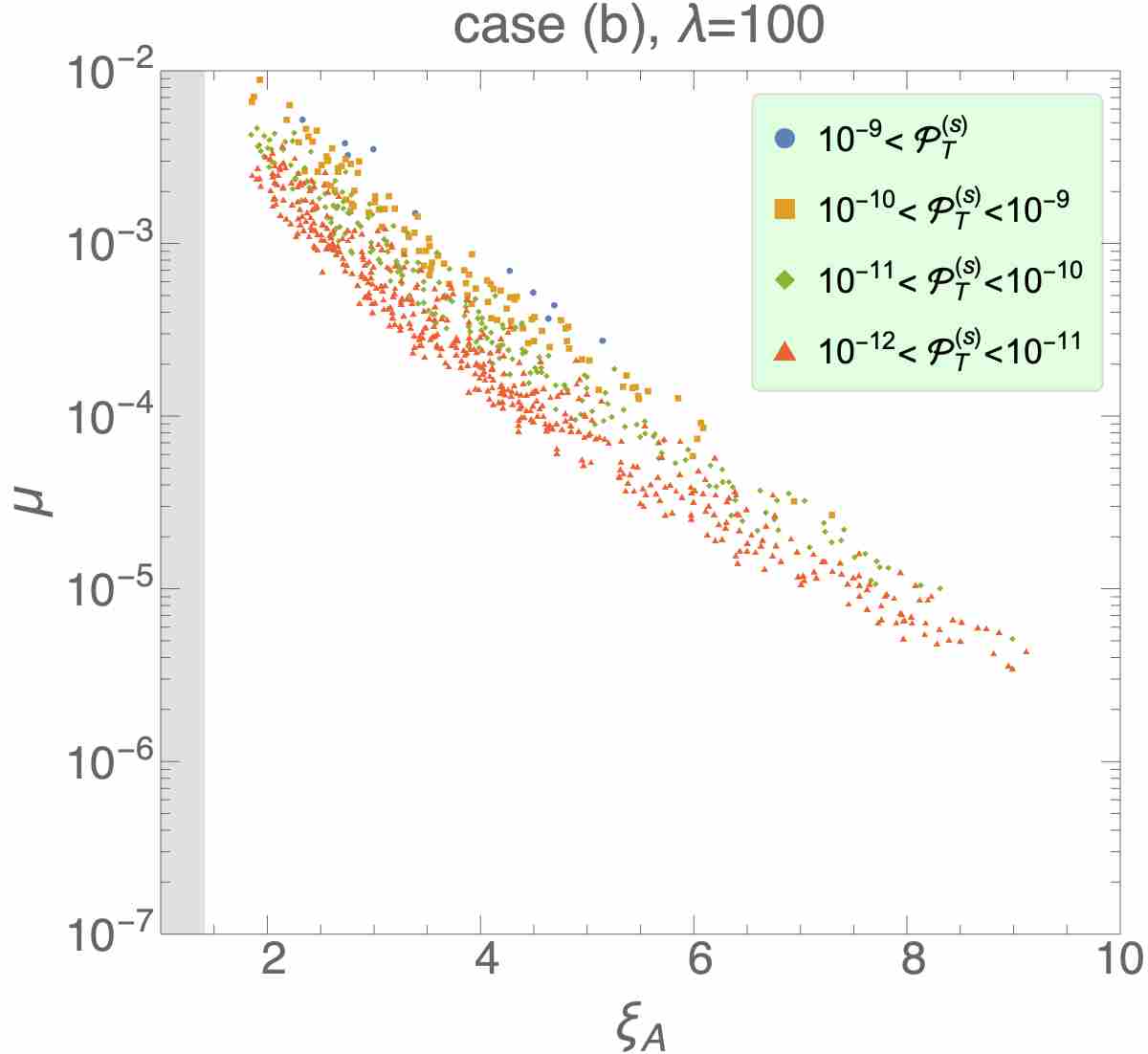}
    \caption{Same as Fig.\,\ref{fig:GW_lam_100_g} but on $(\xi_{A0}(\chi_{0.5}),H)$
      , $(\xi_{A0}(\chi_{0.5}),f)$, and $(\xi_{A0}(\chi_{0.5}),\mu)$ planes.  }
  \label{fig:GW_lam_100_Hmu}
 \end{center}
\end{figure}

Fig.\,\ref{fig:GW_lam_100_Hmu} shows the scatter plots with different
vertical axes, $H$, $f$ and $\mu$. 
We find larger sourced gravitational
waves for larger values of $H$ and $\mu$. Regarding $f$,
on the other hand, we find $f\sim 10^{-2}$\,--\,$1$ and $f\sim
10^{-1}$\,--\,$1$ for the case (a) and (b), respectively, and the
amplitude of the gravitational waves is independent on $f$ (see Figs.\,\ref{fig:GW_lam_050_g}\,--\,\ref{fig:GW_lam_500_Hmu} in Appendix~\ref{app:figs} for $\lambda=50$ and 500).

We now discuss the viable parameter space for large sourced gravitational waves on the scales constrained by CMB observations. We require the amplitude of the scalar
perturbation ${\cal P}_\zeta$ coincide with the observed value ${\cal
  P}_\zeta^{\rm obs}\simeq 2.1\times
10^{-9}$~\cite{Planck:2018vyg}. In the current model,
${\cal P}_\zeta$ is given by
\begin{align}
  {\cal P}_\zeta&\simeq \frac{H^2\epsilon_\phi}{8\pi^2 \epsilon_H^2}
  \simeq \frac{H^2\epsilon_\phi}{8\pi^2 (\epsilon_\phi+\epsilon_B)^2}\,,
\end{align}
where $\epsilon_H=\epsilon_\phi+\epsilon_\chi+\epsilon_B+\epsilon_E$ and 
\begin{align}
  \epsilon_H=-\frac{\dot{H}}{H^2}\,,~~
  \epsilon_\phi=\frac{\dot{\phi}^2}{2H^2}\,,~~
  \epsilon_\chi=\frac{\dot{\chi}^2}{2H^2}\,,~~
  \epsilon_B=\frac{H^2\xi_A^4}{g_A^2}\,,~~
  \epsilon_E=\frac{H^2(\xi_A+\xi_A')^2}{g_A^2}\,.
\end{align}
Here we have checked that $\epsilon_\chi$ and $\epsilon_E$ are
subdominant in $\epsilon_H$. Therefore,
\begin{align}
  \frac{g_A^2}{8\pi^2\xi_A^4}
  \frac{\epsilon_B/\epsilon_\phi}{(1+\epsilon_B/\epsilon_\phi)^2}
  ={\cal P}_\zeta^{\rm obs}\,,
  \label{eq:Pzeta=Pobs}
\end{align}
is satisfied, and a lower bound on $g_A$,
\begin{align}
  g_A\ge g_A^{\rm min}\simeq \sqrt{32\pi^2 {\cal P}_\zeta^{\rm obs}}\xi_A^2\,,
\label{eq:g_Alow}
\end{align}
should be satisfied~\cite{Papageorgiou:2019ecb} for $\epsilon_B/\epsilon_\phi$ to have a solution. We show this bound by the green dotted line in
Fig.\,\ref{fig:GW_lam_100_g}. We find that this constraint is satisfied in the upper left corner of the parameter space shown in the figure for both cases (a) and (b).

There is another constraint coming from scale modes non-linearly sourced by tensor modes~\cite{Papageorgiou:2018rfx,Papageorgiou:2019ecb}. Namely, the quantum loop correction to the inflaton fluctuation ${\cal R}_{\delta
  \phi}$ given by
\begin{align}
  {\cal R}_{\delta \phi} \simeq
  \frac{5\times 10^{-12}}{(1+\epsilon_B/\epsilon_\phi)^2}
  e^{7\xi_A}\xi_A^{11}N_k^2r_{\rm v}^2\,,
\end{align}
should be small. Here, $r_{\rm v}={\cal
  P}_T^{\rm (v)}/{\cal P}_\zeta^{\rm obs}$ is the tensor-to-scalar ratio of the vacuum contribution and $N_k$ is the number of $e$-folds of inflation during which $\xi_A$ rolls slowly.  This expression is 
  valid for $2.5\le\xi_A\le 3.5.$
Demanding ${\cal R}_{\delta
  \phi}<0.1$, we obtain lower bounds on $g_A$ as a function of
$\xi_A$. Note that $\epsilon_B/\epsilon_\phi$ is determined from
Eq.\,\eqref{eq:Pzeta=Pobs} depending on $\epsilon_B/\epsilon_\phi>1$
or $\epsilon_B/\epsilon_\phi<1$. In Fig.\,\ref{fig:GW_lam_100_g}, we
show the constraints from ${\cal R}_{\delta \phi}<0.1$ for
$r_{\rm v}=0.1$ (dot-dashed purple line) and $0.01$ (dashed purple line), assuming $N_k=10$ and $\epsilon_B/\epsilon_\phi>1$. This restricts the parameter space further in the upper left corner.

Can the sourced contribution exceed the vacuum one? 
In Fig.\,\ref{fig:GW_lam_100_g}, we show the parameter space in which
the ratio of the sourced and vacuum tensor modes,
$R_s^{\rm est}$, defined by
\begin{align}
   R_{\rm s}^{\rm est}\equiv
  \frac{{\cal P}_{T\,{\rm est}}^{\rm (s)\,max}}{{\cal P}_T^{\rm (v)}}\,,
  \label{eq:Rsest}
\end{align}
is less than unity. Here we use an estimation for ${\cal P}^{(s)\,{\rm max}}_{T\,{\rm
    est}}$ given in Eq.\,\eqref{eq:PTes}, instead of the numerical solution.
We find a parameter space in which $R_s^{\rm est}>1$ and all the constraints discussed above are satisfied. This is the parameter space most relevant to future CMB experiments~\cite{SimonsObservatory:2018koc,CMB-S4:2016ple,Hazumi:2019lys,NASAPICO:2019thw}, which we focus now.

\begin{figure}
  \begin{center}
    \includegraphics[width=7cm]{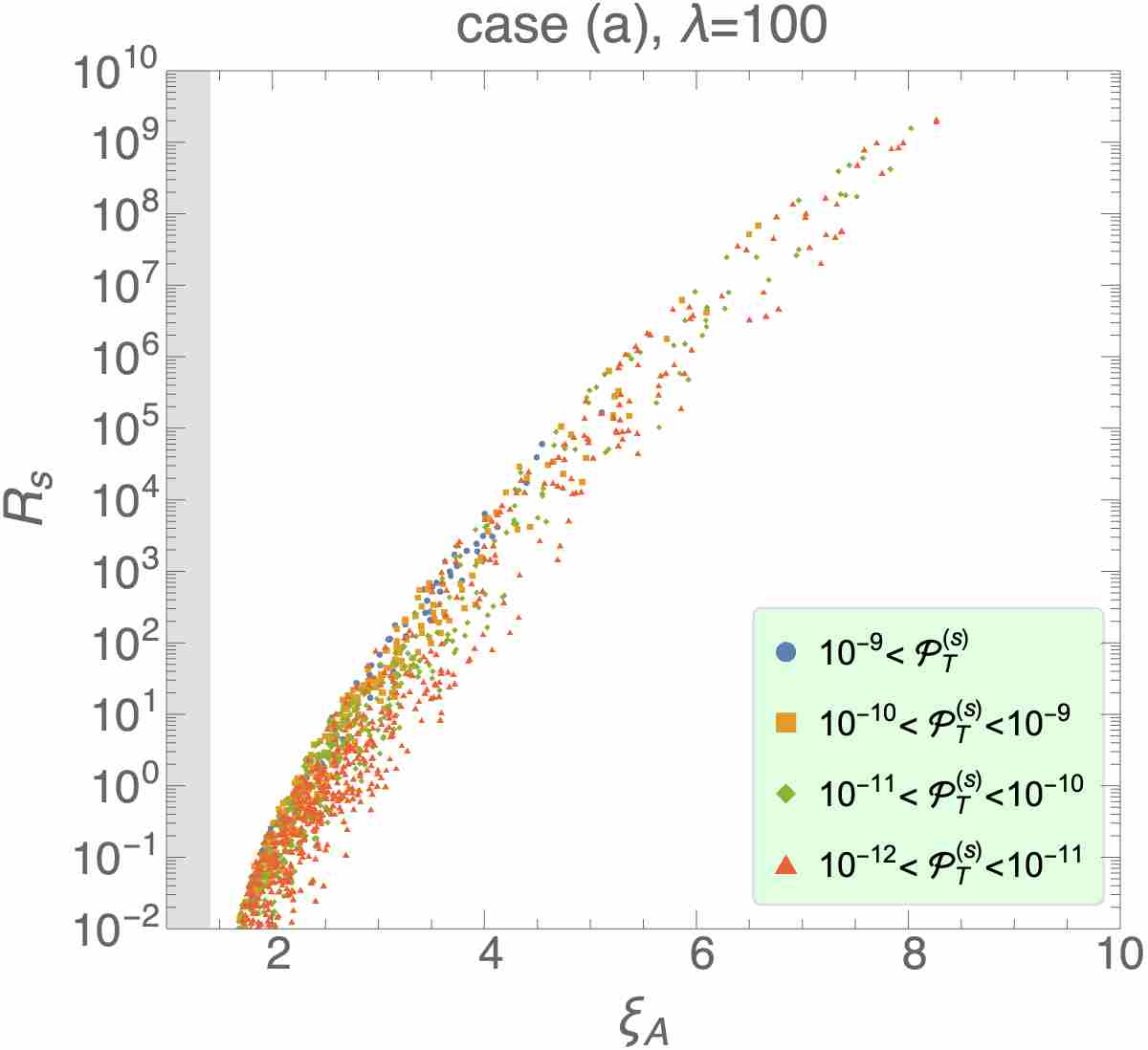}
    \includegraphics[width=7cm]{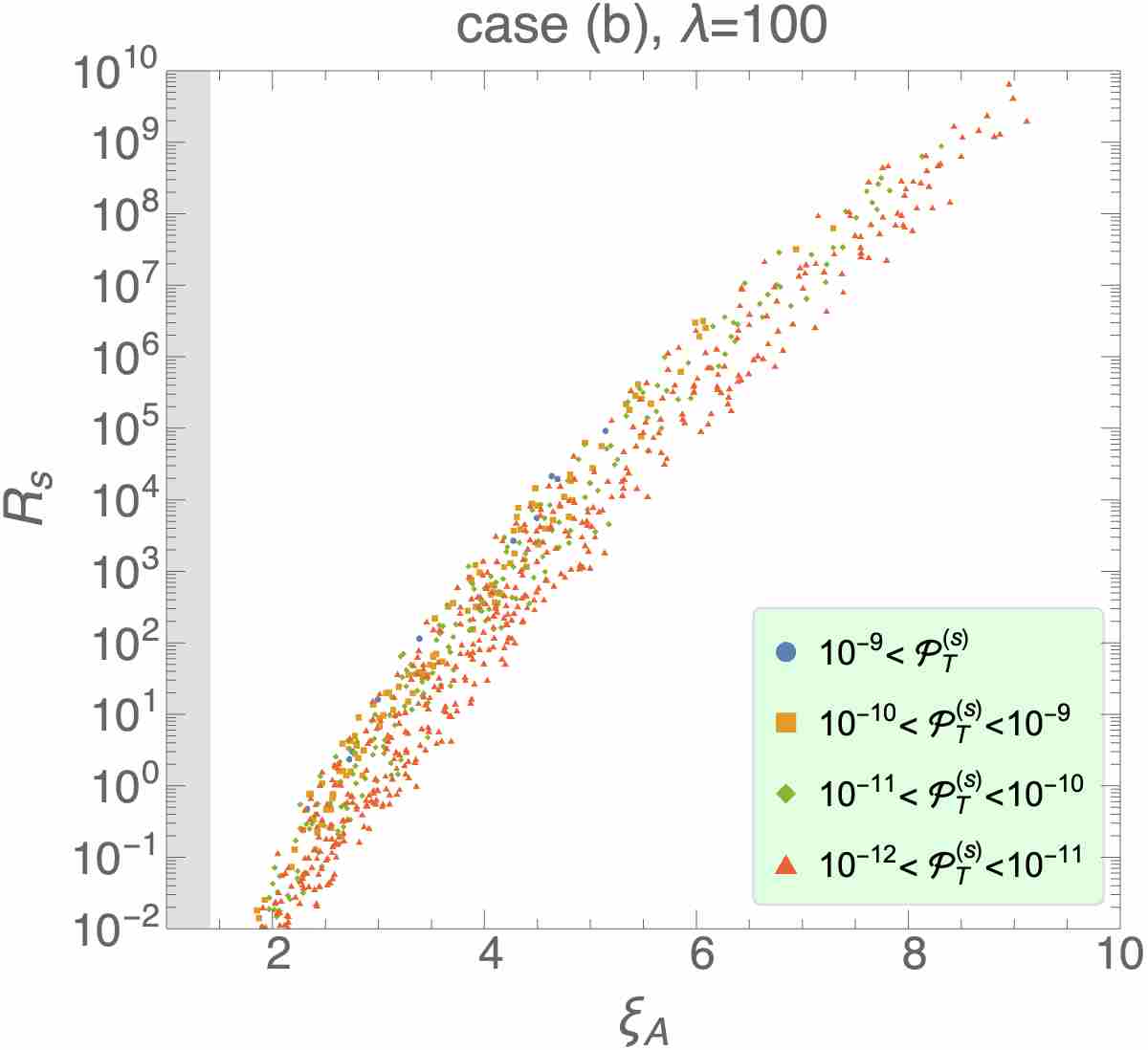}
    \includegraphics[width=7cm]{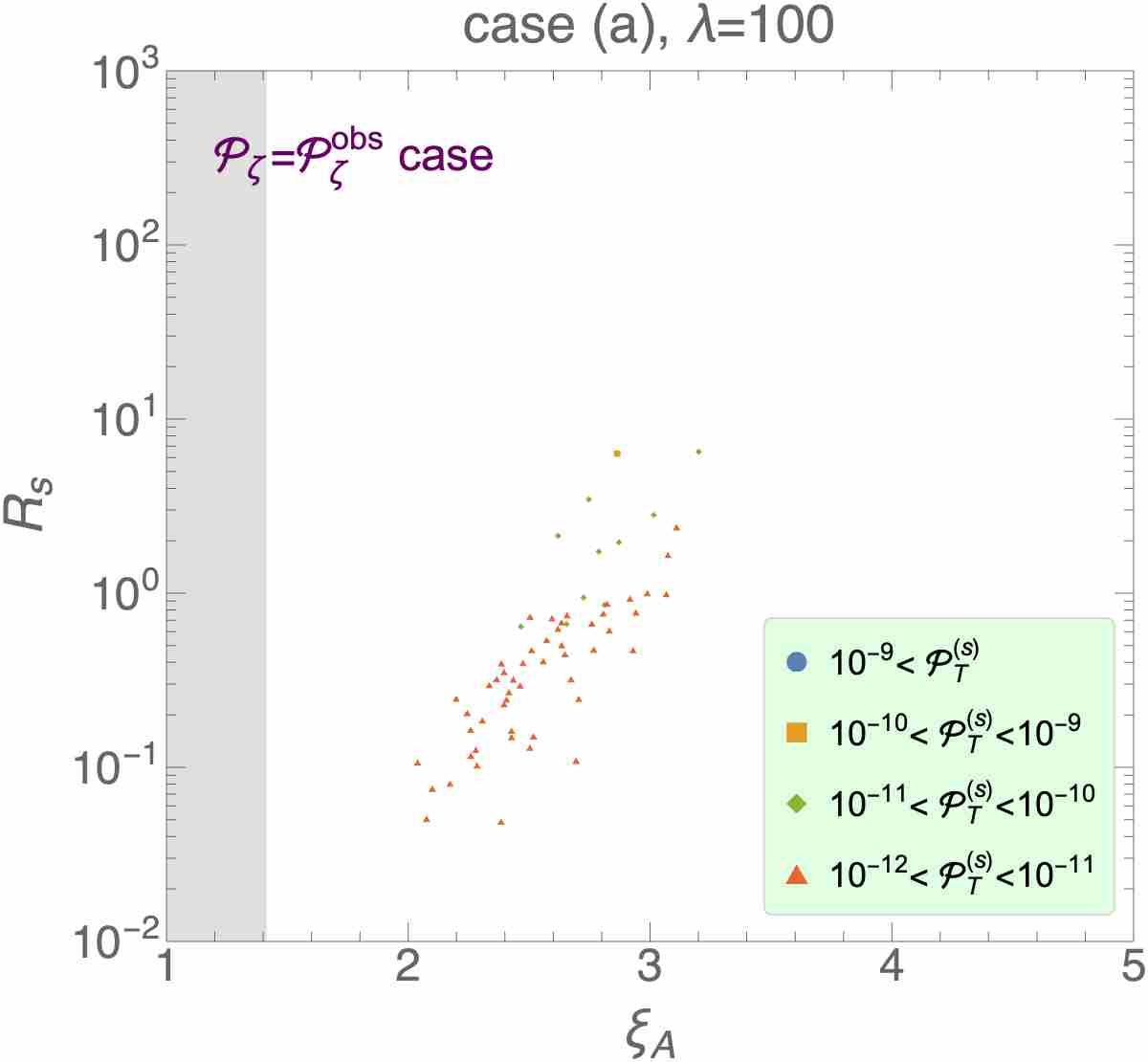}   
    \includegraphics[width=7cm]{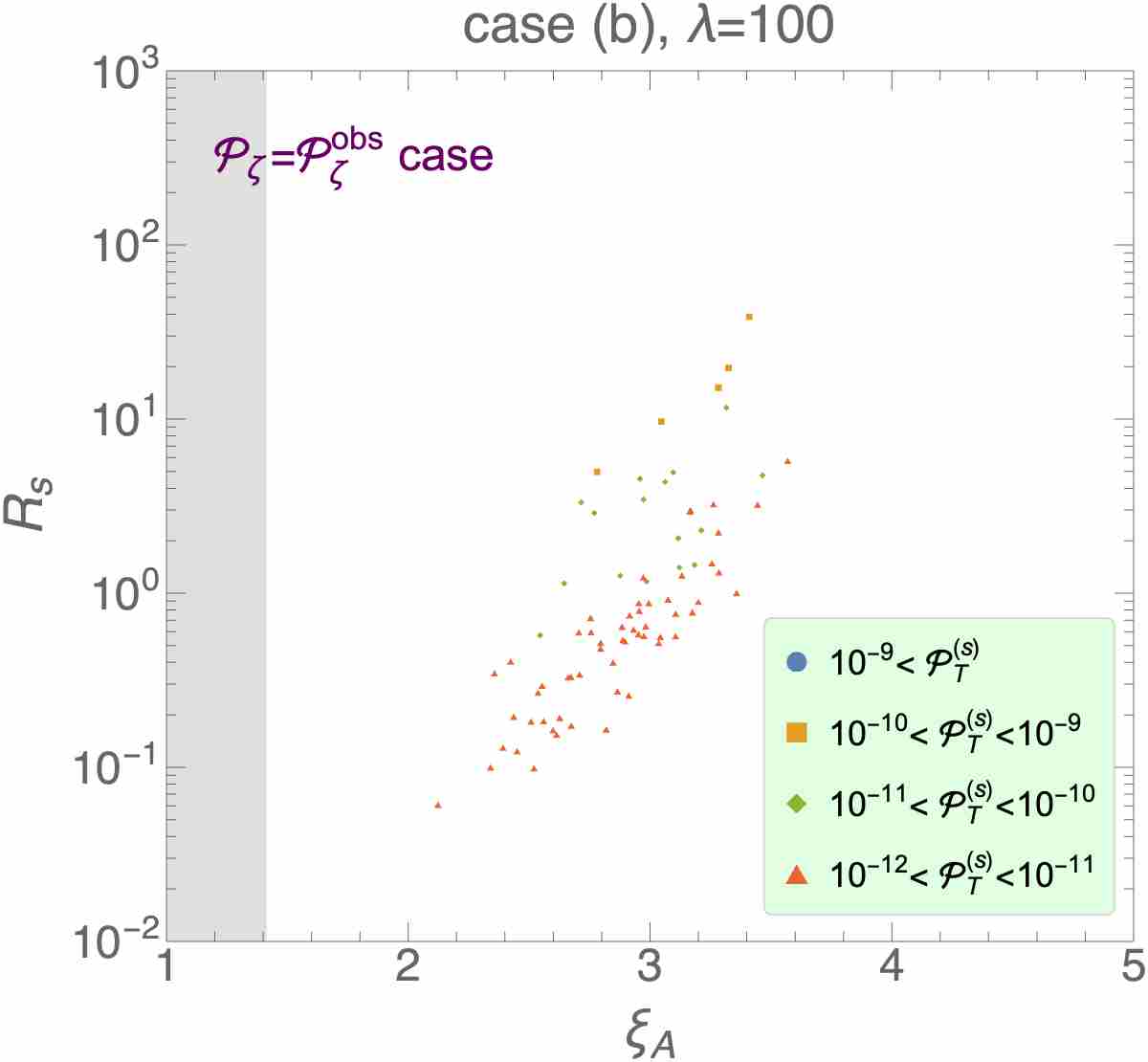}
    \includegraphics[width=7cm]{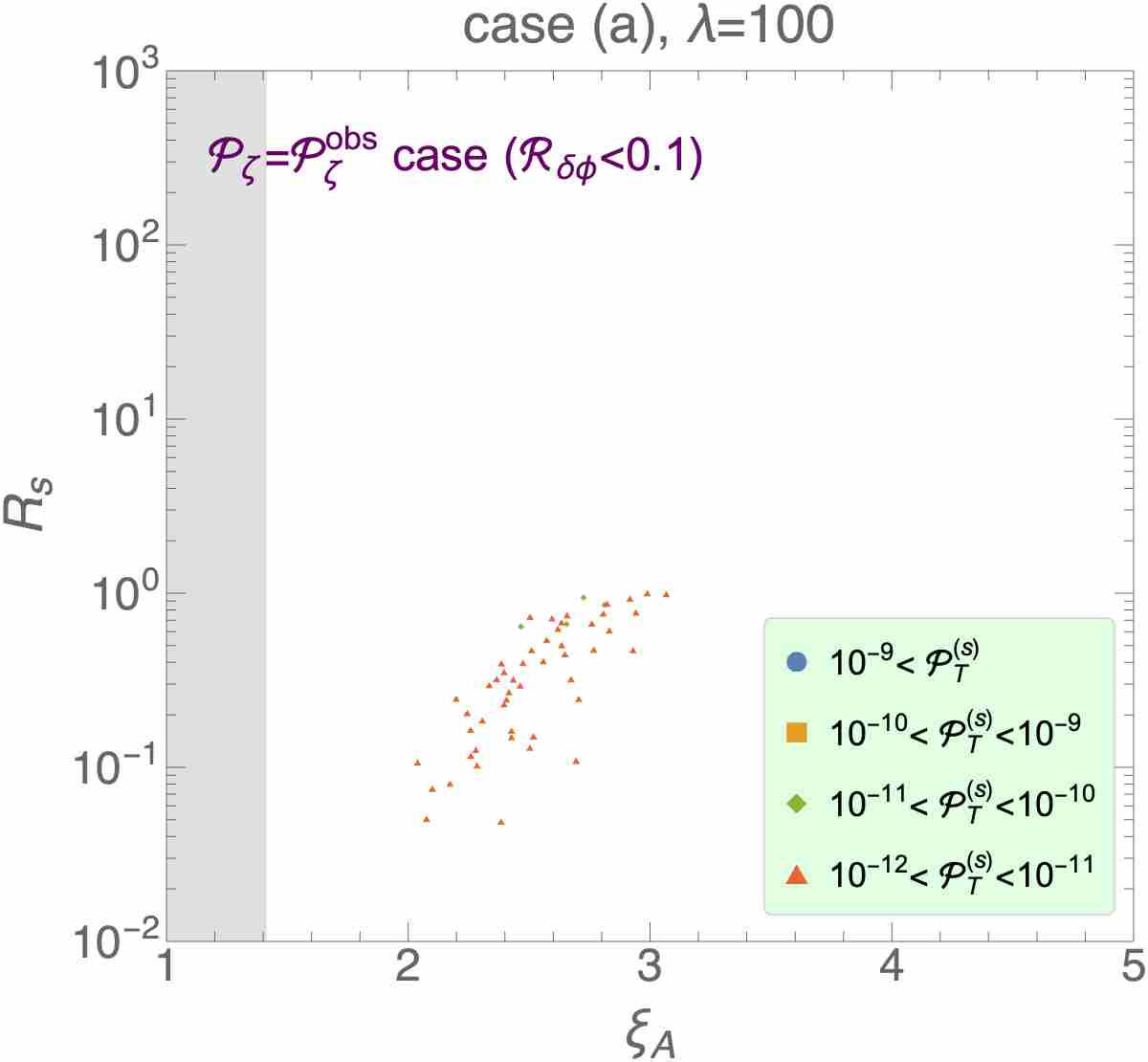}   
    \includegraphics[width=7cm]{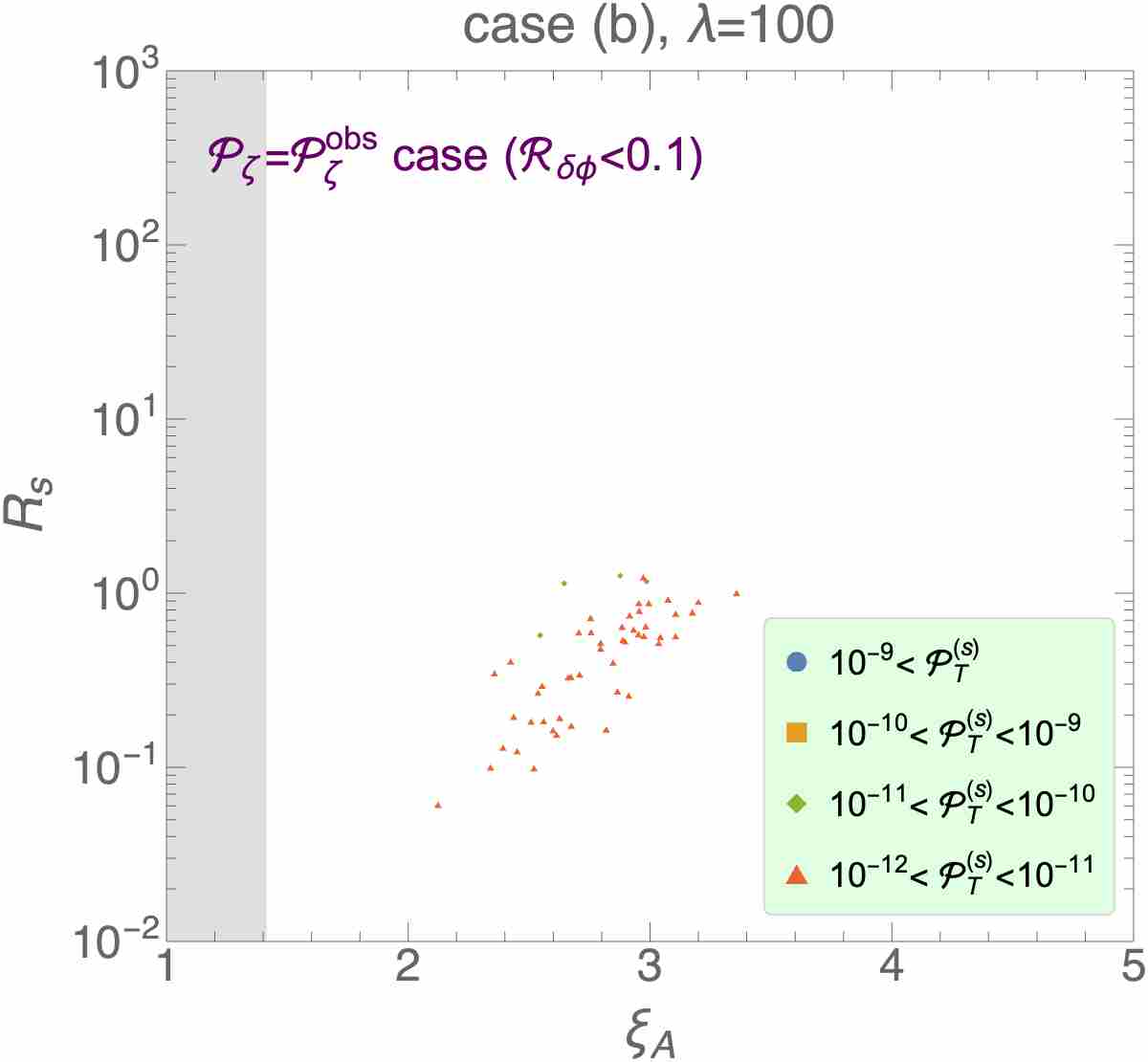}
    \caption{Same as Fig.\,\ref{fig:GW_lam_100_g} but on
      $(\xi_A,R_s)$ plane (top). The middle  and bottom panels are the same as the top ones
      but for ${\cal P}_\zeta={\cal P}_\zeta^{\rm obs}$ and $r_{\rm v}<0.036$ (middle) and
    ${\cal P}_\zeta={\cal P}_\zeta^{\rm obs}$, ${\cal R}_{\delta \phi}<0.1$ and $r_{\rm v}<0.036$ (bottom). }
  \label{fig:GW_lam_100_Rs}
 \end{center}
\end{figure}

We compute the correct ratio of the sourced and vacuum
tensor modes,
\begin{align}
  R_{\rm s}\equiv
  \frac{{\cal P}_{T}^{\rm (s)\,max}}{{\cal P}_T^{\rm (v)}}\,,
\end{align}
 by solving the tensor mode equations numerically, rather than using an estimate given in Eq.\,\eqref{eq:PTes}.
The results are shown in Fig.\,\ref{fig:GW_lam_100_Rs} for $\lambda=100$ (see Figs.\,\ref{fig:GW_lam_050_Rs} and \ref{fig:GW_lam_500_Rs} in Appendix~\ref{app:figs} for $\lambda=50$ and $500$, respectively).
We find that the parameter space giving $R_{\rm s}>1$ decreases as we impose more constraints coming from observations.
The scalar power spectrum amplitude constraint, ${\cal
  P}_\zeta^{\rm obs}\simeq 2.1\times
10^{-9}$~\cite{Planck:2018vyg}, and the upper bound on the vacuum tensor-to-scalar ratio, $r_\mathrm{v}<0.036$~\cite{BICEP:2021xfz},
allow for $R_{\rm s}\sim \order{10}$. This is the most robust constraint applicable to CMB observations at all multipoles. The tensor power spectrum from the gauge field with a cosine potential has a bump as shown in Fig.~\ref{fig:PT_lam_100}, whereas the vacuum contribution is nearly scale invariant. Therefore, the vacuum contribution must satisfy the current upper bound on $r$ from the CMB polarization measurement at $\ell\simeq 80$. On the other hand, the sourced contribution can avoid $\ell\simeq 80$ and make a significant contribution at lower (e.g., ``reionization bump'' at $\ell\lesssim 10$) or higher multipoles, without violating the current upper bound on $r$.

The non-linearly sourced scalar modes reduce the allowed value of $R_{\rm s}$ further to $\order{1}$. This constraint applies to CMB measurements at high multipoles, where the power spectrum and non-Gaussiainity of scalar modes are constrained precisely. The constraint is much weaker at $\ell\lesssim 10$ due to large cosmic variance of CMB temperature anisotropy data: for ${\cal R}_{\delta \phi}<1$, we find that $R_{\rm s}\simeq 5$ is still allowed. 
We thus conclude that $R_{\rm s}\gtrsim 1$ is possible on the CMB scales without violating slow-roll dynamics and the current observational constraints on scalar and tensor modes.

These stringent constraints do not apply to other probes of primordial gravitational waves such as pulsar timing arrays and laser/atomic interferometers, as they probe much higher frequencies than accessible by CMB observations. Therefore, $R_\mathrm{s}\gg 1$ is possible for these probes~\cite{Thorne:2017jft,Campeti:2020xwn}.

\section{Conclusions}
\label{sec:conclusion}

We have studied the axion-gauge field dynamics with backreaction
in the spectator axion-SU(2) model of Ref.~\cite{Dimastrogiovanni:2016fuu}.
We solved the equations of motion for the axion and gauge fields with backreaction.
Unlike the previous study based on the order-of-magnitude estimation or a numerical calculation for a particular choice of the parameters, we provided the accurate condition, ${\cal I}<0.1$ (Eq.~\eqref{eq:R}), which guarantees stable slow-roll dynamics in the comprehensive parameter space.

We found a new slow-roll solution, ``case (b)'' (Eq.~\eqref{eq:xiA0_2}), in addition to the known one, ``case (a)'' (Eq.~\eqref{eq:xiA0_1}). The new solution shares similar features with the known one, but quantitative results are different. We then computed the power spectra of gravitational waves sourced by
the gauge field. We found that the sourced gravitational
waves can be much larger than the vacuum contribution, $R_{\rm s}\gg 1$, in a wide parameter space without violating slow-roll dynamics.

If we further assume that the axion-gauge dynamics occurs at around the last 50\,--\,60 $e$-folds before the end of inflation, which is relevant to CMB experiments, 
 the parameter space that gives $R_{\rm s}>1$  is constrained by the observed power spectra of scalar and tensor modes and the scalar modes non-linearly sourced by tensor modes. We found that $R_{\rm s}\sim \order{10}$ is possible for the observed scalar and tensor amplitudes, and $R_{\rm s}\sim\order{1}$ is still possible for the additional constraint on non-linearly sourced scalar modes at high multipoles. At low multipoles $R_{\rm s}\simeq 5$ would be allowed. 
 
 Such sourced gravitational waves are scale-dependent, chiral and non-Gaussian, and  can be probed in future CMB experiments~\cite{SimonsObservatory:2018koc,CMB-S4:2016ple,Hazumi:2019lys,NASAPICO:2019thw}. Measurements at low multipoles, which are accessible only to space-borne experiments \cite{Hazumi:2019lys,NASAPICO:2019thw}, are especially important because the sourced contribution can exceed the vacuum contribution. A large enhancement of the sourced contribution is also possible for non-CMB probes of gravitational waves such as pulsar timing arrays and laser/atomic interferometers~\cite{Thorne:2017jft,Campeti:2020xwn}.
  Discovery of the sourced gravitational waves opens up new windows into the particle physics behind inflation, with profound implications for fundamental physics.

\section*{Acknowledgments}
We thank A. Maleknejad and K. Schmitz for discussion and comments on the draft. KI and EK thank S. Ando and the institutes of Physics and Astronomy at the University of Amsterdam and the Vrije Universiteit Amsterdam for hospitality, where this work was initiated. EK thanks the Stichting Van der Waals Fonds for the Johannes Diderik van der Waals rotating chair which enabled the visit. The work of KI is supported by JSPS KAKENHI Grant No.~JP18H05542, JP20H01894, and JSPS Core-to-Core Program Grant No.~JPJSCCA20200002. The work of EK is supported in part by JSPS KAKENHI Grant No.~JP20H05850 and JP20H05859, and the Deutsche Forschungsgemeinschaft (DFG, German Research Foundation) under Germany's Excellence Strategy - EXC-2094 - 390783311. The work of IO is supported by JSPS Overseas Research Fellowship. The Kavli IPMU is supported by World Premier International Research Center Initiative (WPI), MEXT, Japan.

\appendix

\section{Additional figures}
\label{app:figs}

In this appendix, we give the results for $\lambda=50$ and 500. All the figures are shown in the same 
manner as those in the main text. The additional parameters are given in the figure caption when needed. 

\begin{figure}
  \begin{center}
    \includegraphics[width=7cm]{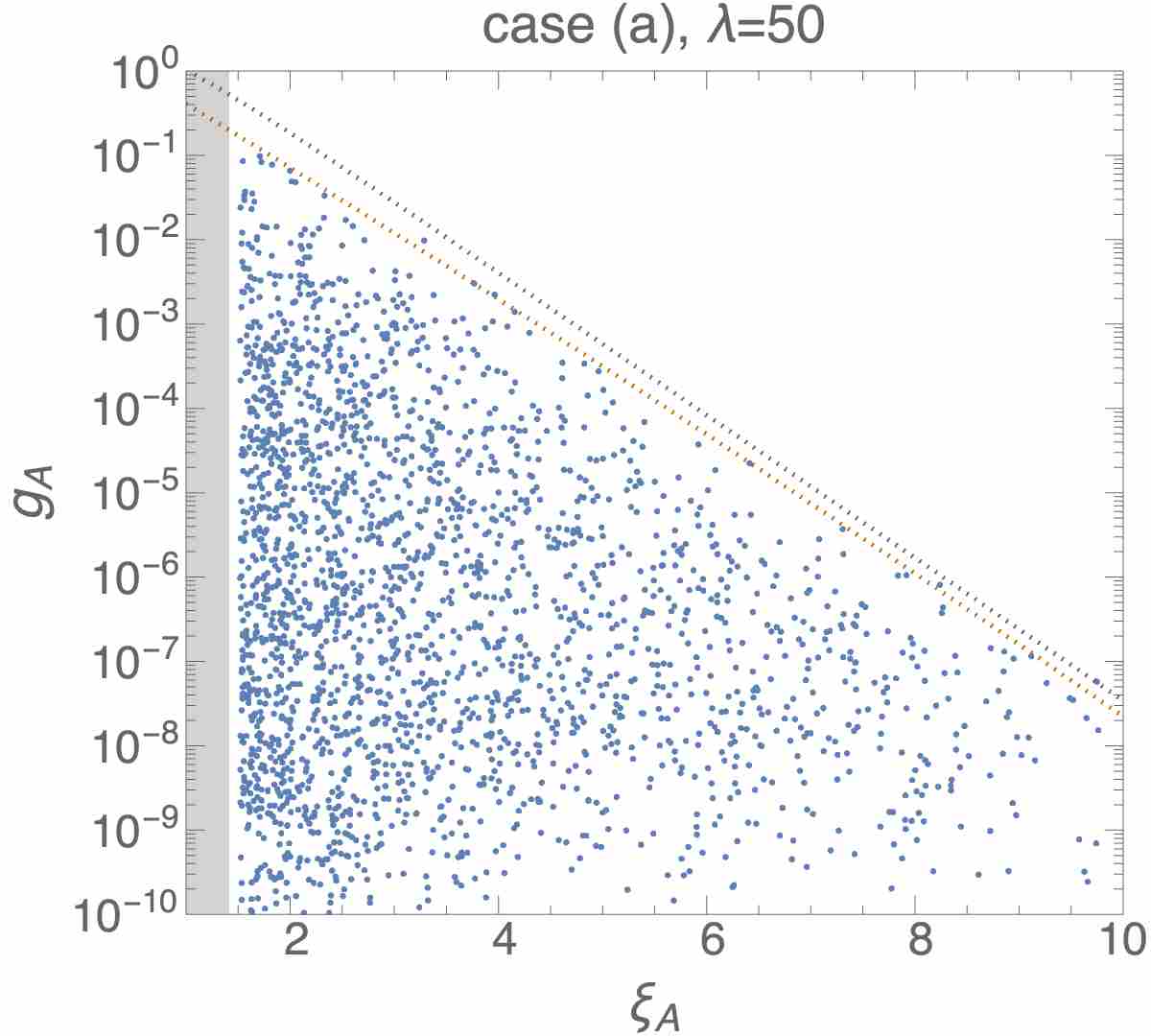}
    \includegraphics[width=7cm]{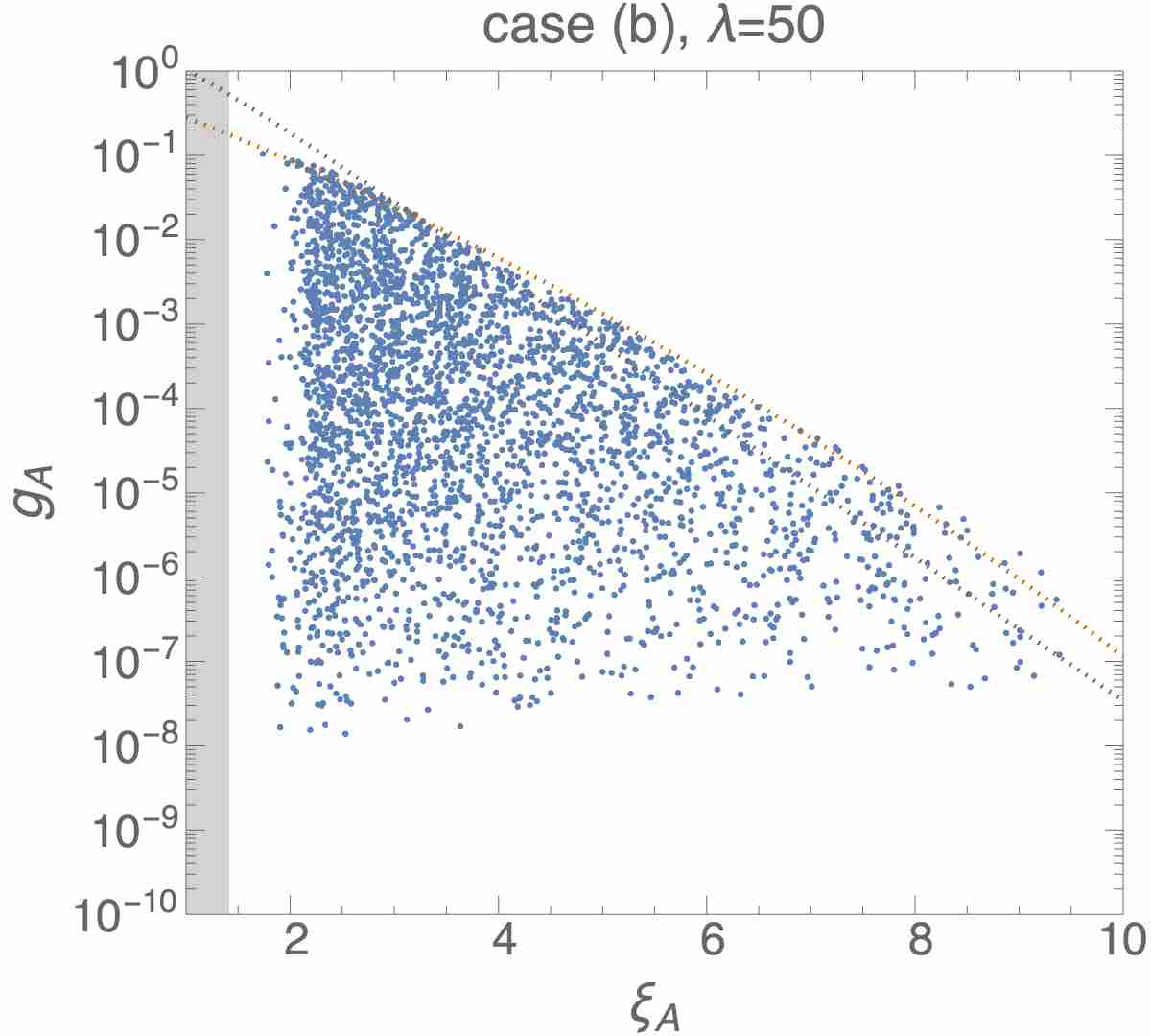}
    \includegraphics[width=7cm]{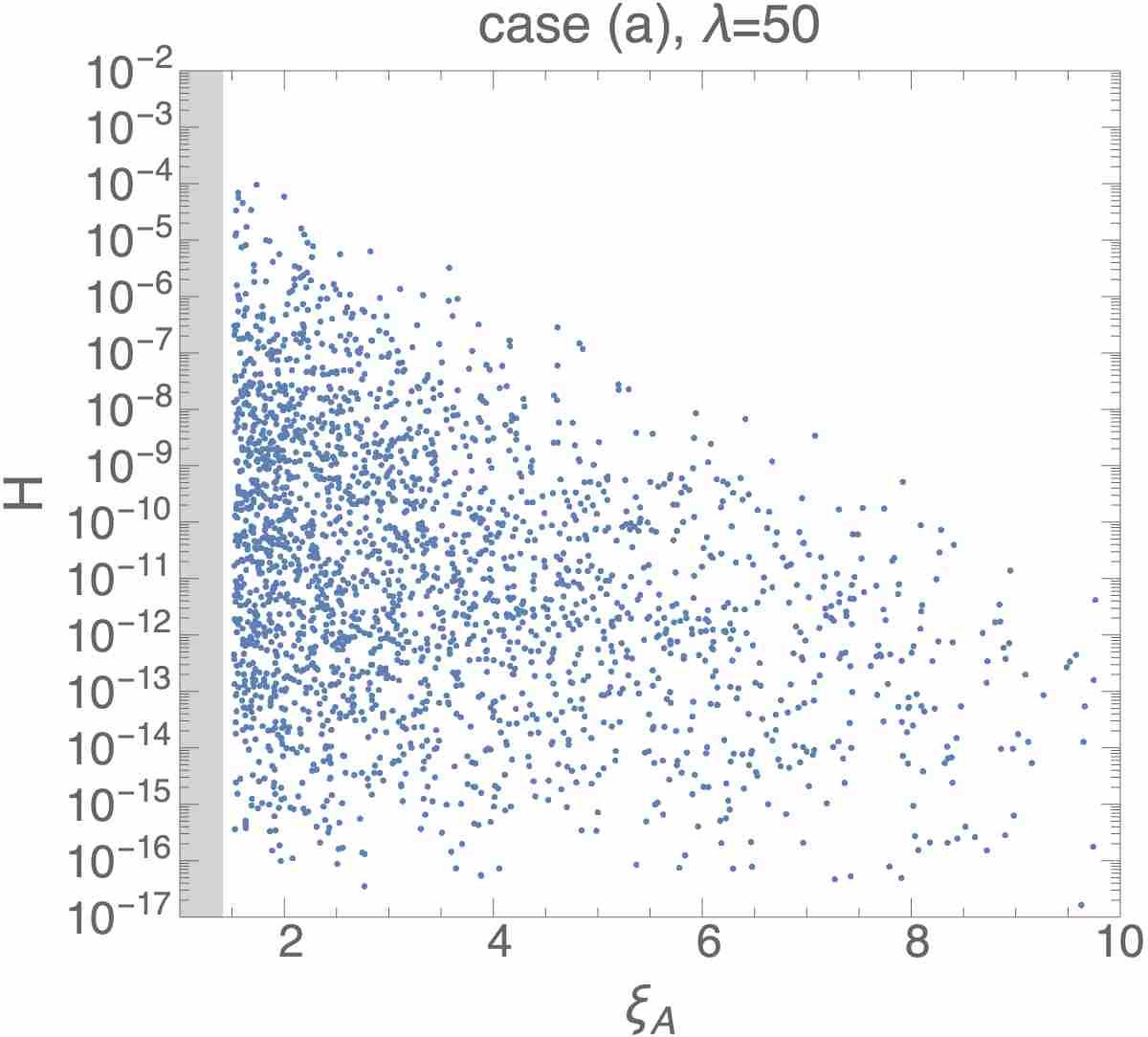}
    \includegraphics[width=7cm]{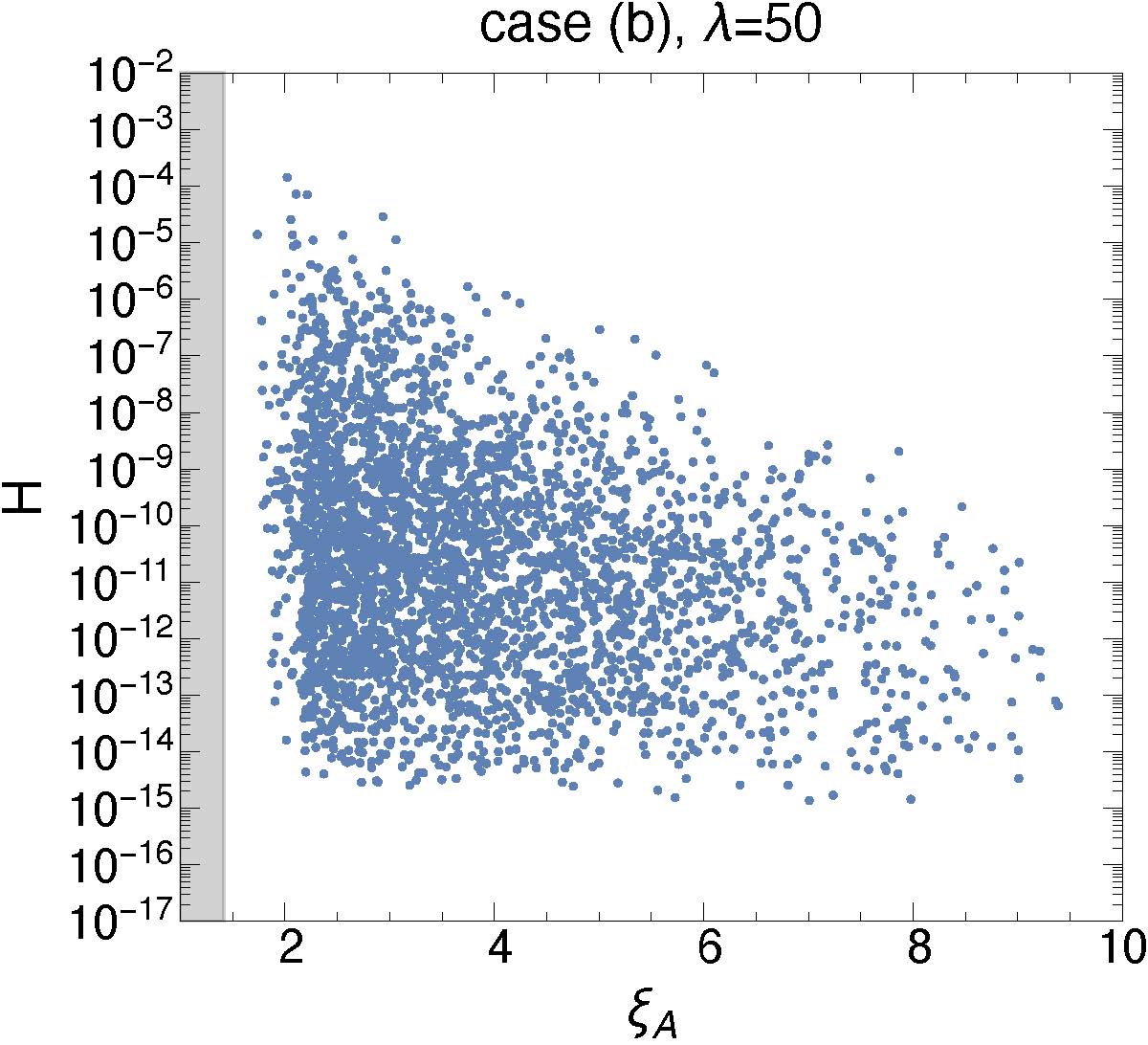}
    \includegraphics[width=7cm]{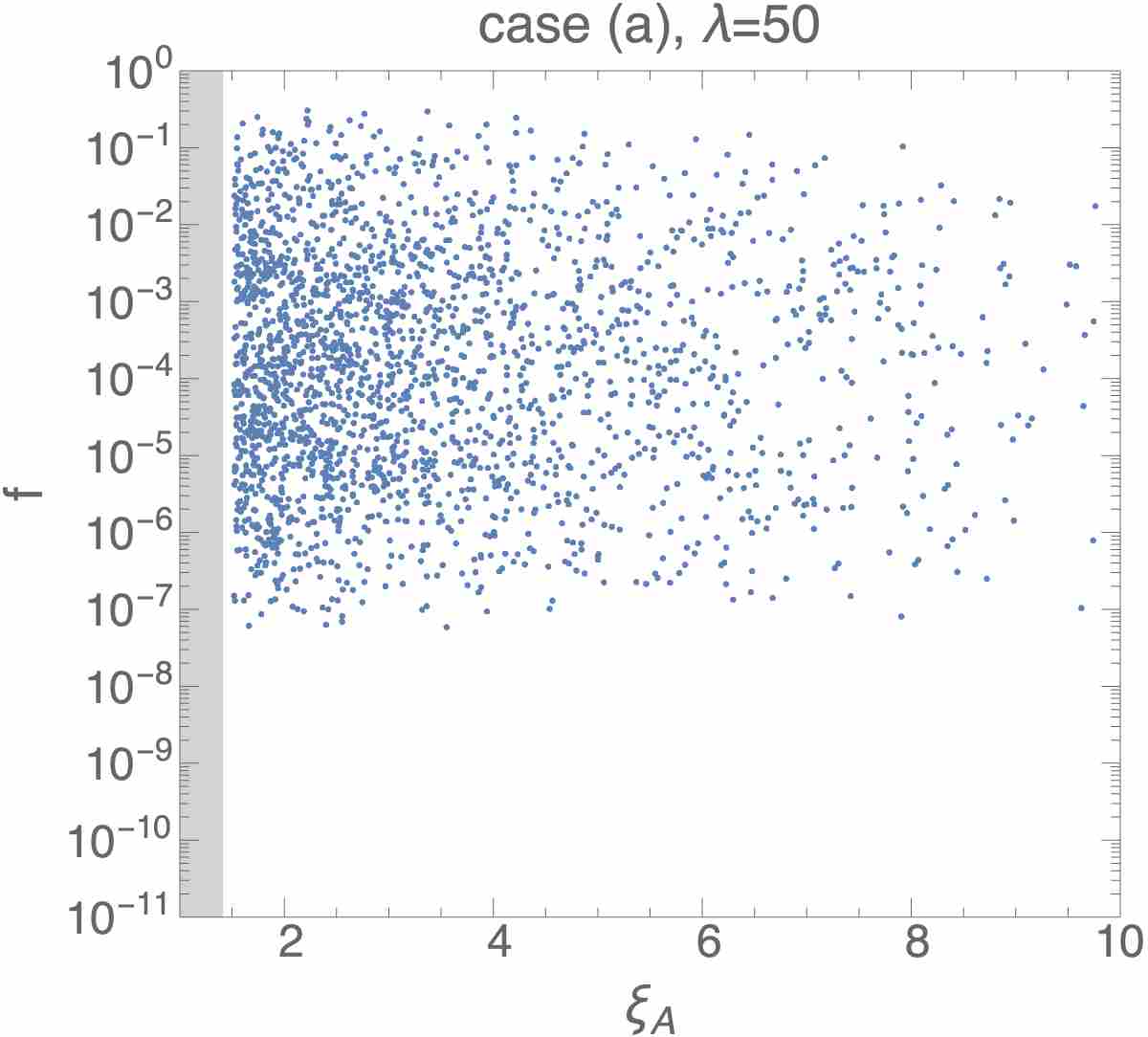}   
    \includegraphics[width=7cm]{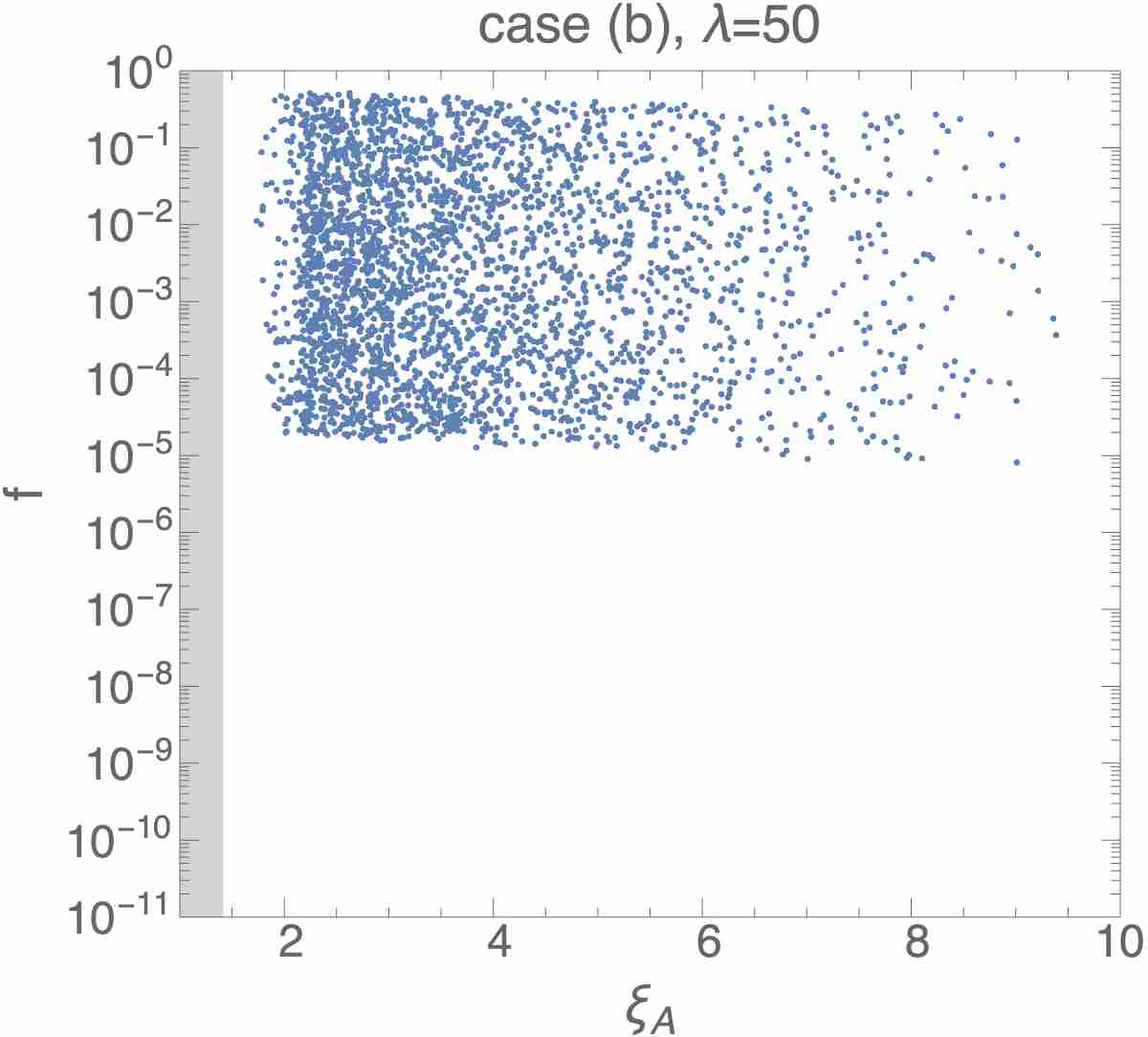}
    \caption{Same as Figs.\,\ref{fig:lam_100_g} and \ref{fig:lam_100_Hf} but for $\lambda=50$.}
  \label{fig:lam_050_gHf}
 \end{center}
\end{figure}

\begin{figure}
  \begin{center}
    \includegraphics[width=7cm]{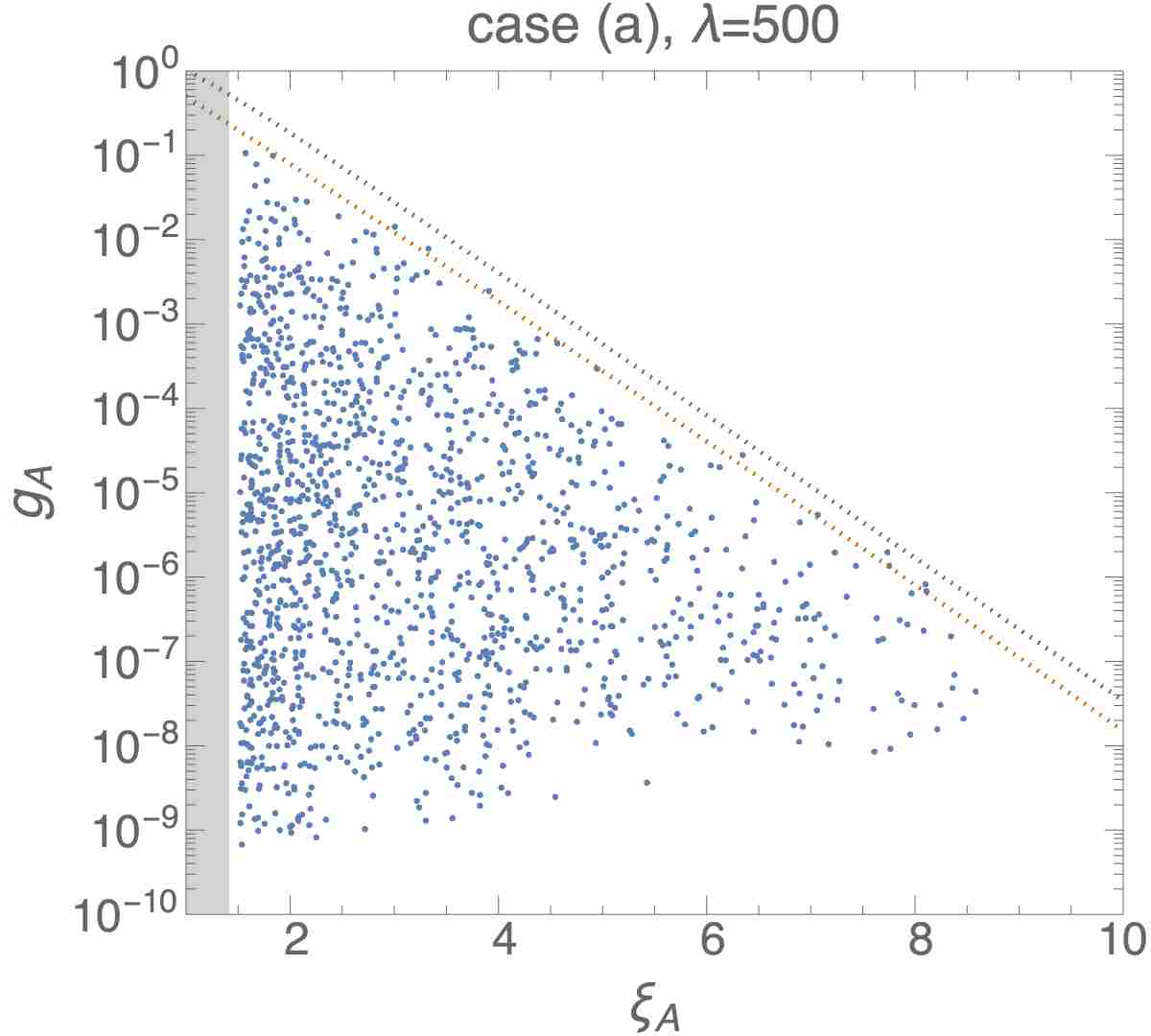}
    \includegraphics[width=7cm]{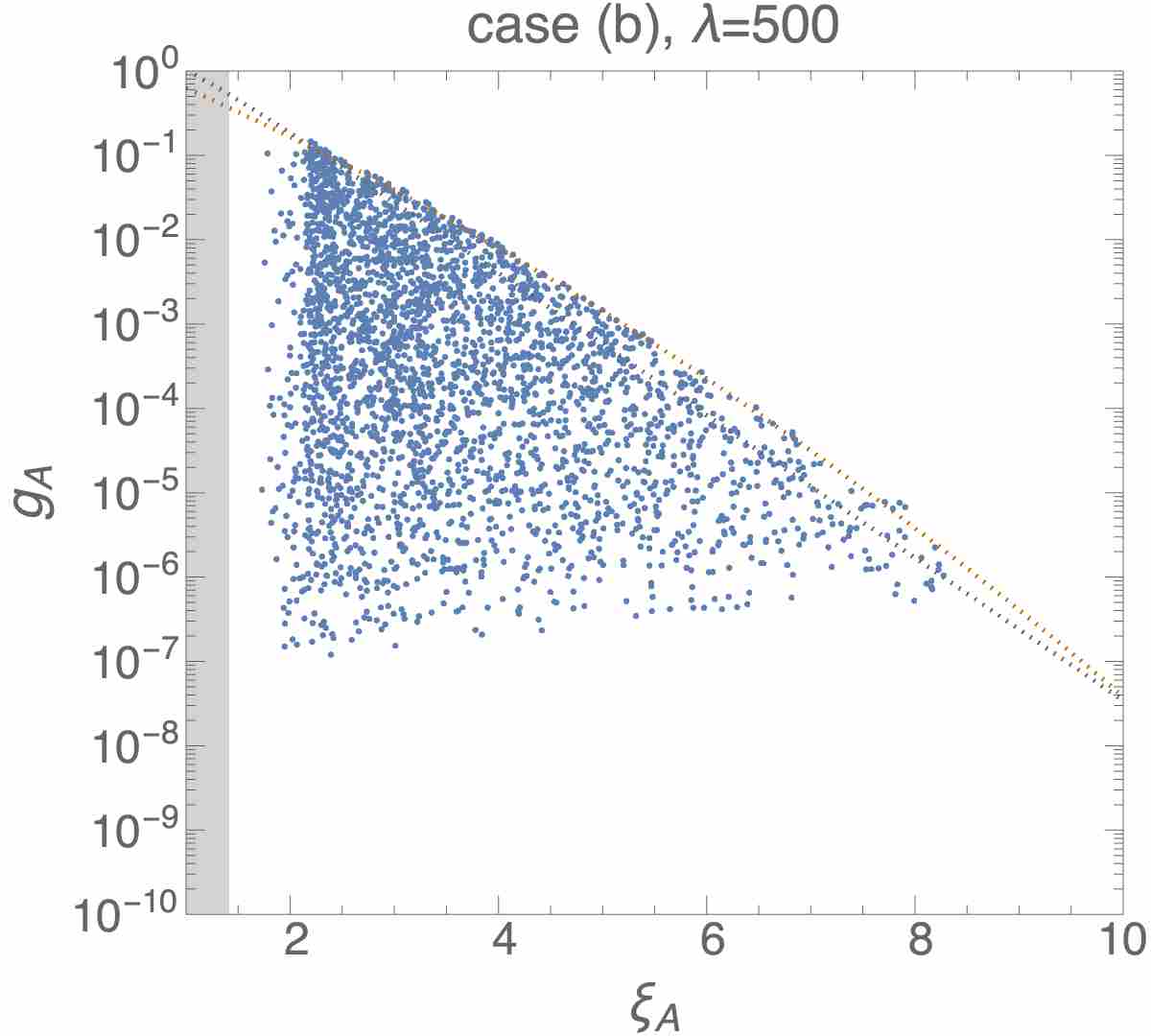}
    \includegraphics[width=7cm]{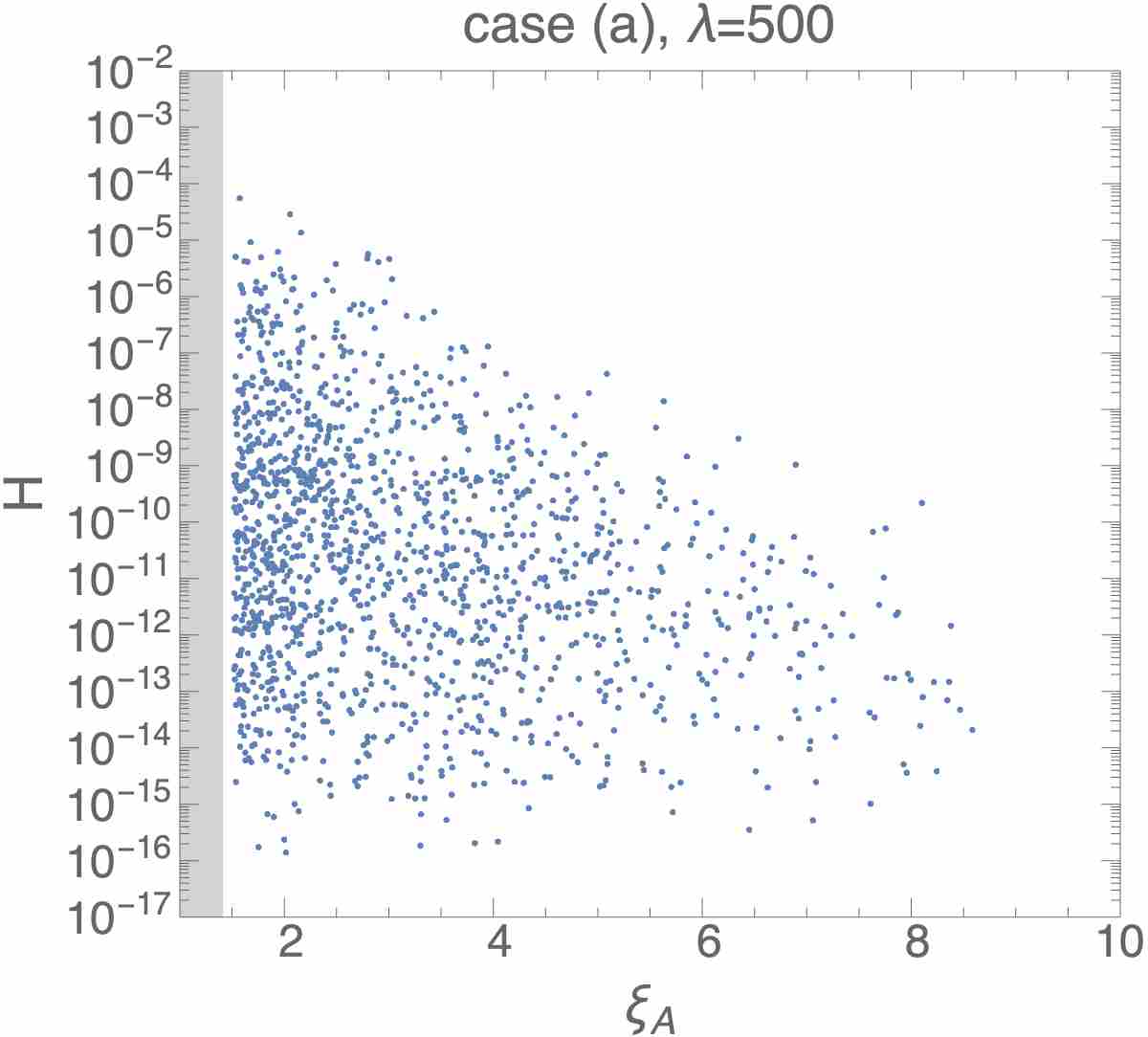}
    \includegraphics[width=7cm]{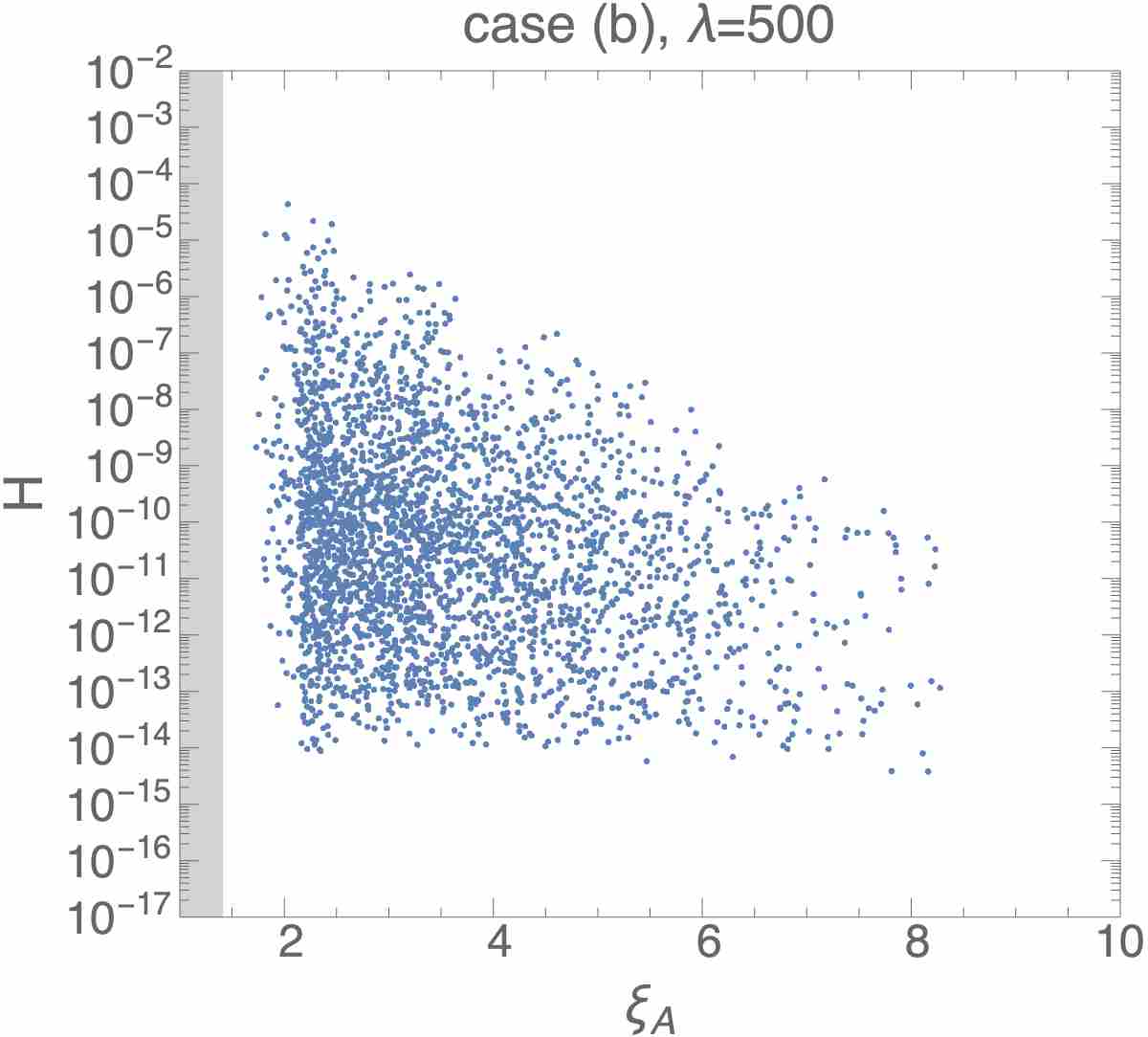}
    \includegraphics[width=7cm]{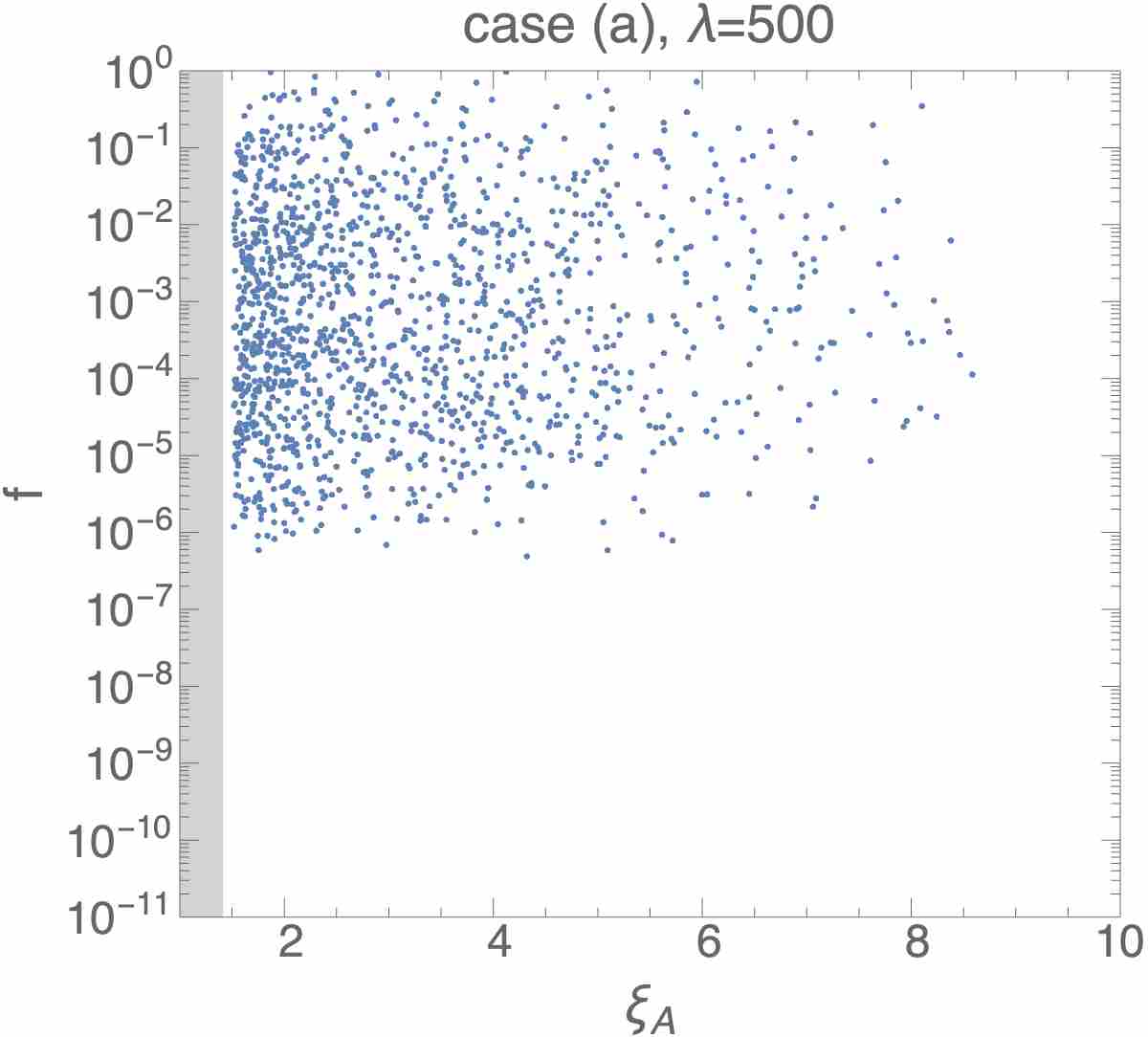} 
    \includegraphics[width=7cm]{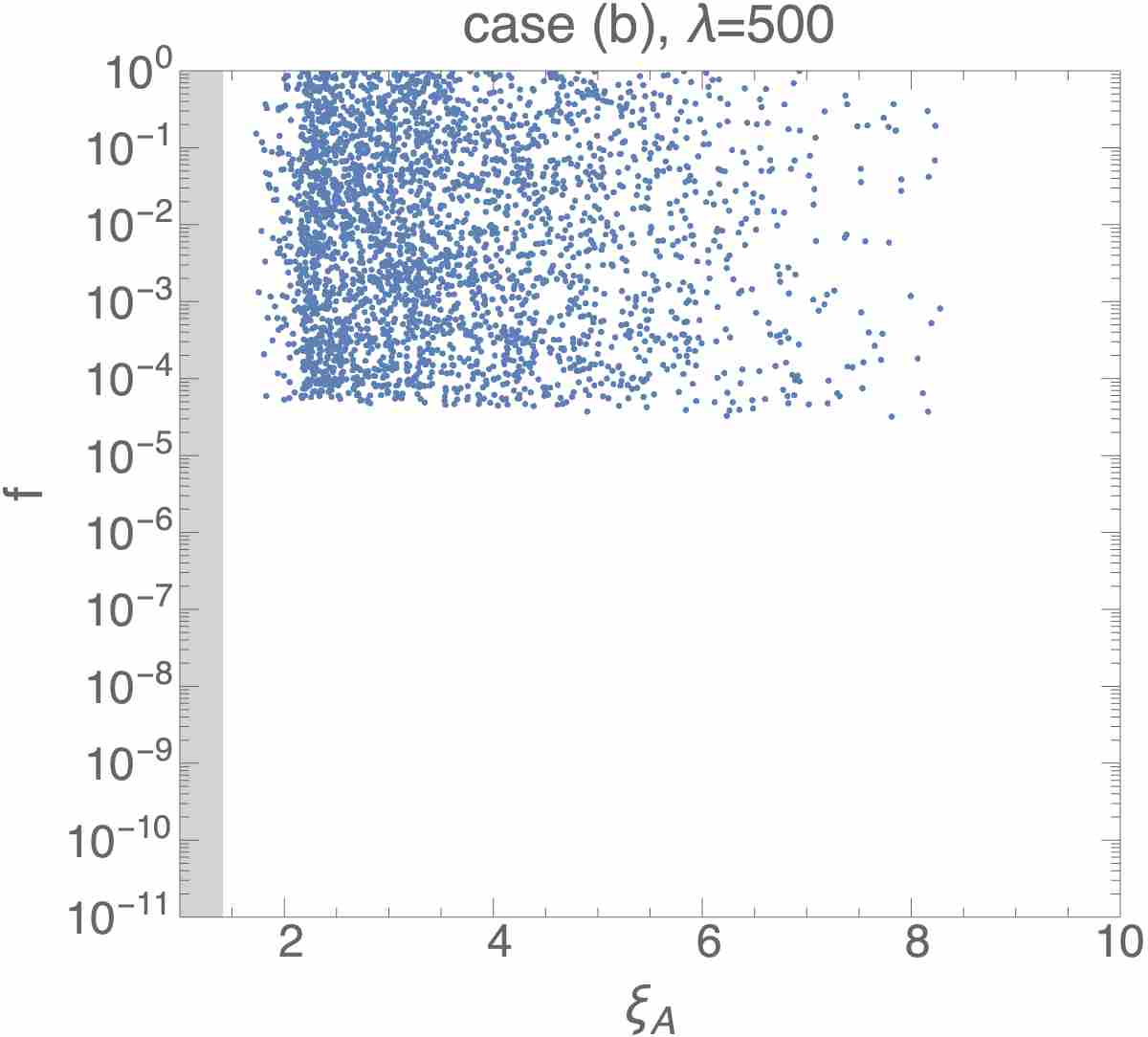}
    \caption{Same as Figs.\,\ref{fig:lam_100_g} and  \ref{fig:lam_100_Hf} but for $\lambda=500$.}
  \label{fig:lam_500_gHf}
 \end{center}
\end{figure}

\begin{figure}
  \begin{center}
    \includegraphics[width=6.5cm]{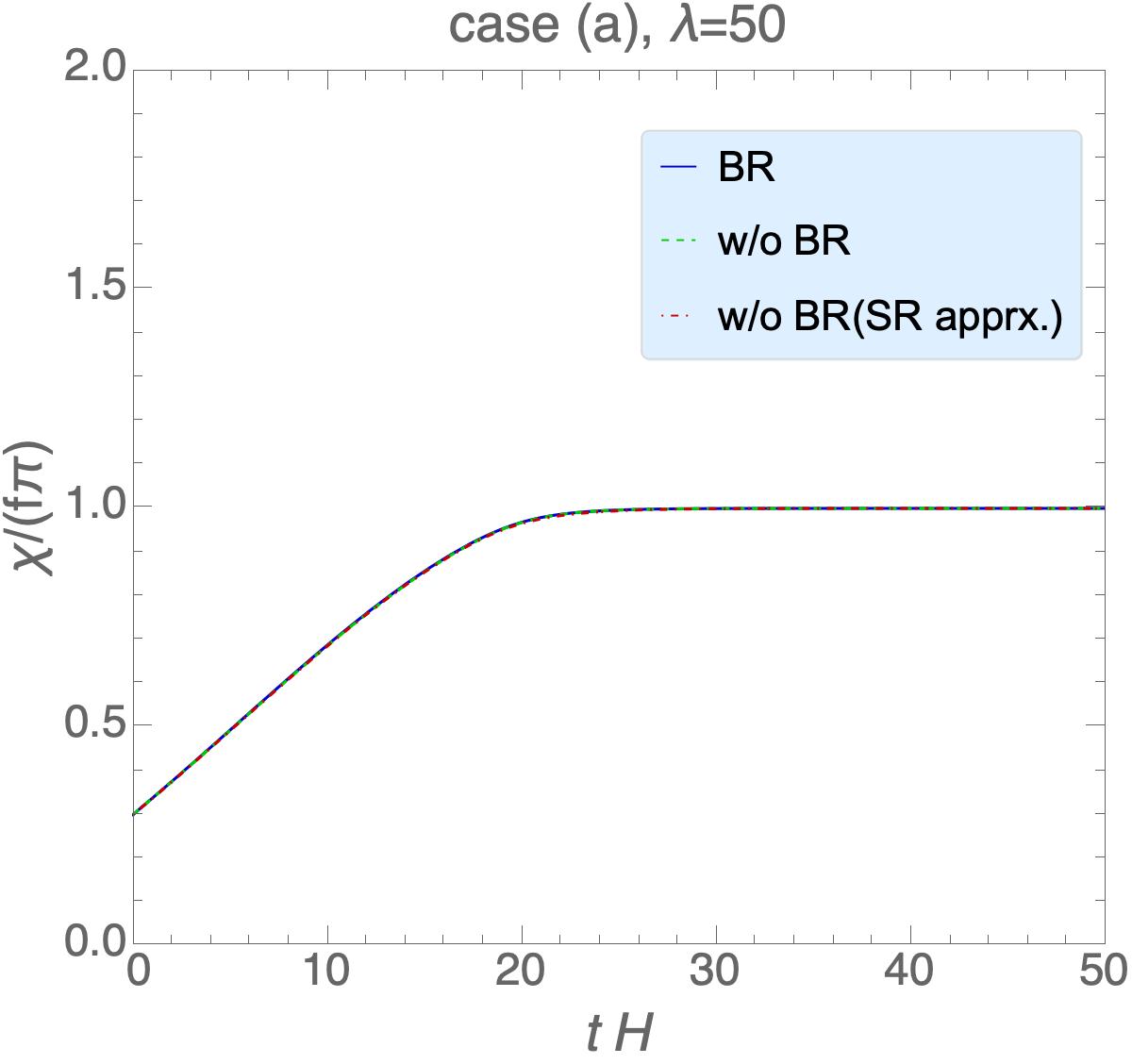}
    \includegraphics[width=6.5cm]{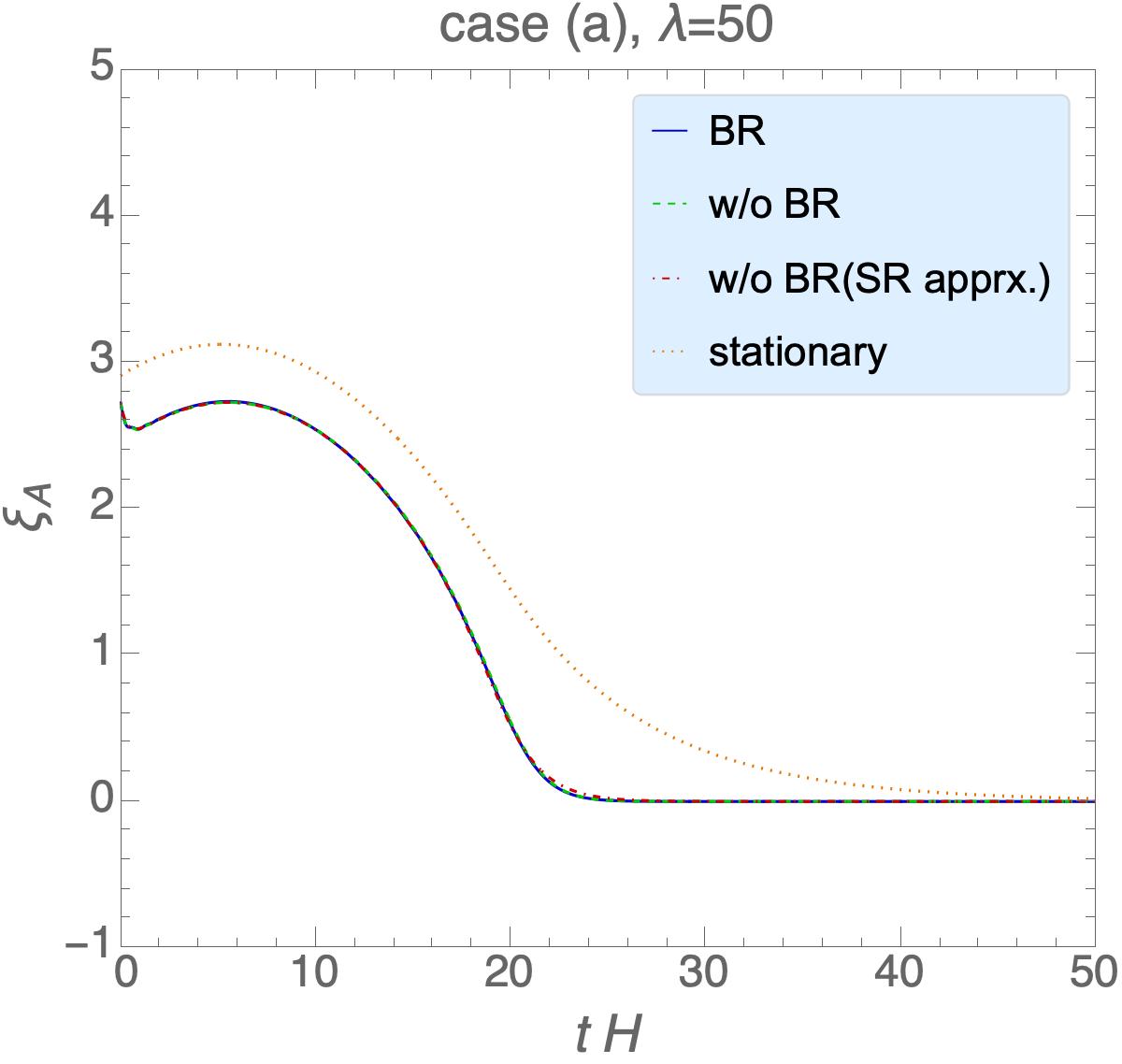}
    \includegraphics[width=6.5cm]{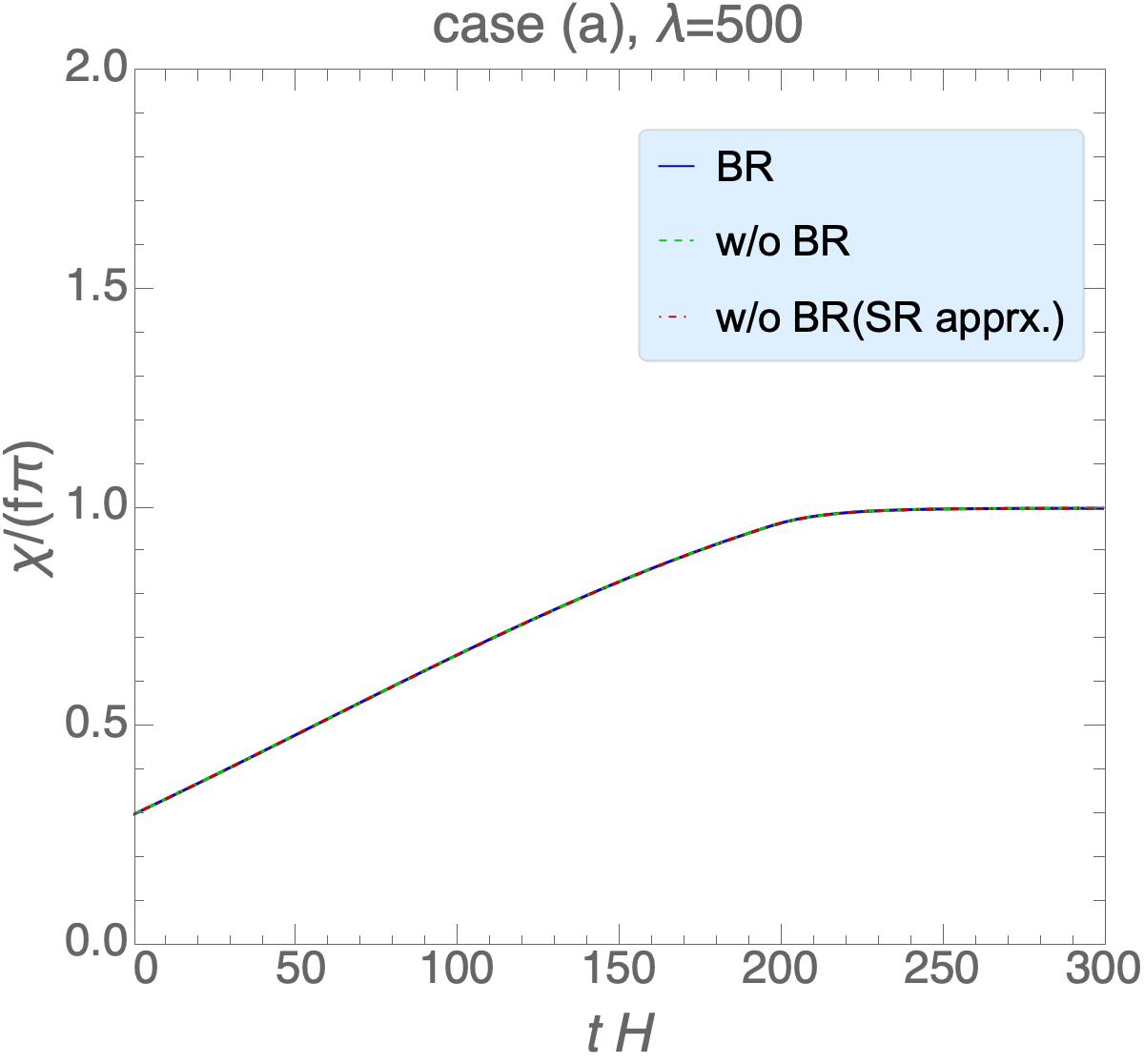}
    \includegraphics[width=6.5cm]{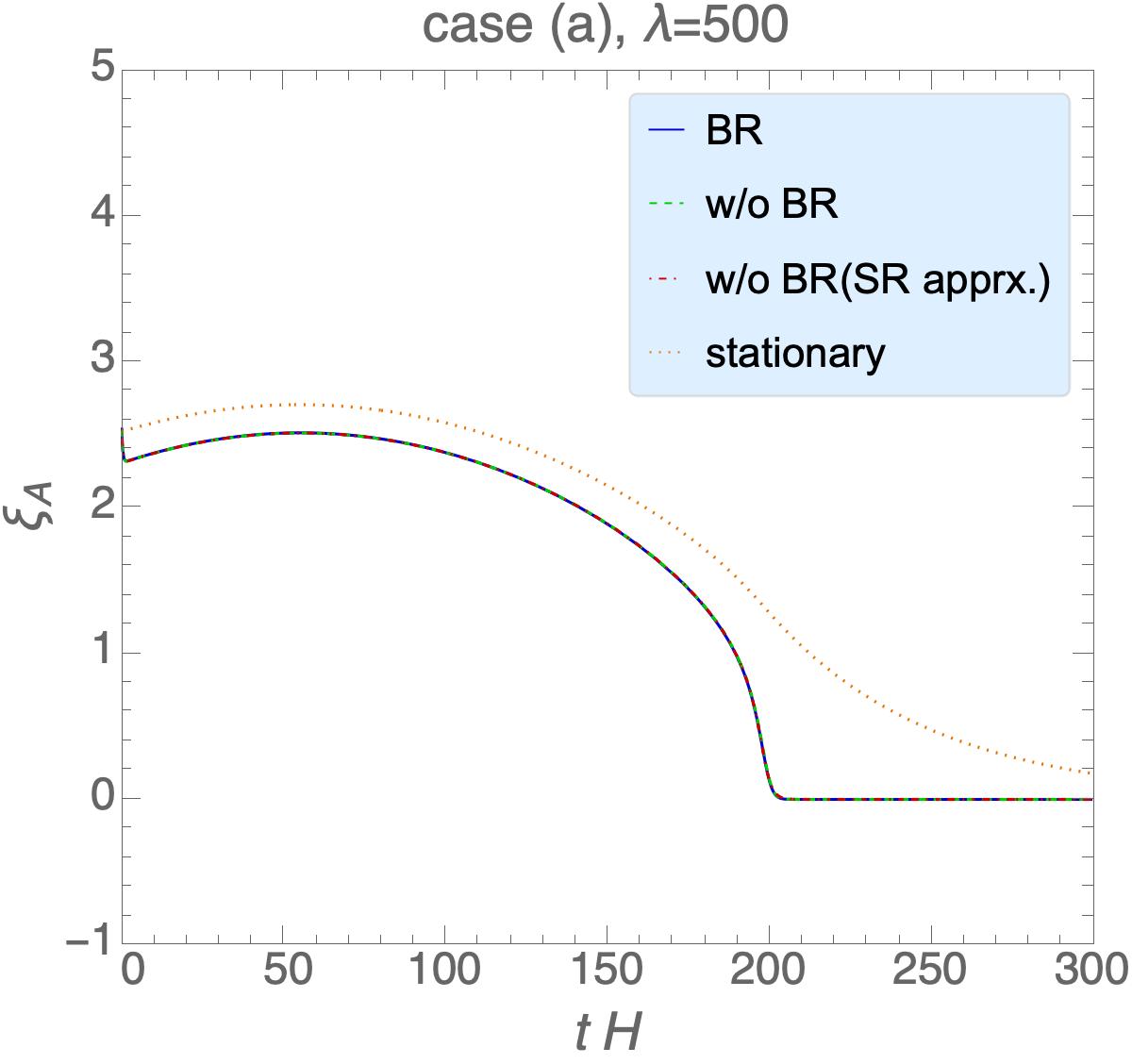}
     \caption{Same as Fig.\,\ref{fig:dy_a_lam100} but for $\lambda=50$, $g=9.3\times 10^{-3}$, $f=5.8\times 10^{-2}$, $H=8.5\times
     10^{-6}$ and $\mu=7.2\times 10^{-4}$ (top), and $\lambda=500$, $g=6.1\times 10^{-3}$, $f=7.4\times 10^{-1}$, $H=1.1\times
     10^{-5}$ and $\mu=1.8\times 10^{-3}$ (bottom).  }
  \label{fig:dy_a_lam050_500}
 \end{center}
\end{figure}

\begin{figure}
  \begin{center}
    \includegraphics[width=6.5cm]{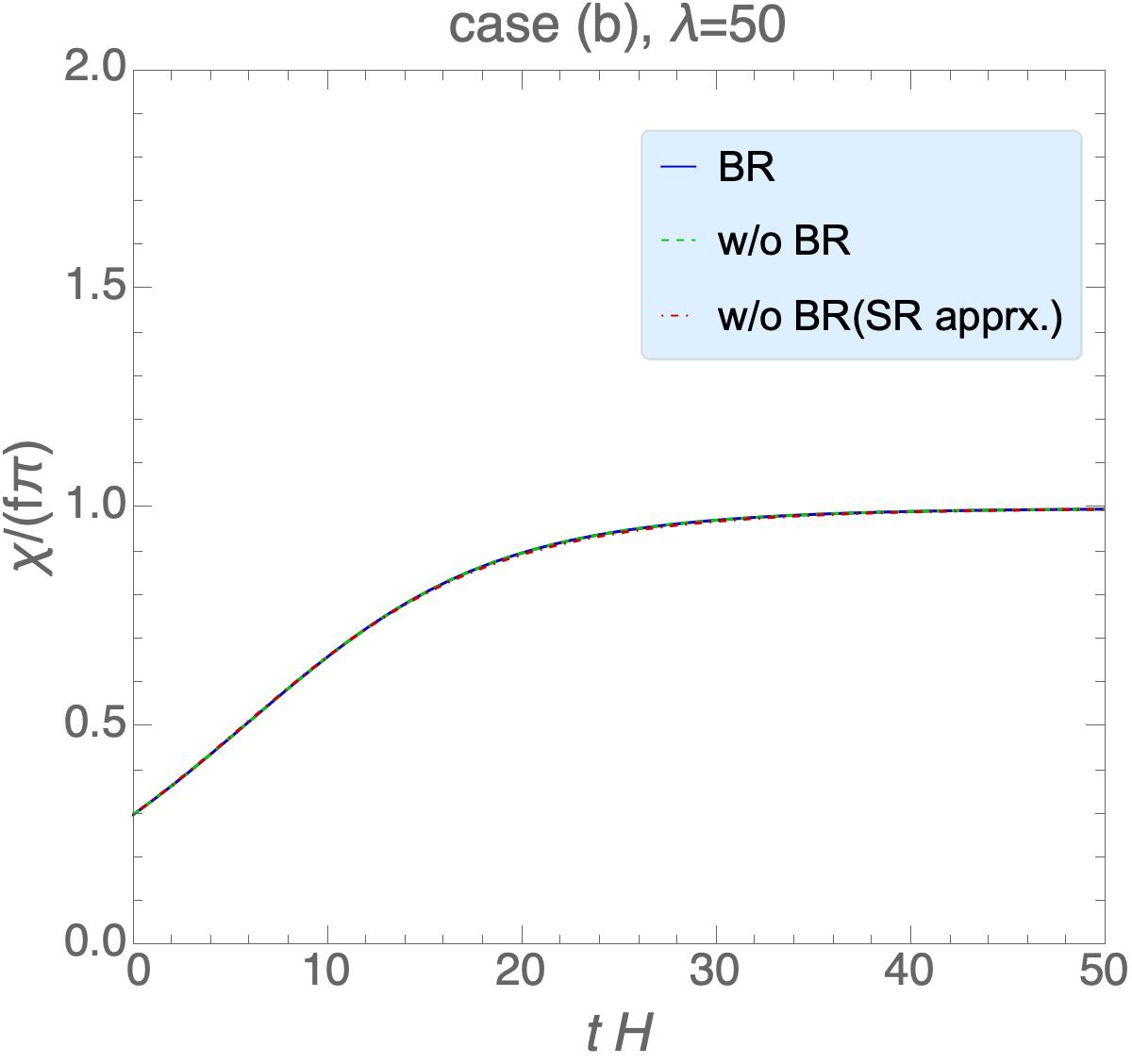}
    \includegraphics[width=6.5cm]{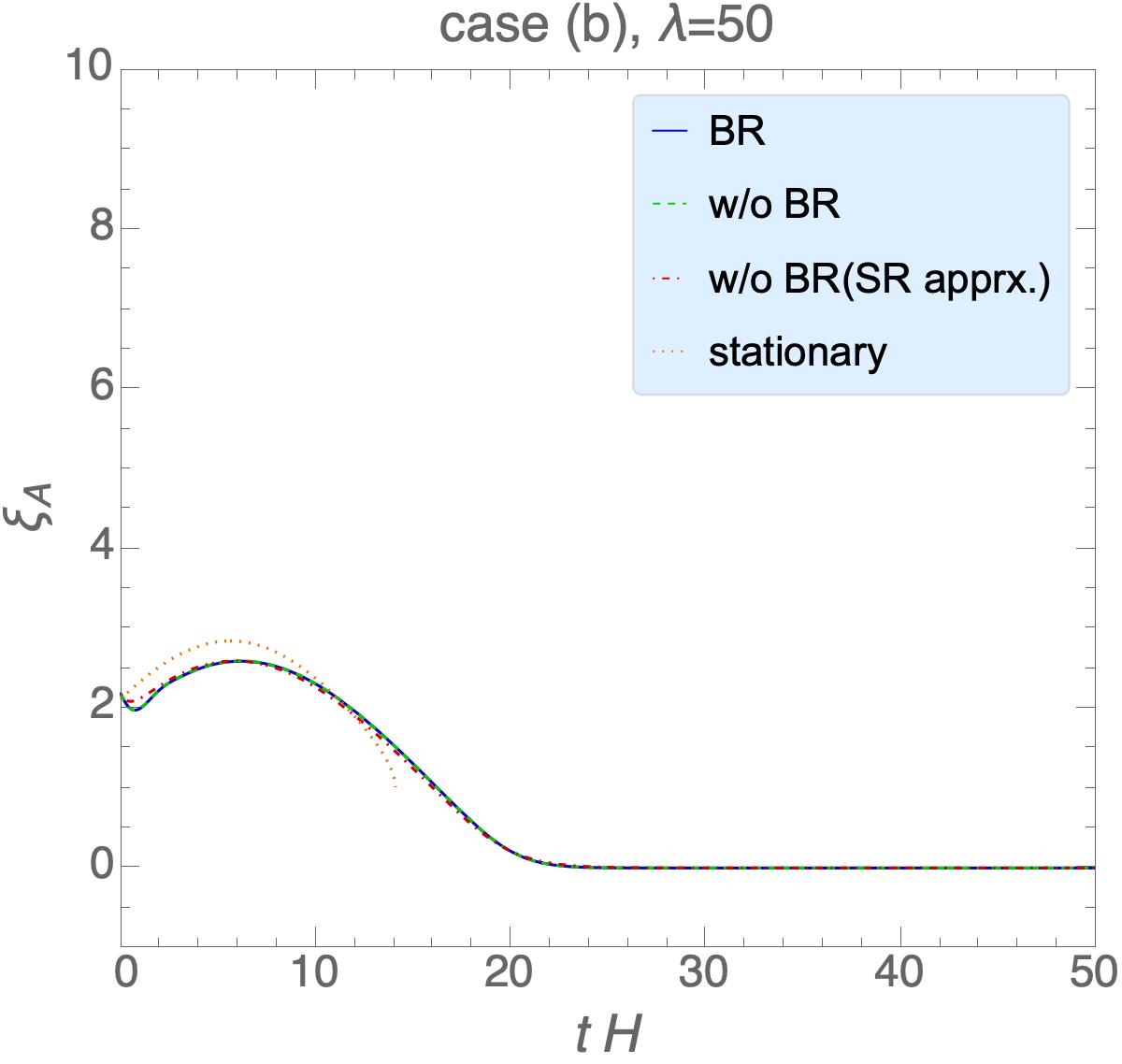}
    \includegraphics[width=6.5cm]{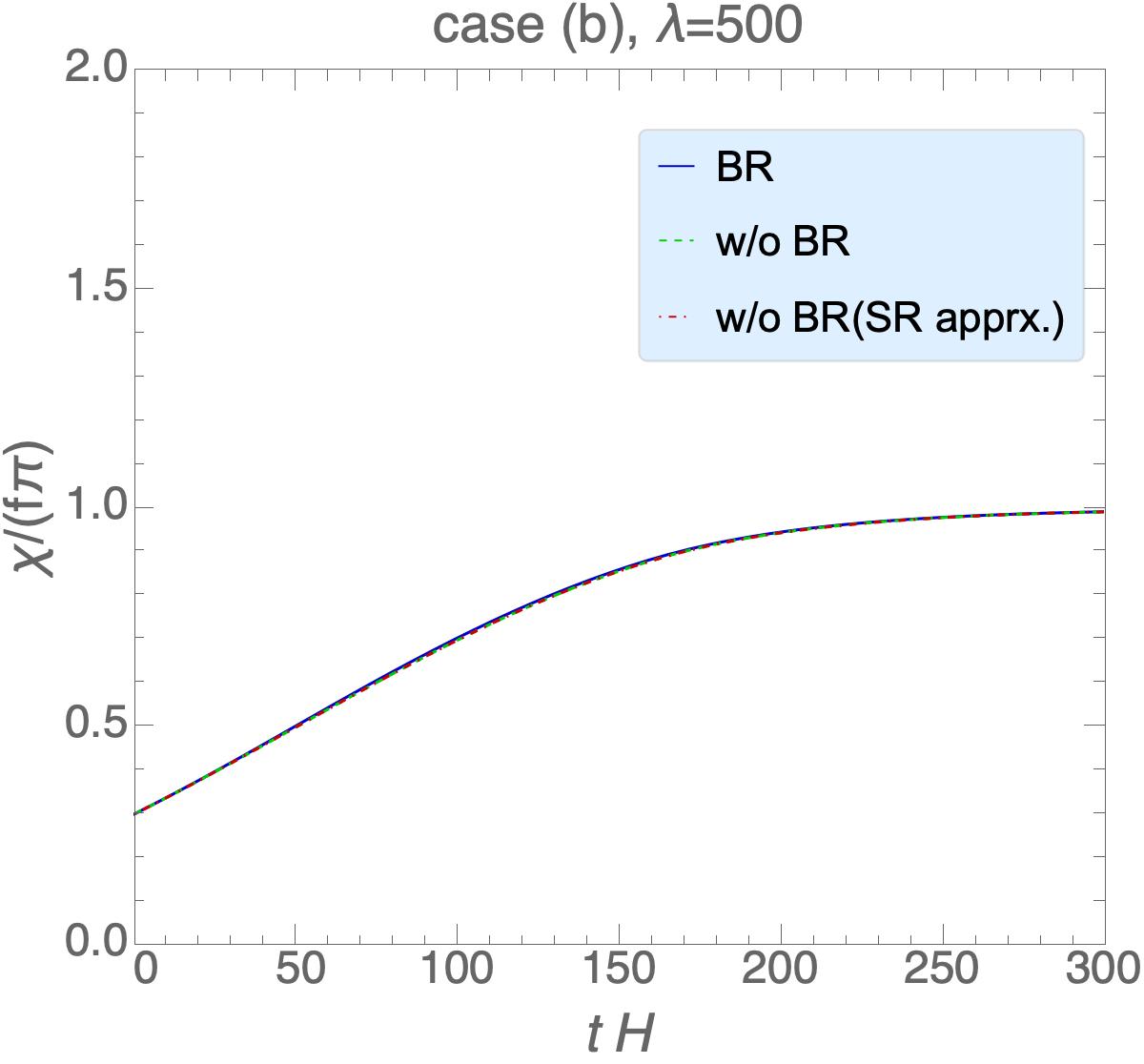}
    \includegraphics[width=6.5cm]{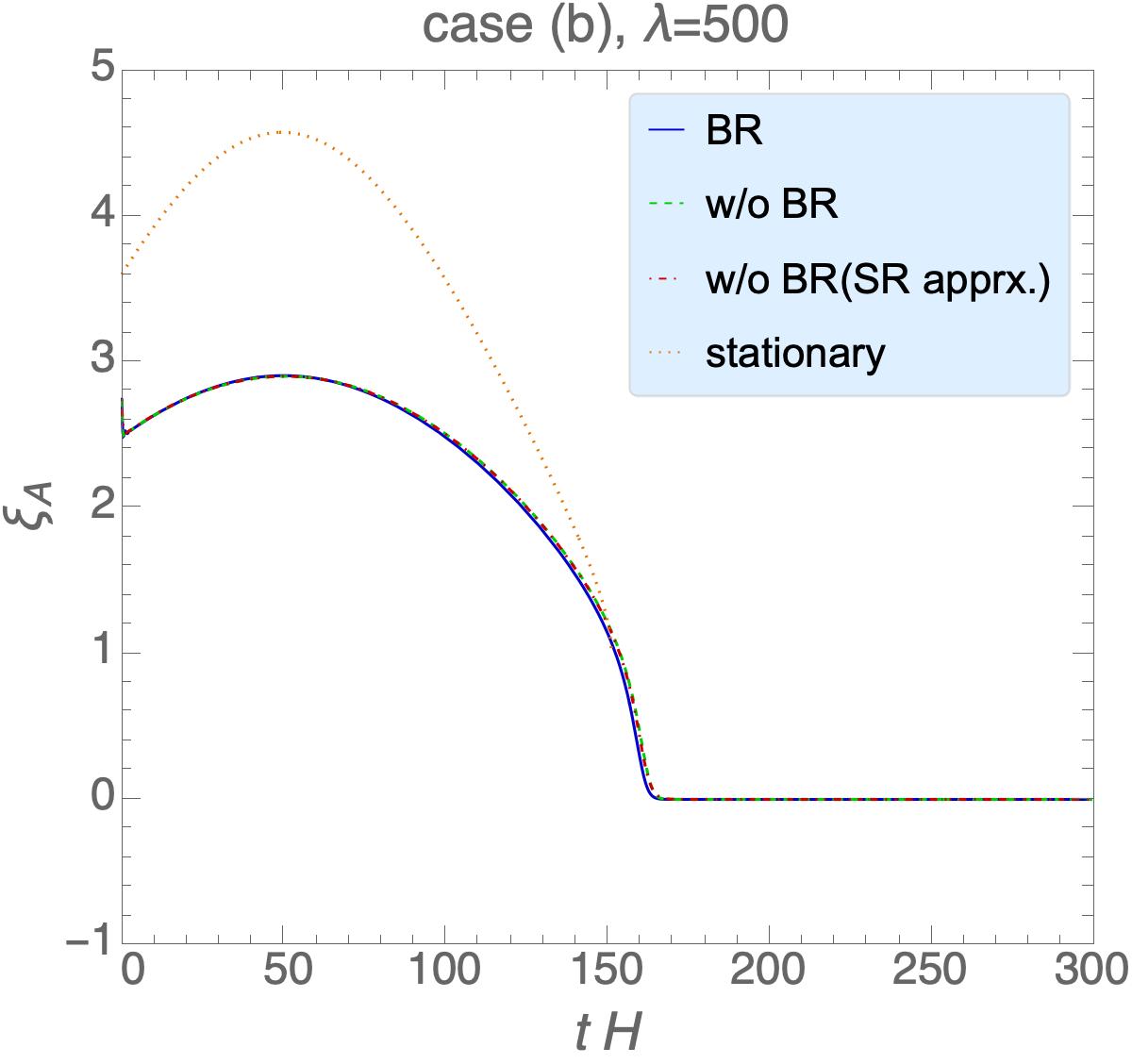}
     \caption{Same as Fig.\,\ref{fig:dy_b_lam100} but for $\lambda=50$, $g=7.5\times 10^{-3}$, $f=4.3\times 10^{-1}$, $H=1.0\times
     10^{-5}$ and $\mu=1.7\times 10^{-3}$ (top), and $\lambda=500$, $g=1.5\times 10^{-2}$, $f=8.6\times 10^{-1}$, $H=9.1\times
     10^{-6}$ and $\mu=1.4\times 10^{-3}$ (bottom). }
  \label{fig:dy_b_lam050_500}
 \end{center}
\end{figure}

\begin{figure}
  \begin{center}
    \includegraphics[width=6.5cm]{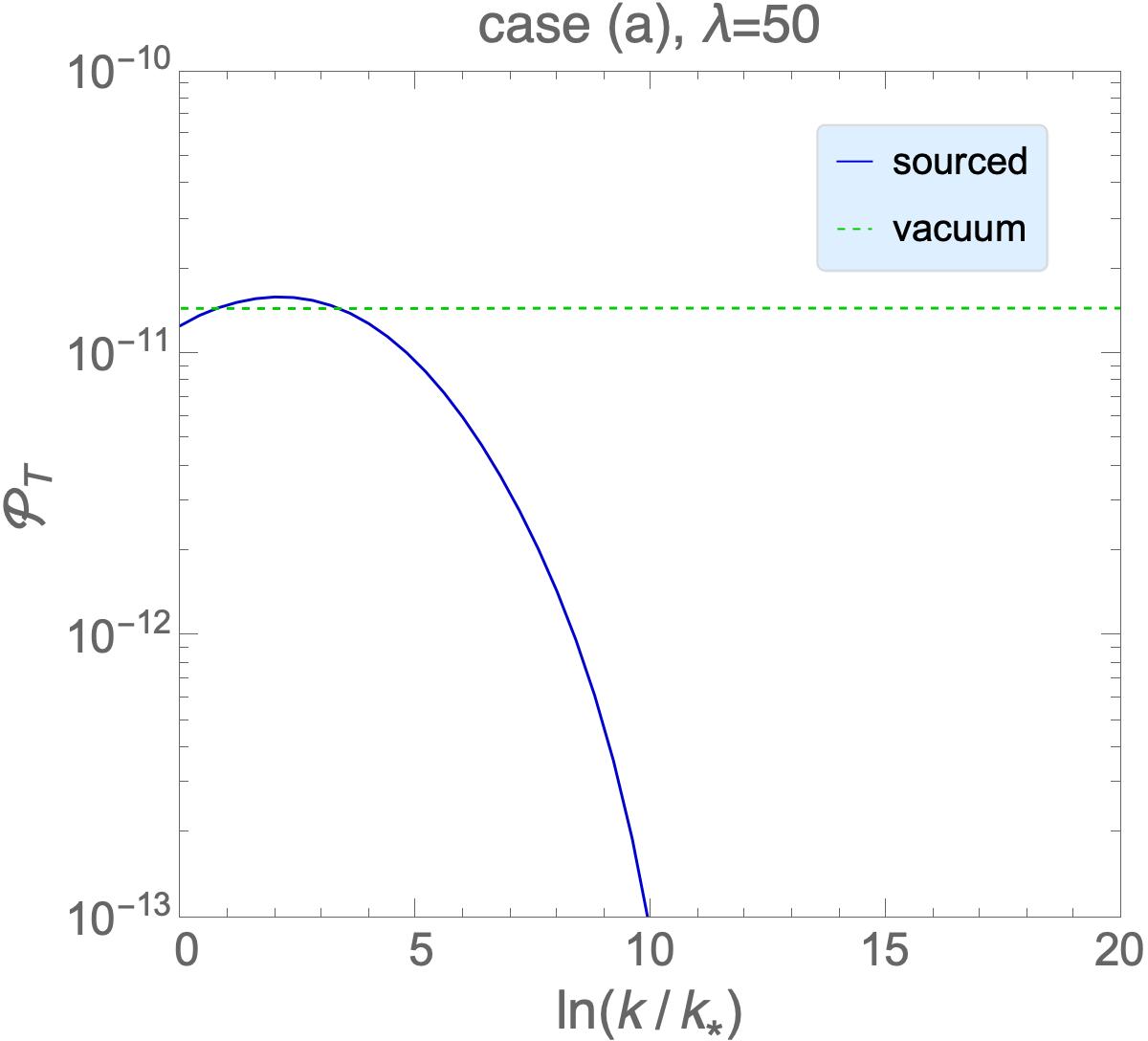}
    \includegraphics[width=6.5cm]{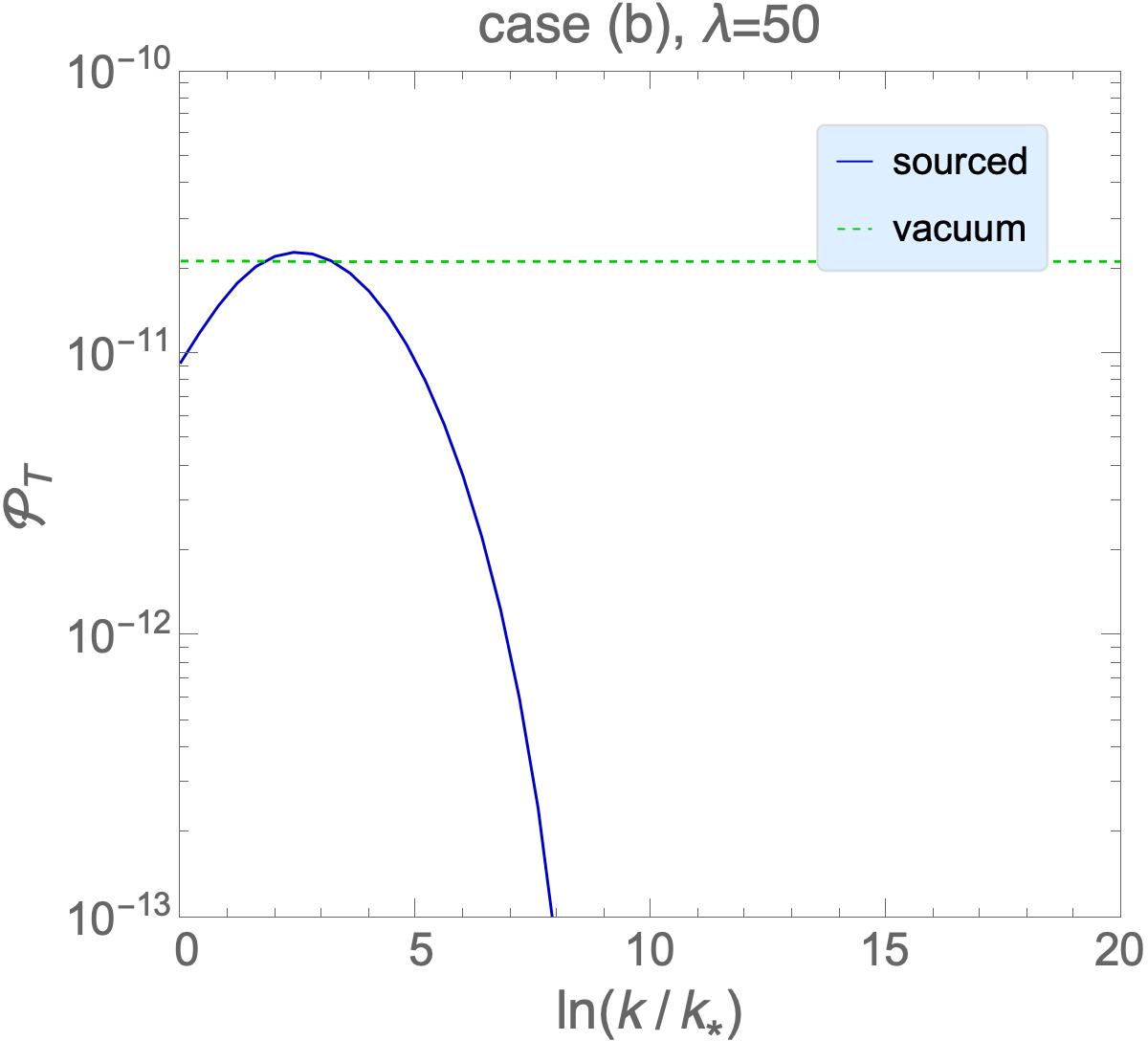}
    \includegraphics[width=6.5cm]{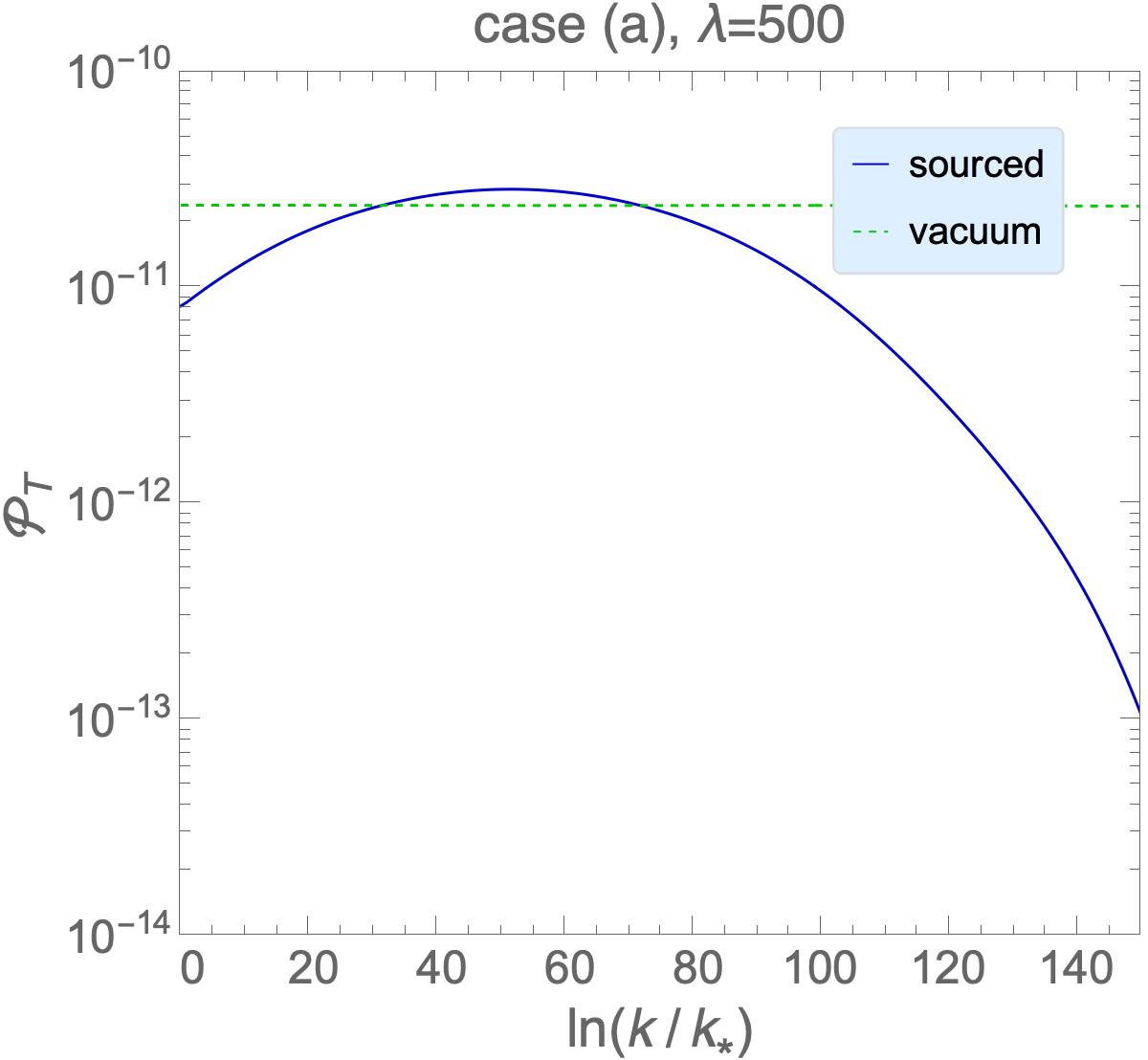}
    \includegraphics[width=6.5cm]{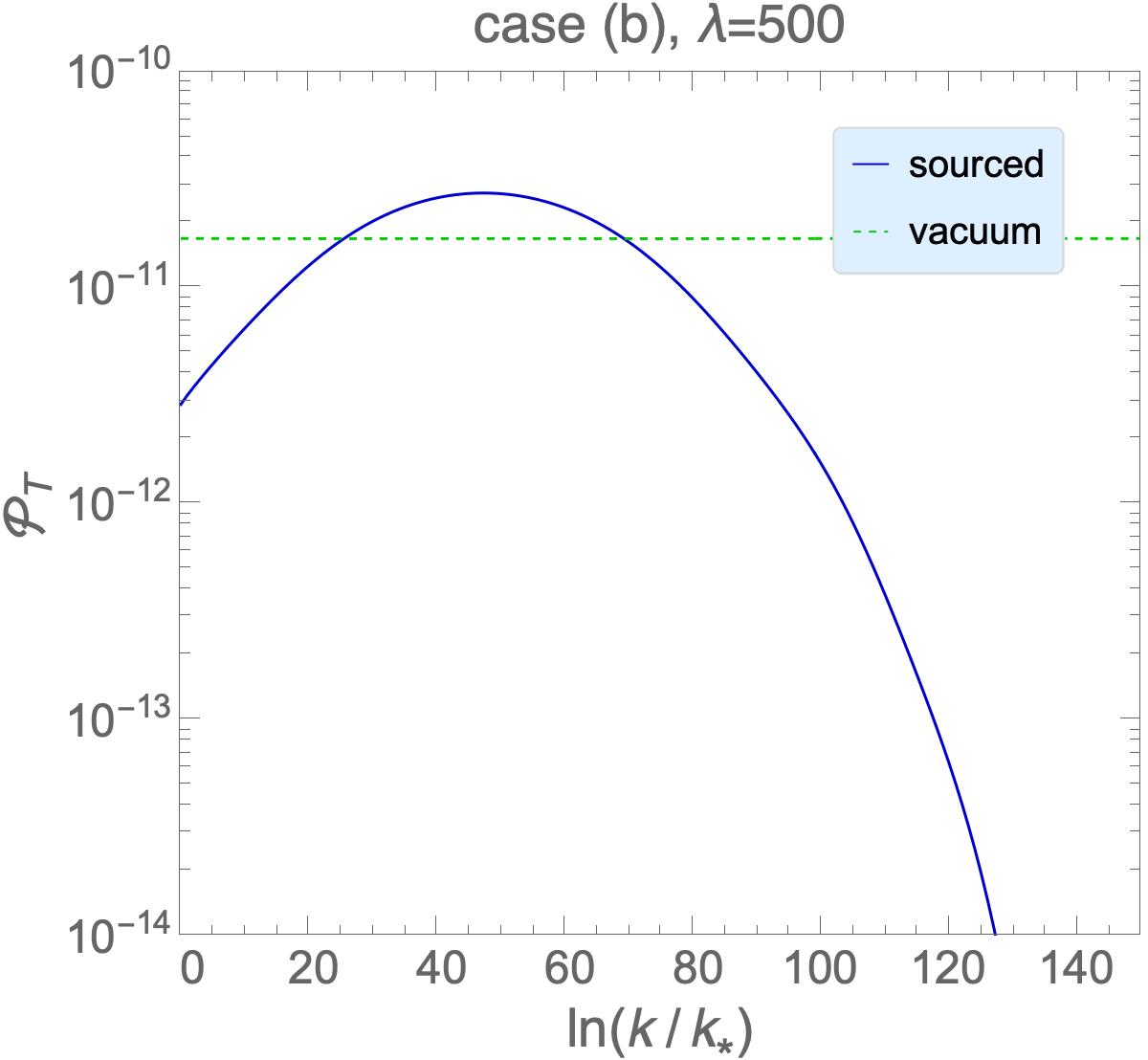}
    \caption{Same as Fig.\,\ref{fig:PT_lam_100} but $\lambda=50$ (top) and 500 (bottom) and the 
    other parameters are the same as Figs.\,\ref{fig:dy_a_lam050_500} and \ref{fig:dy_b_lam050_500}
    for case (a) and (b), respectively.}
  \label{fig:PT_lam_050_500}
 \end{center}
\end{figure}

\begin{figure}
  \begin{center}
    \includegraphics[width=7cm]{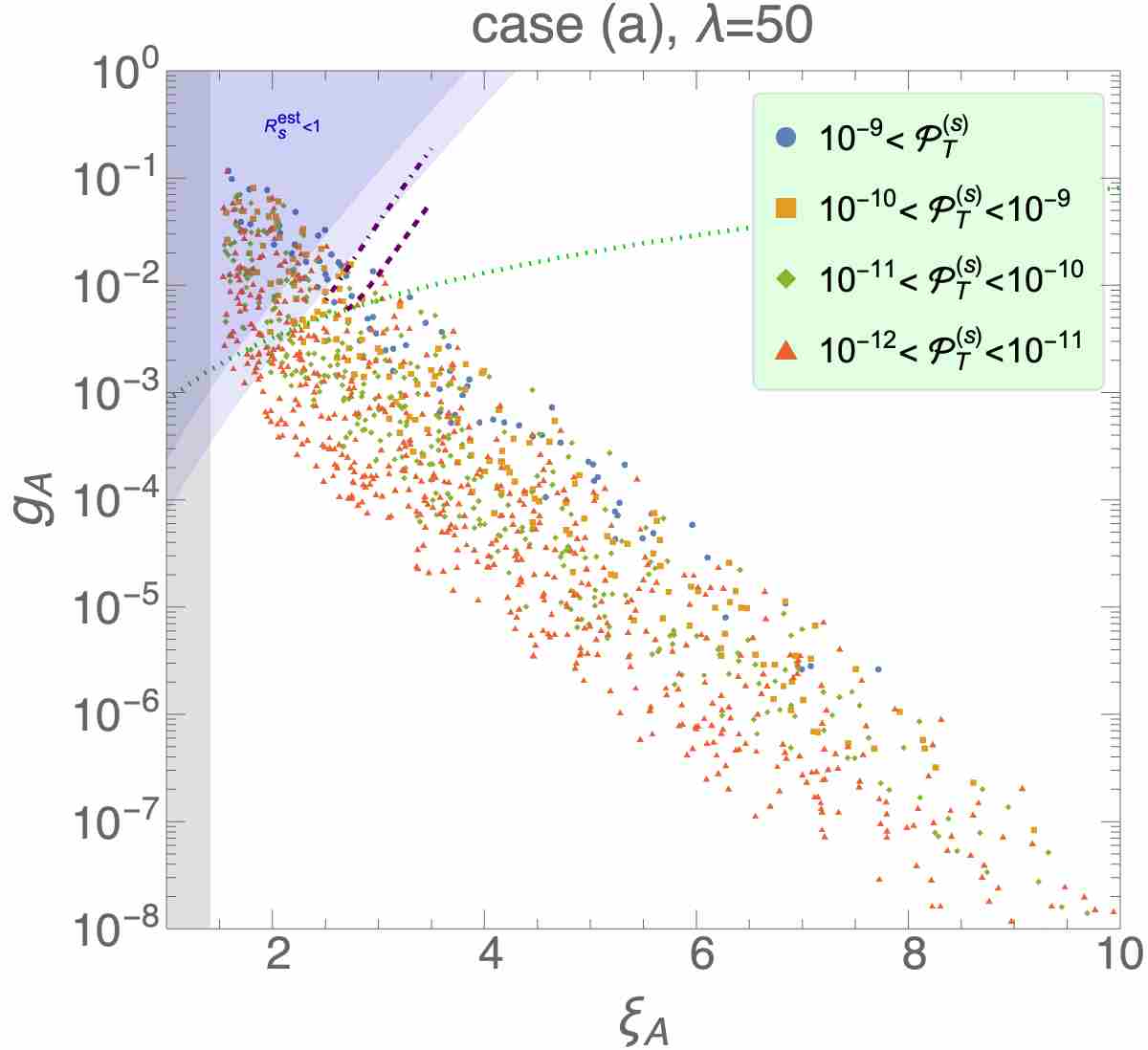}
    \includegraphics[width=7cm]{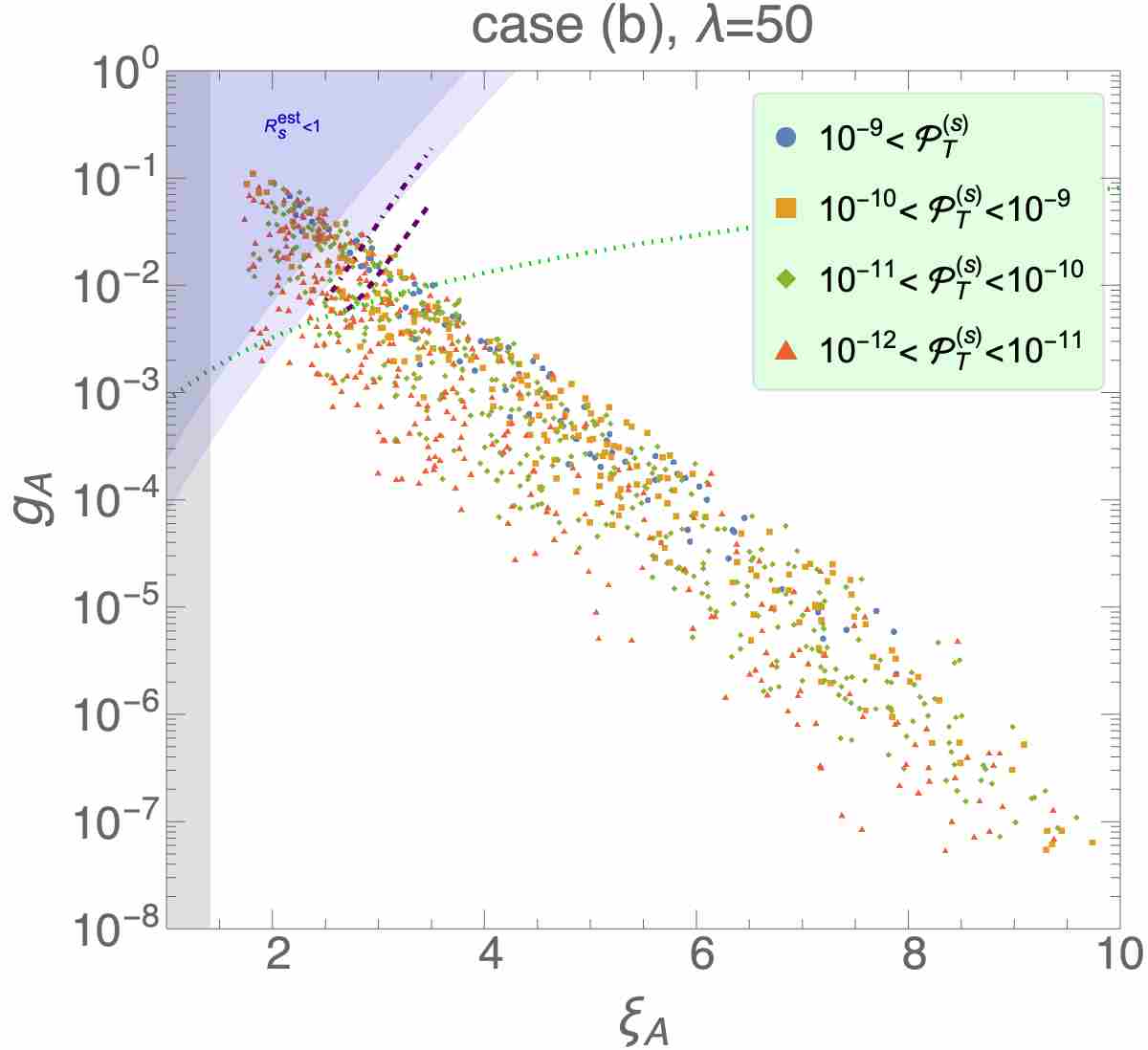}
    \caption{Same as Fig.\,\ref{fig:GW_lam_100_g} but for $\lambda=50$. }
  \label{fig:GW_lam_050_g}
 \end{center}
\end{figure}

\begin{figure}
  \begin{center}
    \includegraphics[width=7cm]{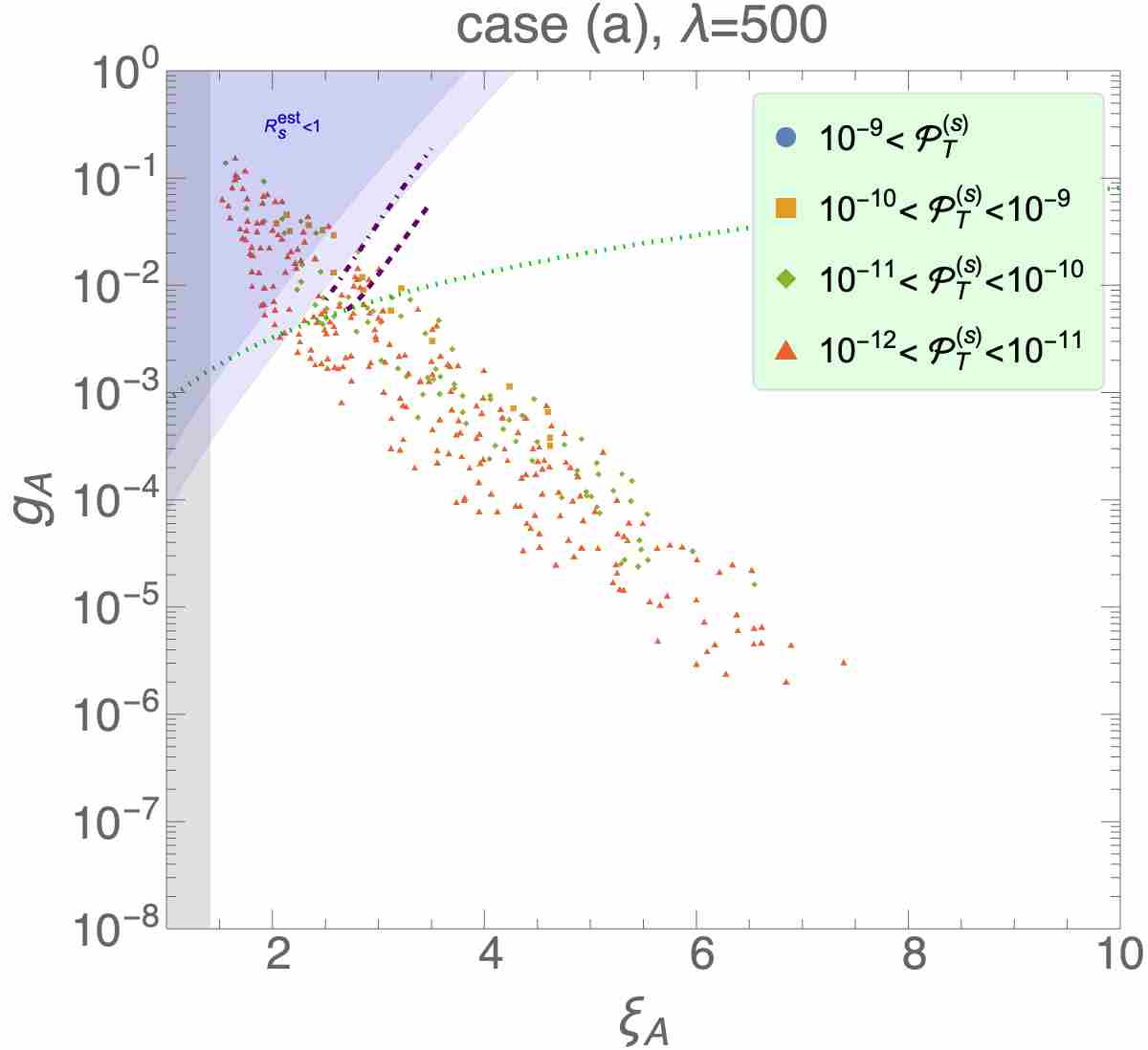}
    \includegraphics[width=7cm]{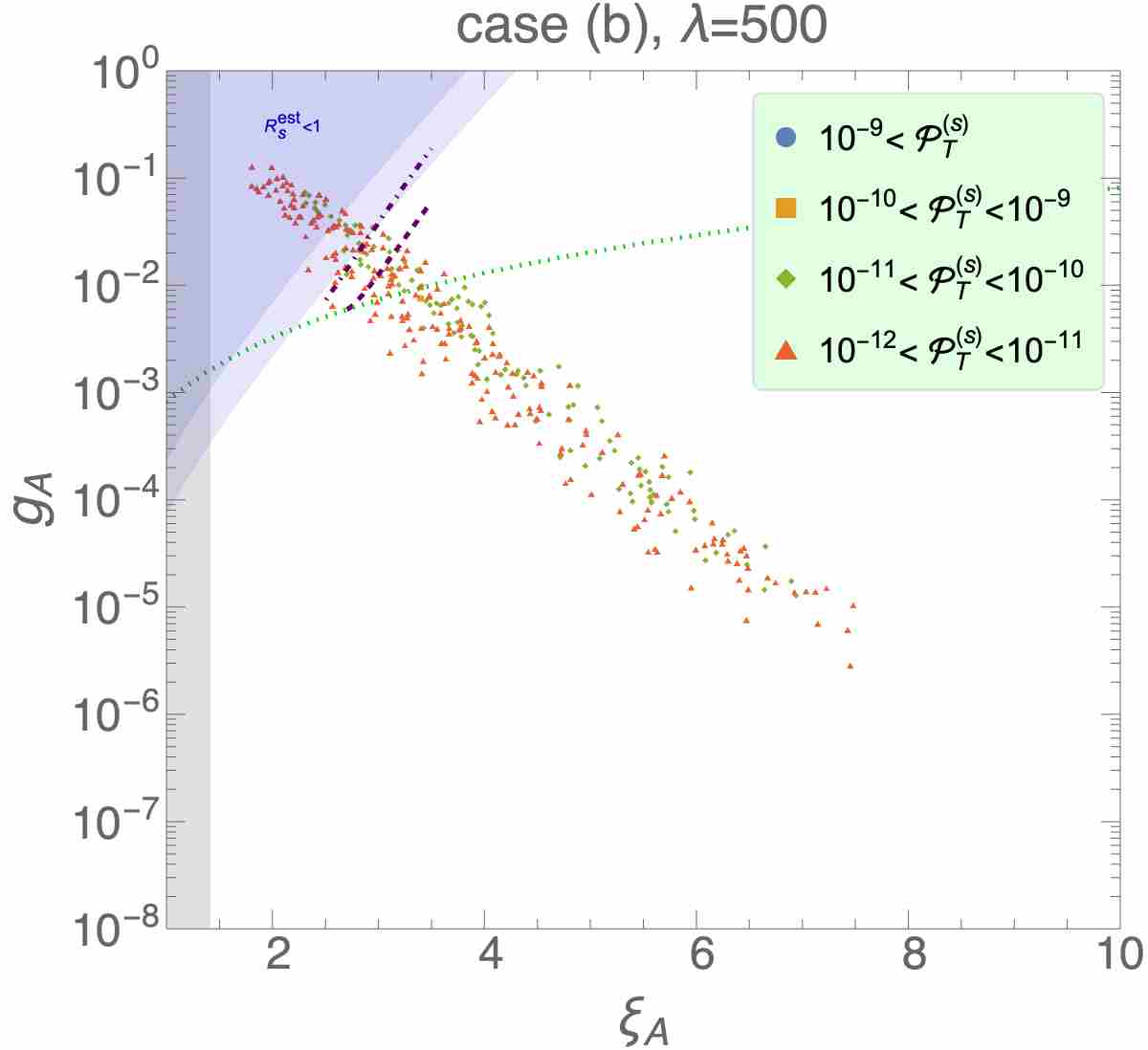}
    \caption{Same as Fig.\,\ref{fig:GW_lam_100_g} but for $\lambda=500$. }
  \label{fig:GW_lam_500_g}
 \end{center}
\end{figure}

\begin{figure}
  \begin{center}
    \includegraphics[width=7cm]{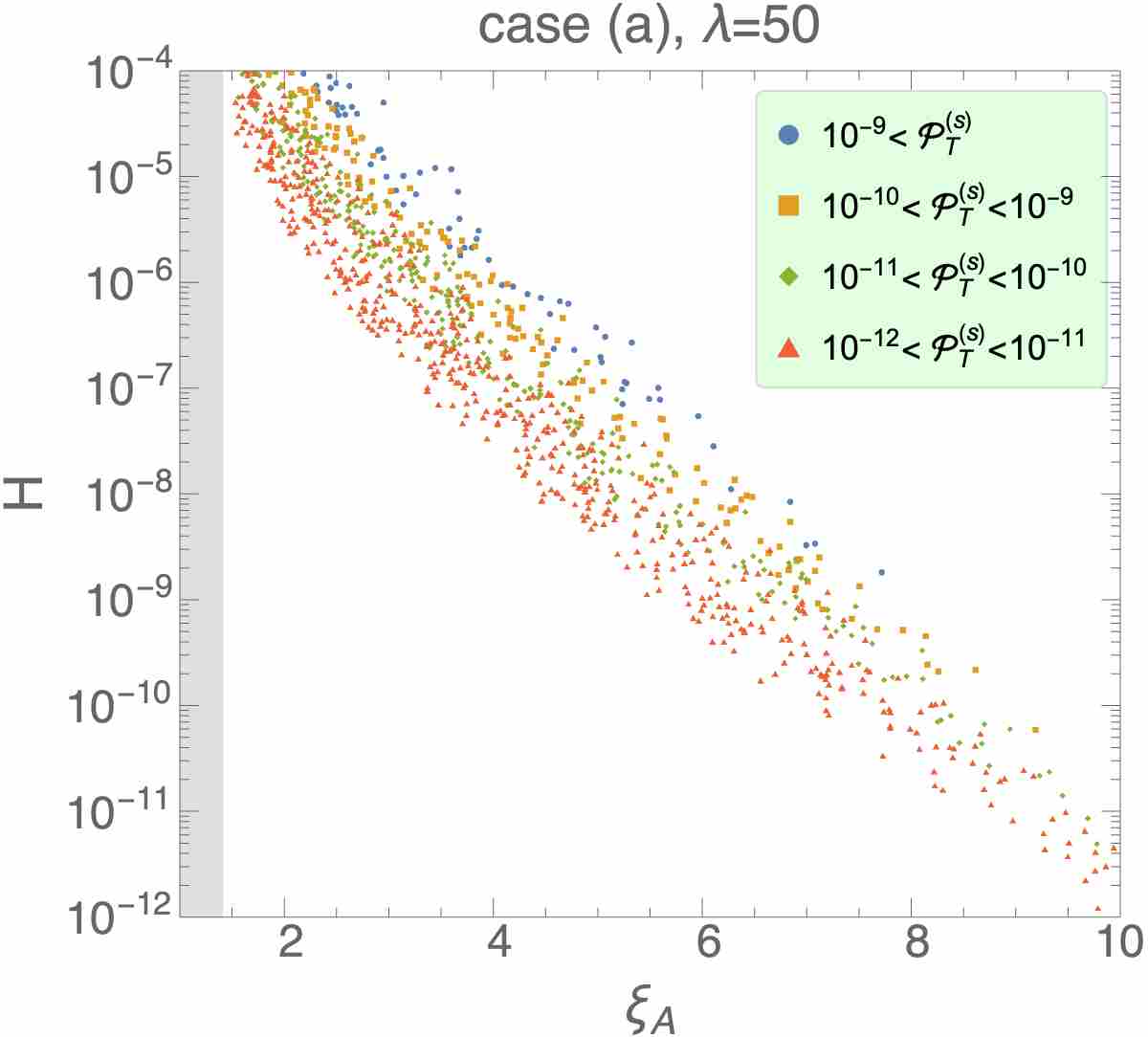}
    \includegraphics[width=7cm]{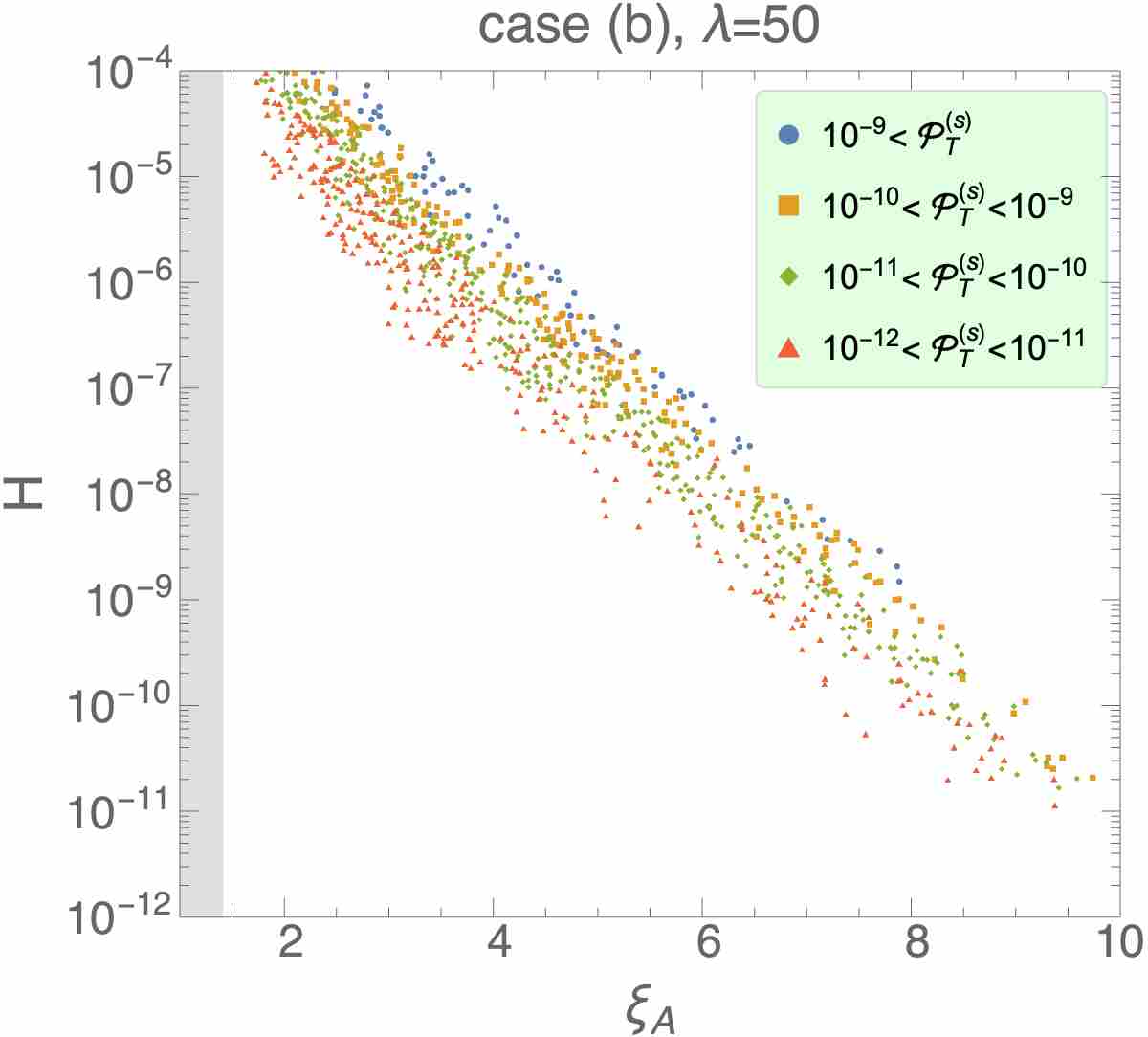}
    \includegraphics[width=7cm]{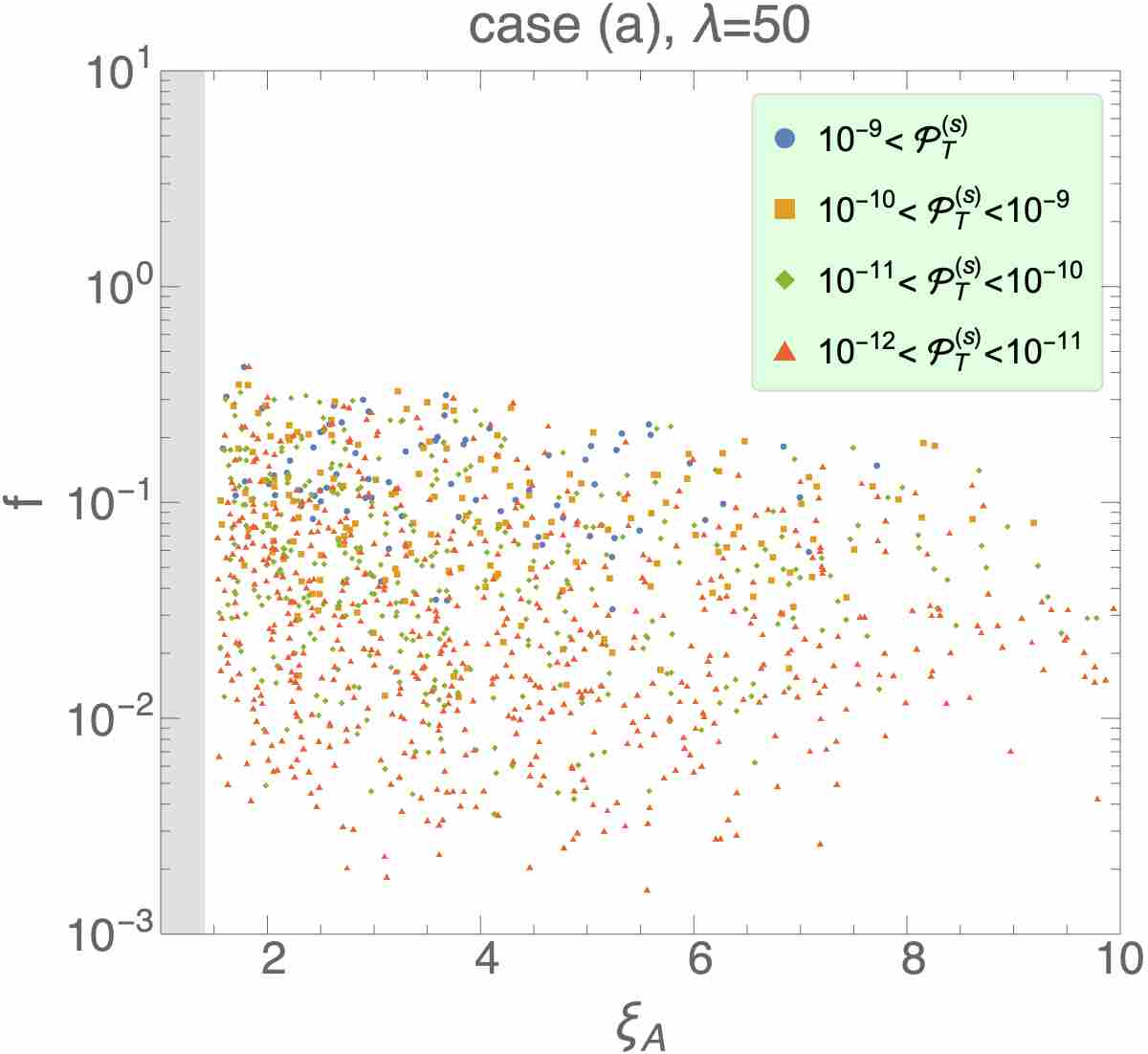}   
    \includegraphics[width=7cm]{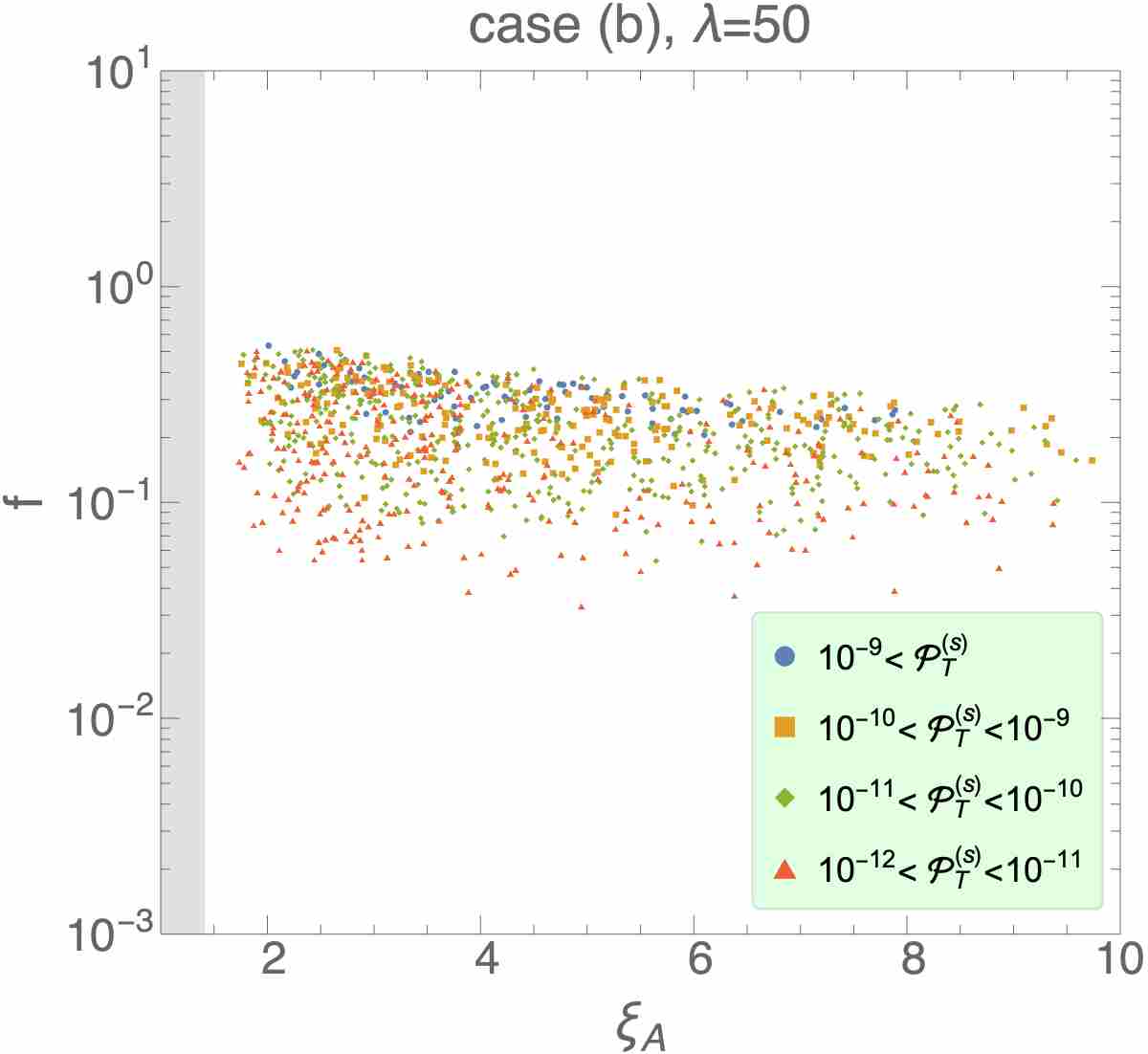}
    \includegraphics[width=7cm]{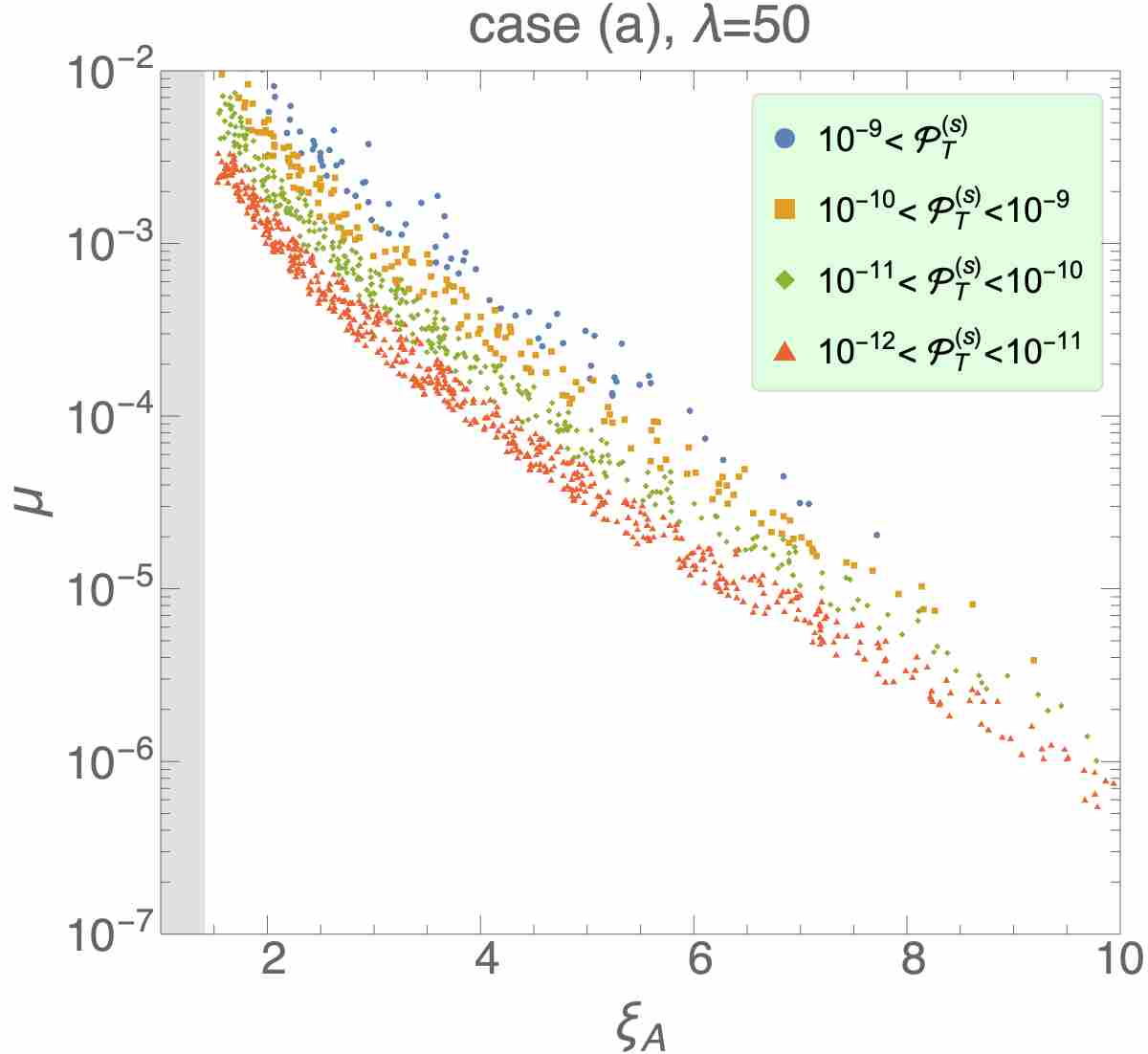}   
    \includegraphics[width=7cm]{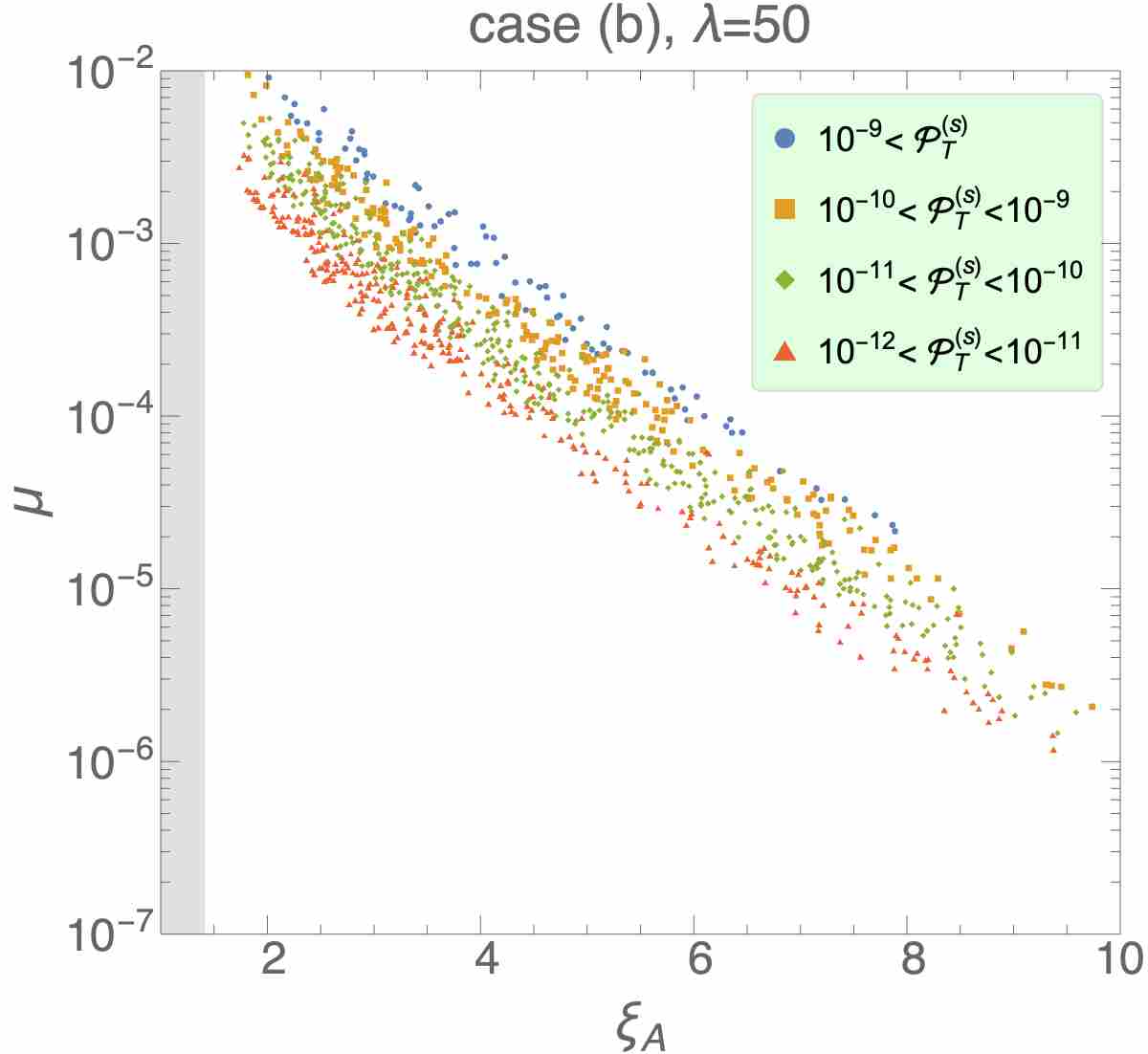}
    \caption{Same as Fig.\,\ref{fig:GW_lam_100_Hmu} but for $\lambda=50$.  }
  \label{fig:GW_lam_050_Hmu}
 \end{center}
\end{figure}

\begin{figure}
  \begin{center}
    \includegraphics[width=7cm]{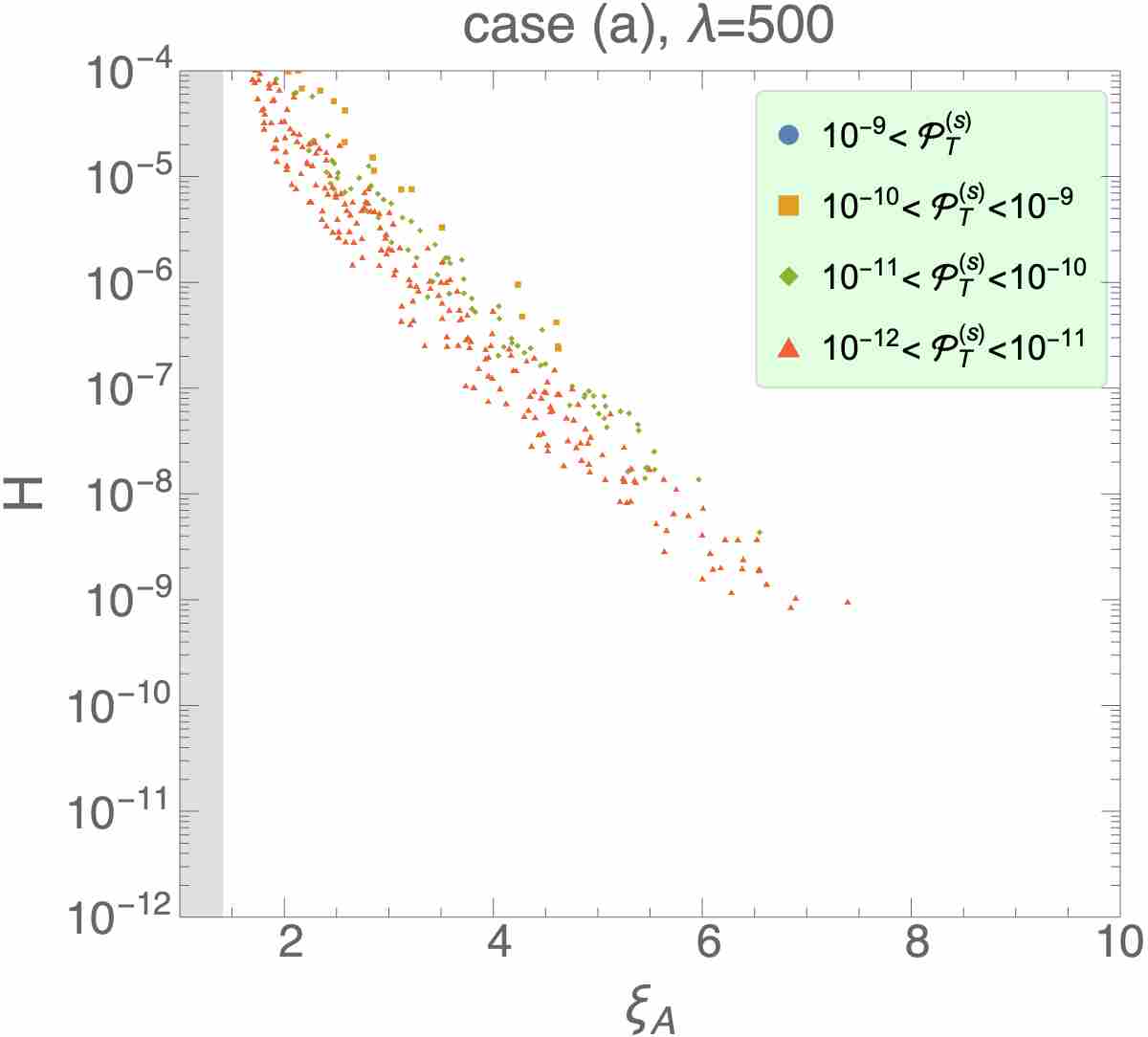}
    \includegraphics[width=7cm]{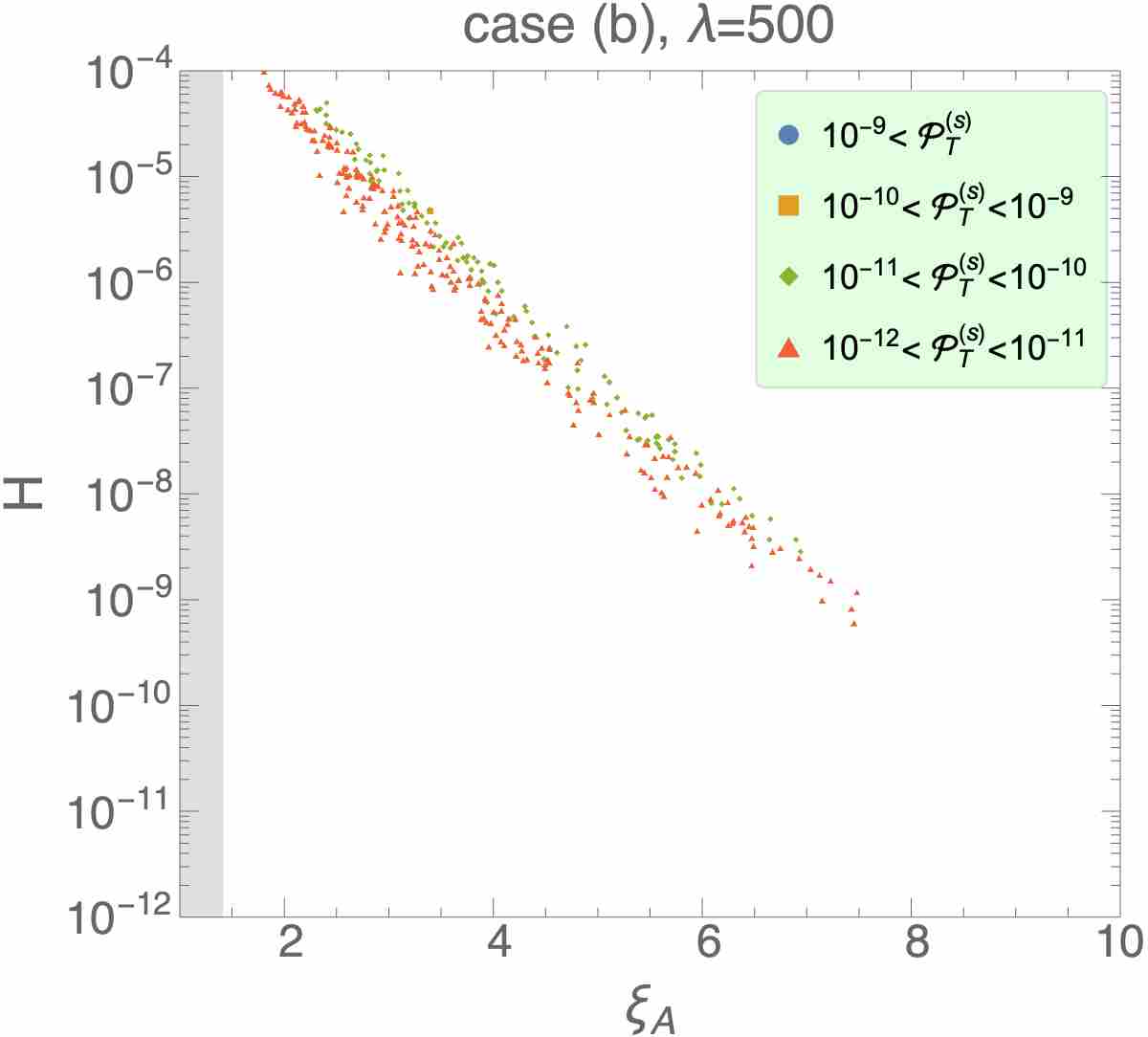}
    \includegraphics[width=7cm]{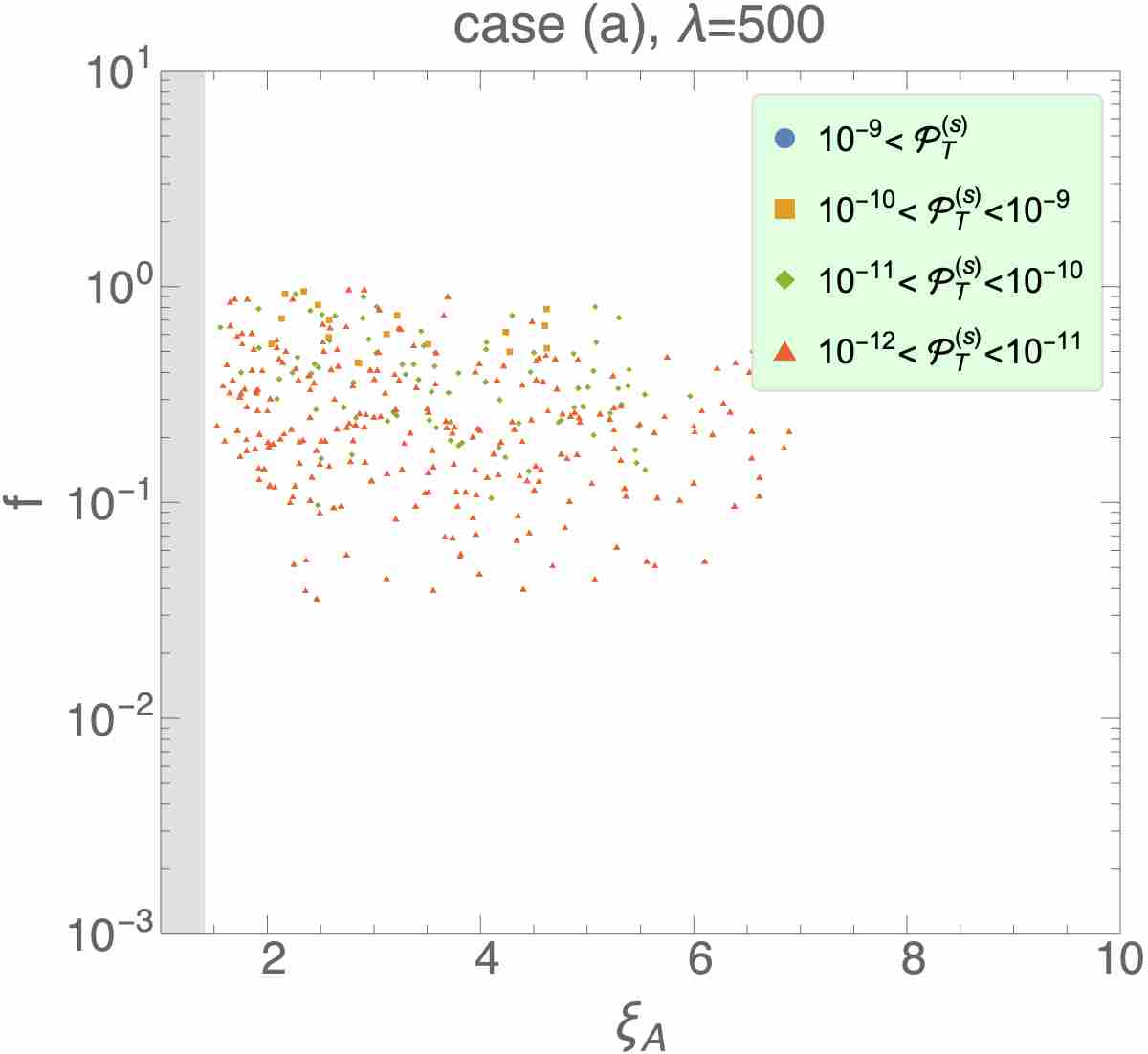}   
    \includegraphics[width=7cm]{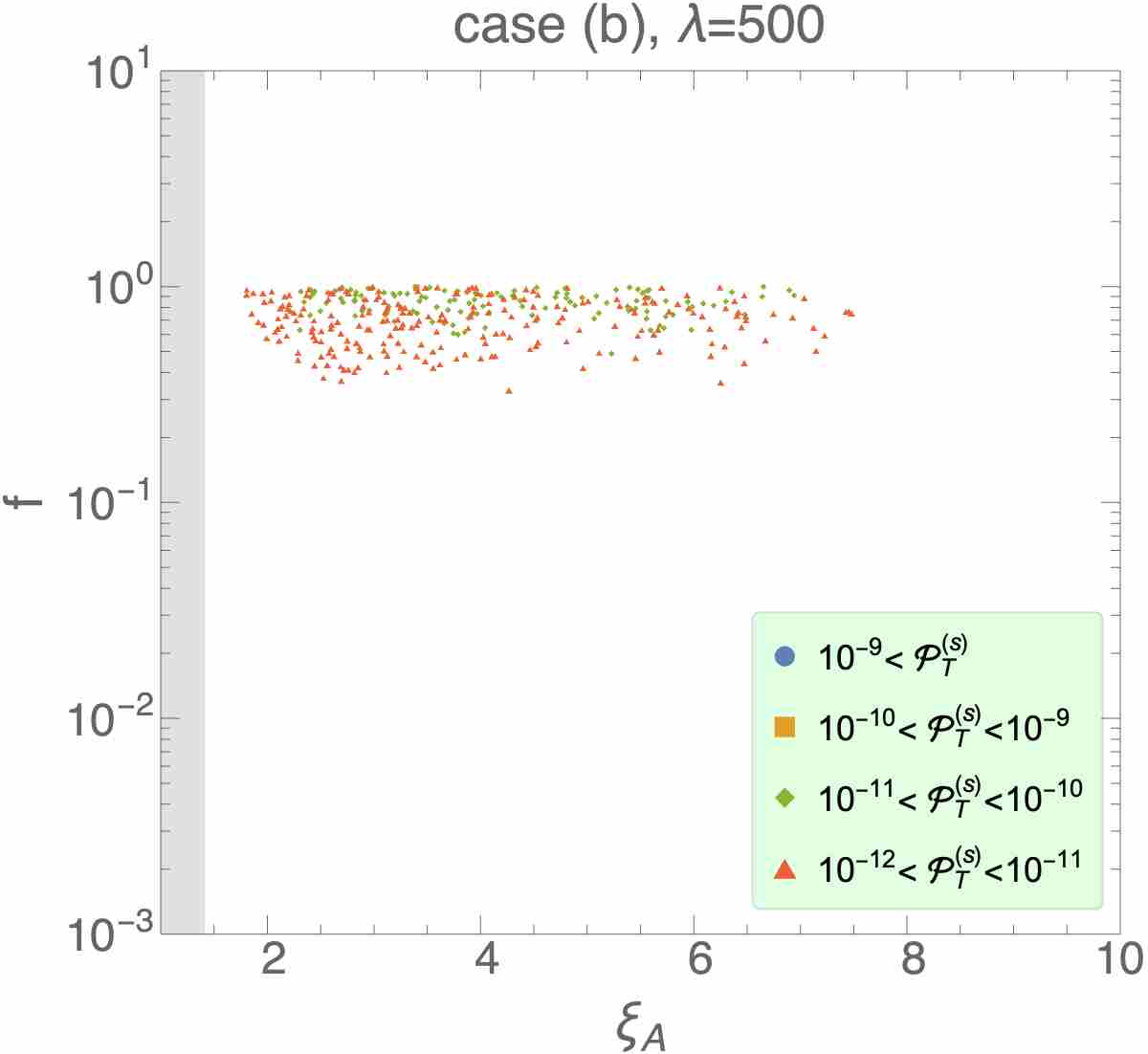}
    \includegraphics[width=7cm]{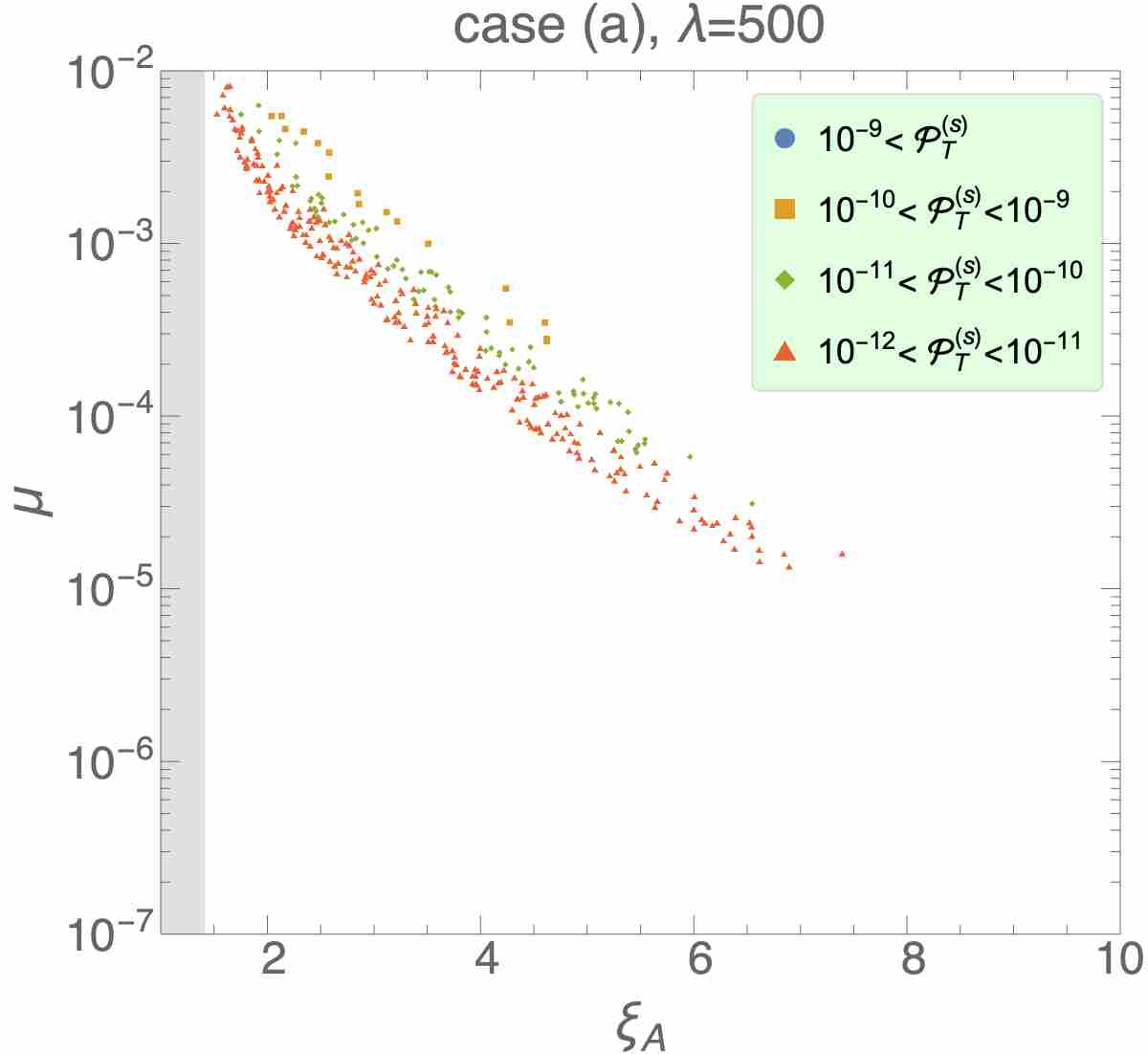}   
    \includegraphics[width=7cm]{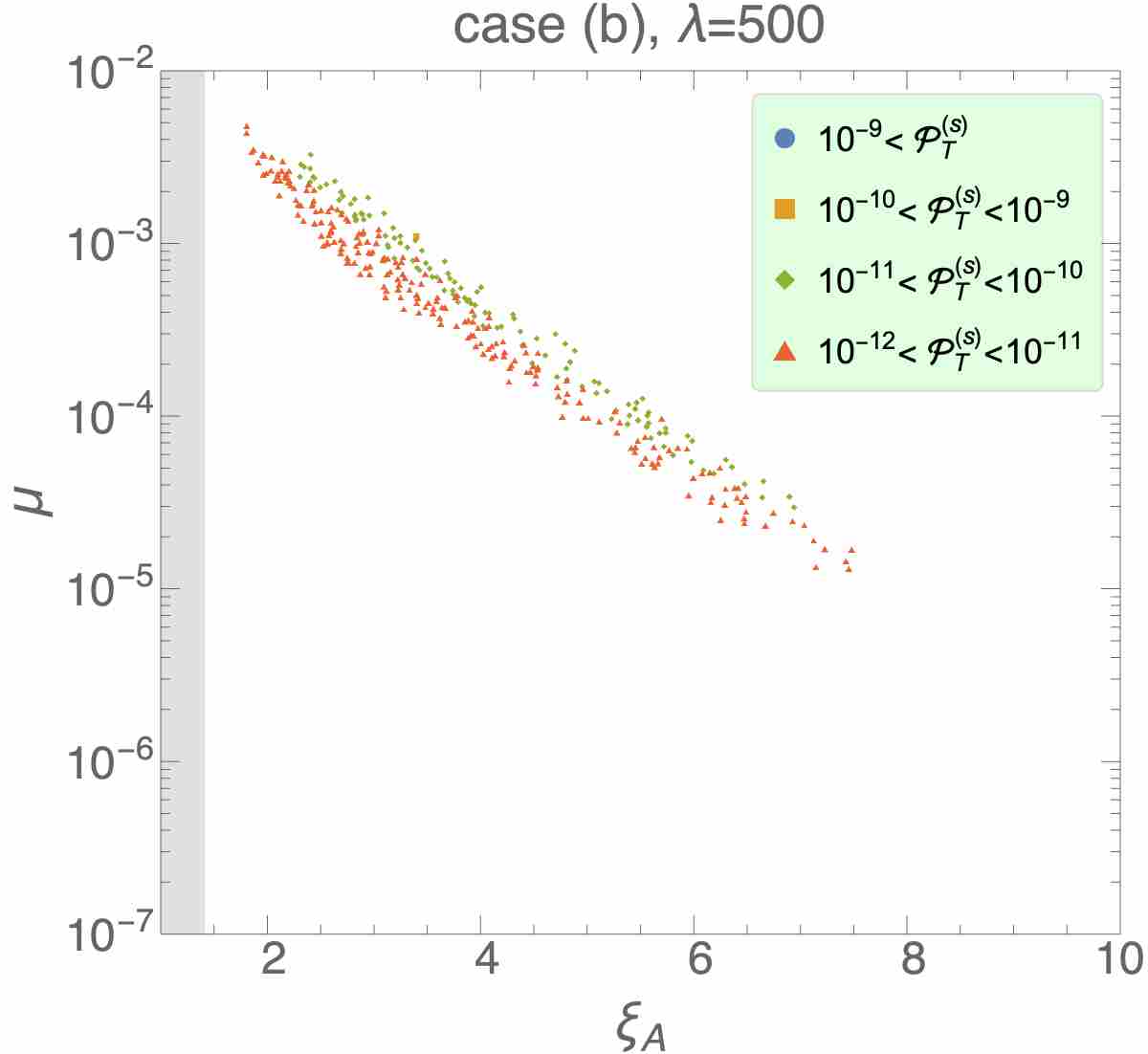}
    \caption{Same as Fig.\,\ref{fig:GW_lam_100_Hmu} but for $\lambda=500$.  }
  \label{fig:GW_lam_500_Hmu}
 \end{center}
\end{figure}

\begin{figure}
  \begin{center}
    \includegraphics[width=7cm]{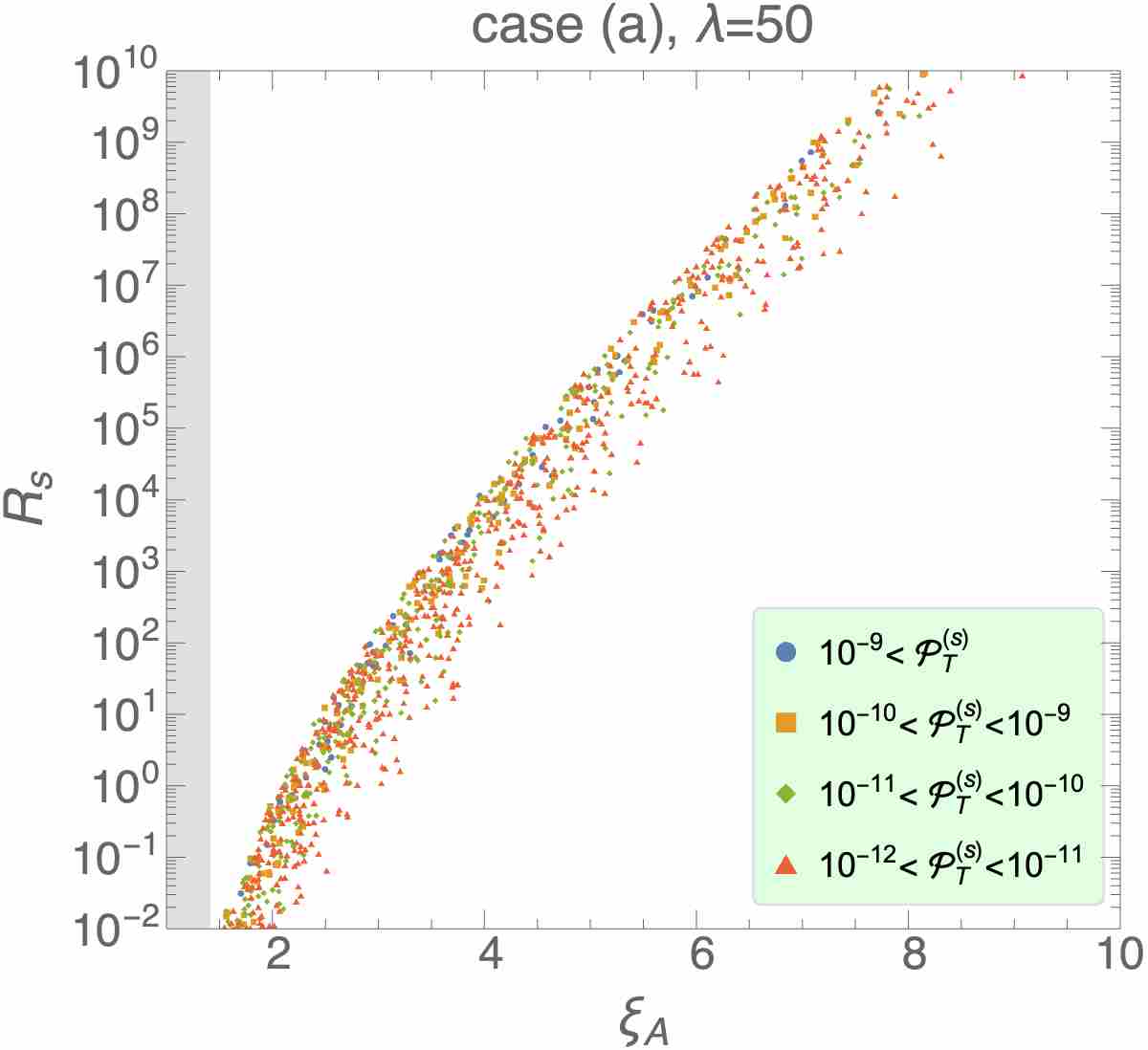}
    \includegraphics[width=7cm]{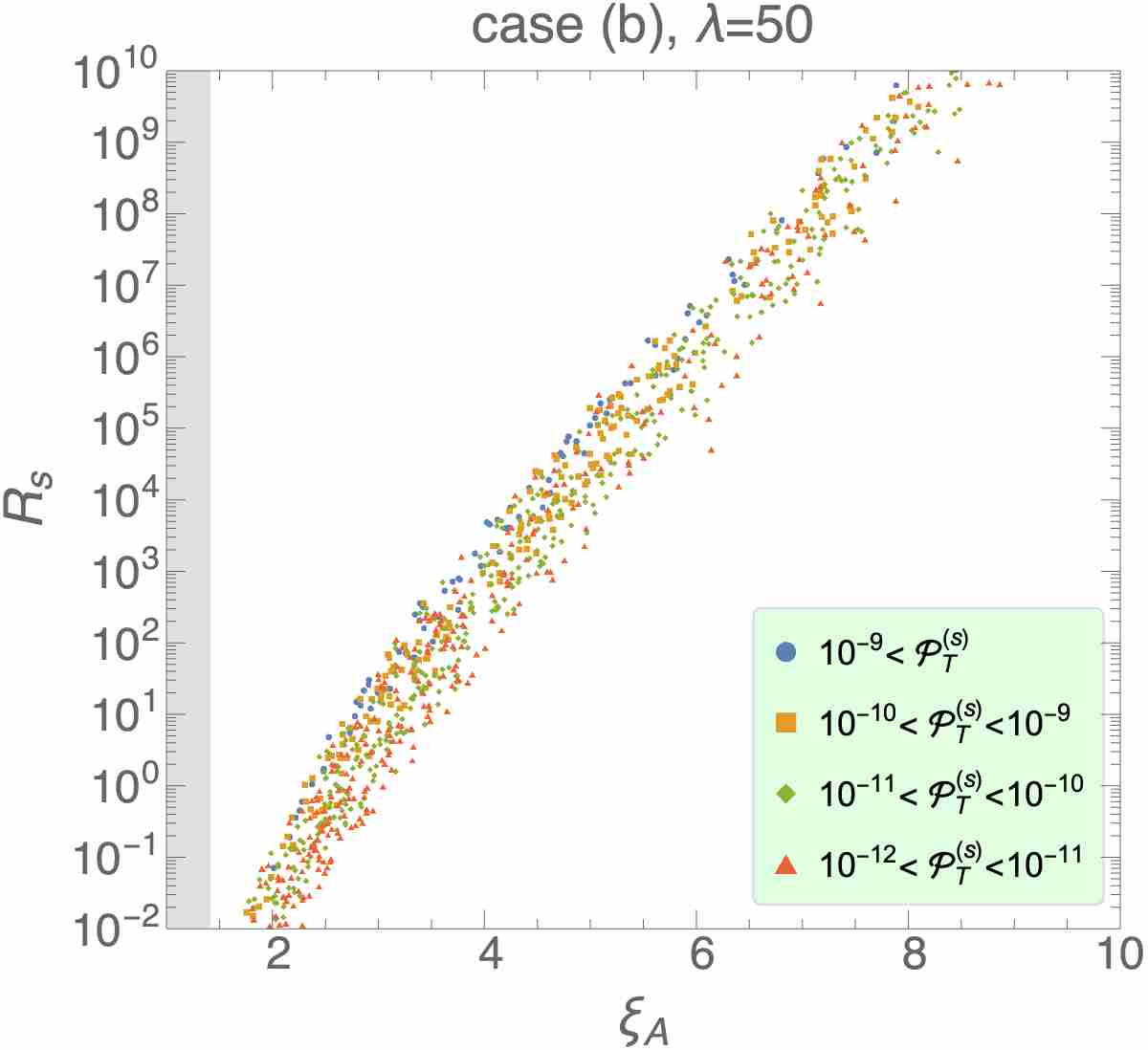}
    \includegraphics[width=7cm]{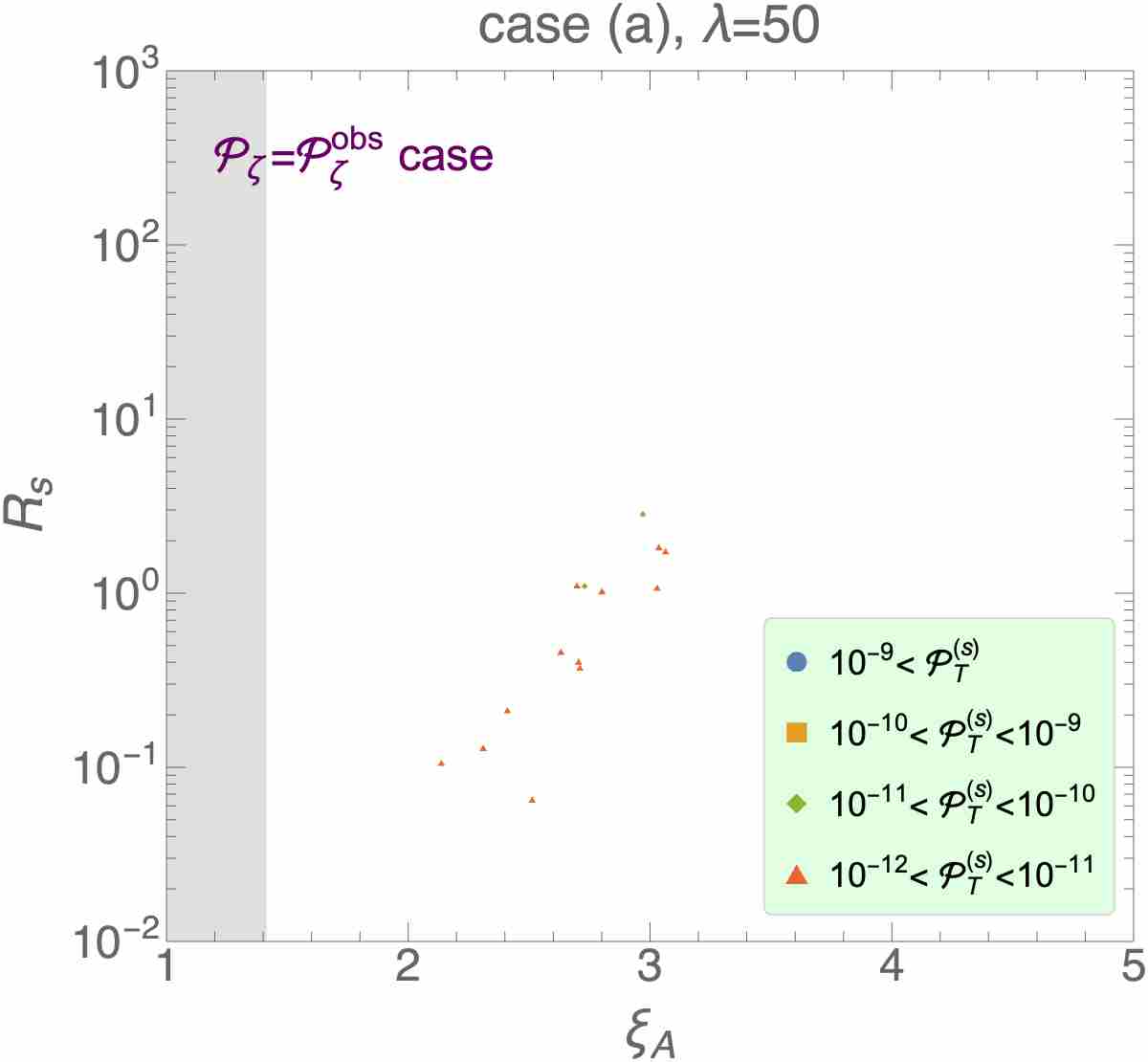}   
    \includegraphics[width=7cm]{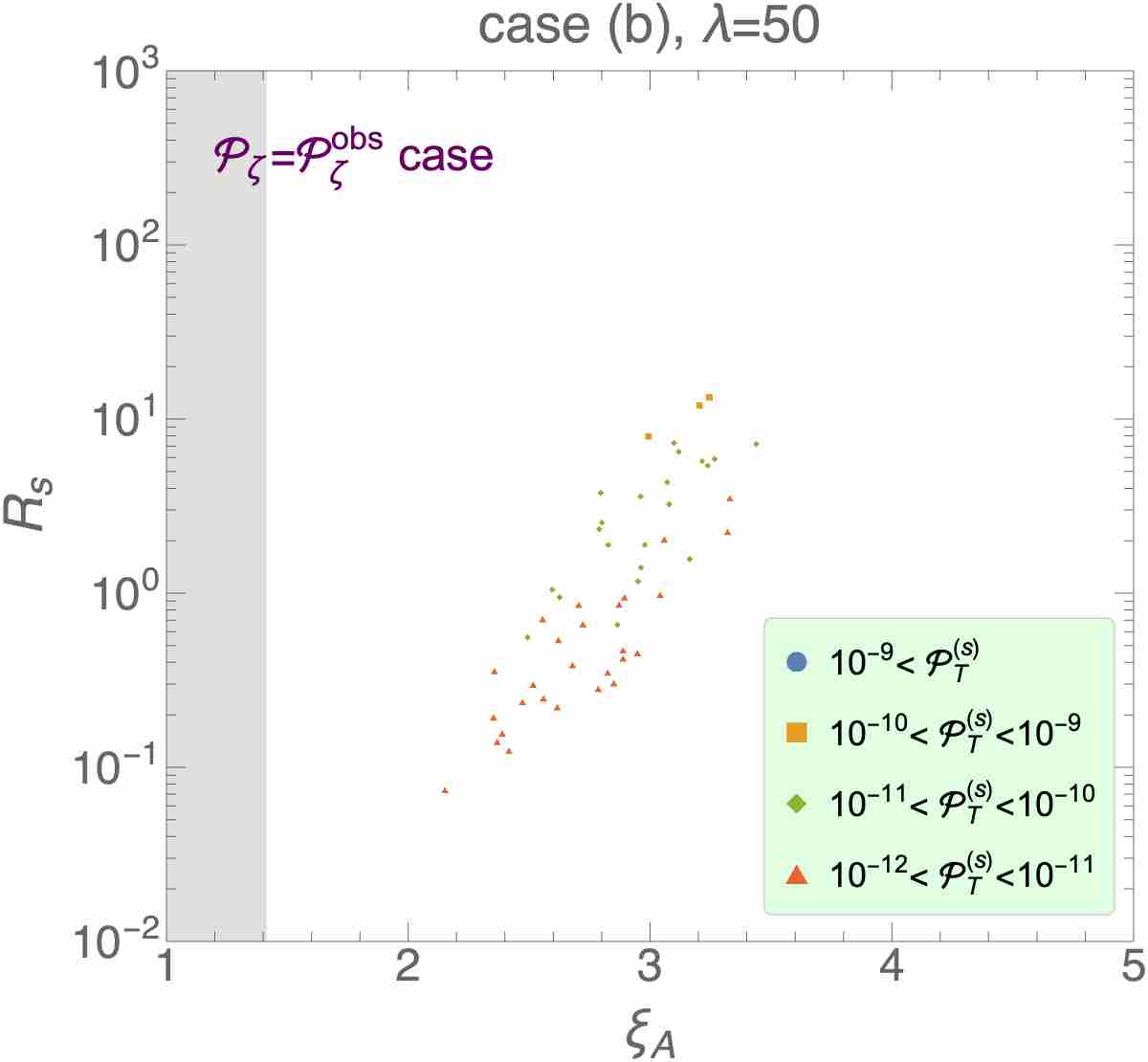}
    \includegraphics[width=7cm]{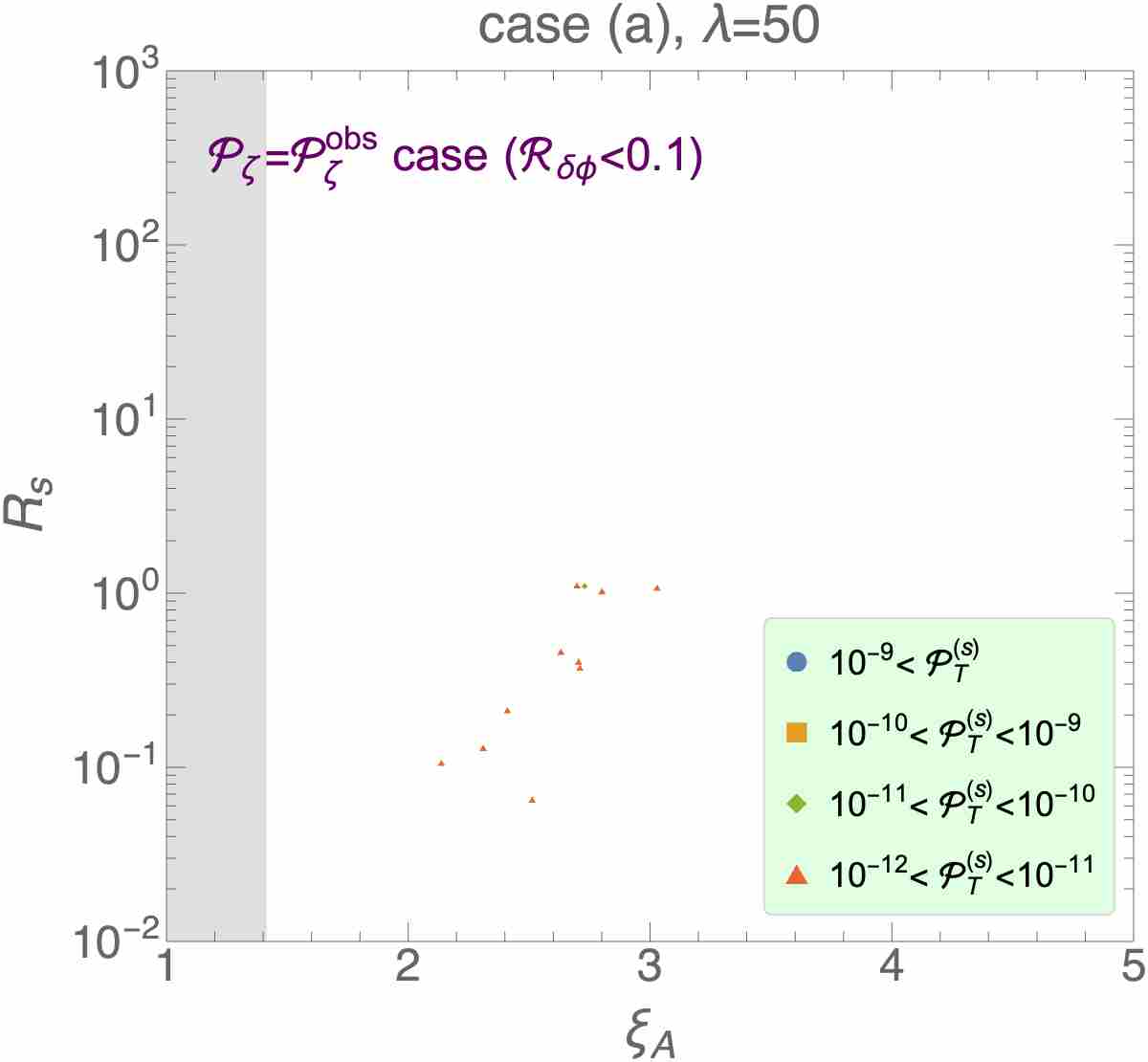}   
    \includegraphics[width=7cm]{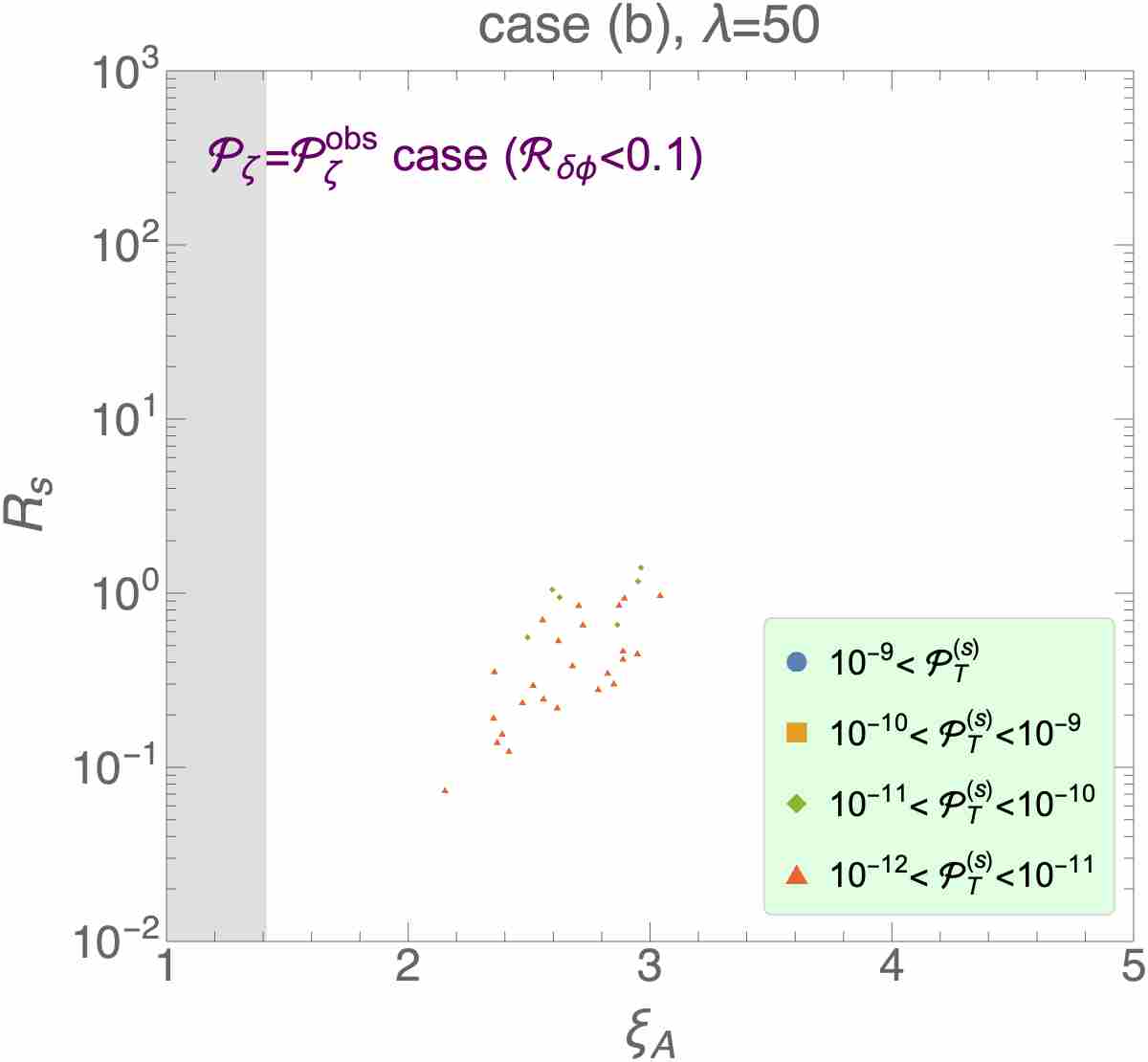}
    \caption{Same as Fig.\,\ref{fig:GW_lam_100_Rs} but for $\lambda=50$. }
  \label{fig:GW_lam_050_Rs}
 \end{center}
\end{figure}

\begin{figure}
  \begin{center}
    \includegraphics[width=7cm]{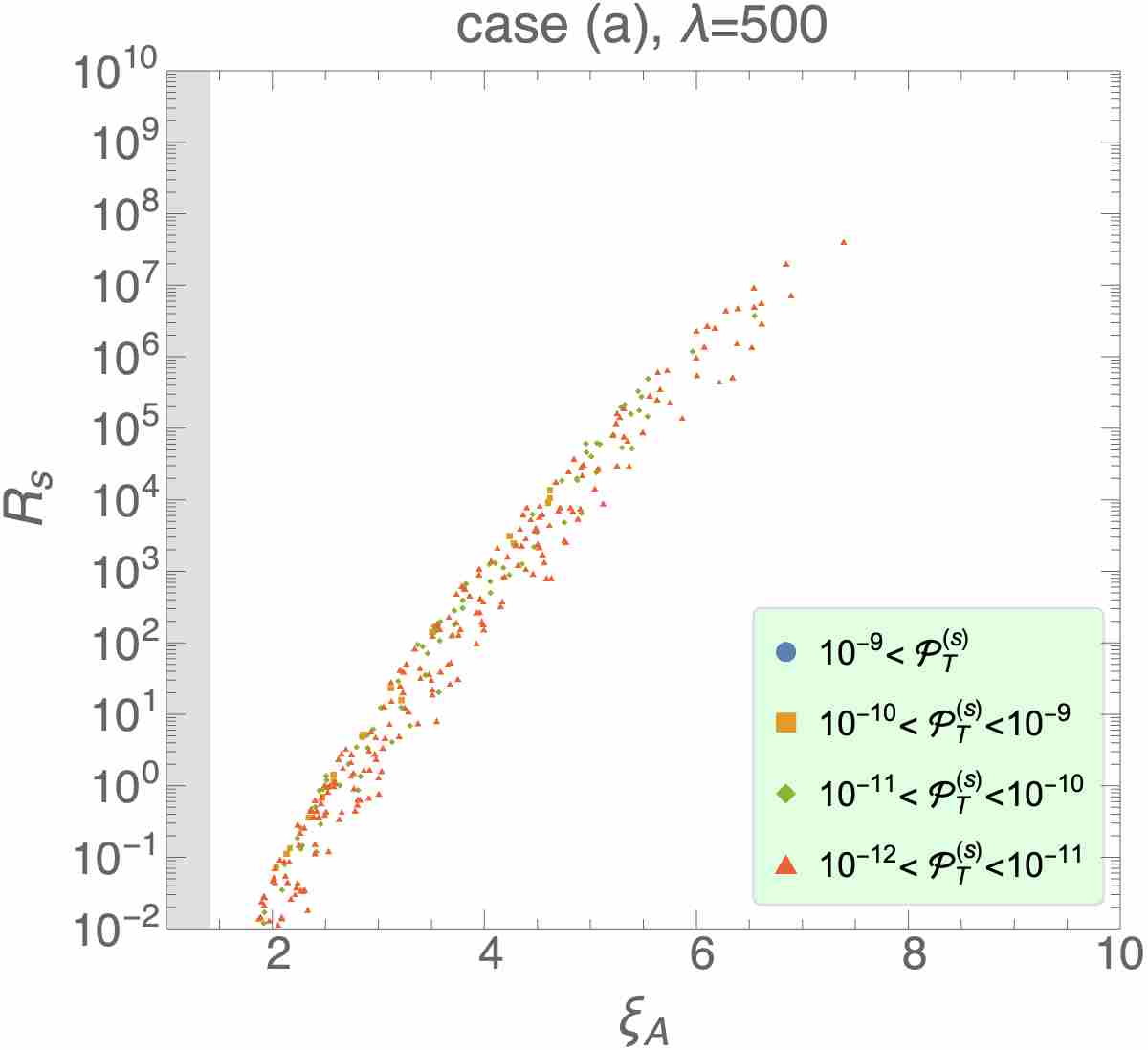}
    \includegraphics[width=7cm]{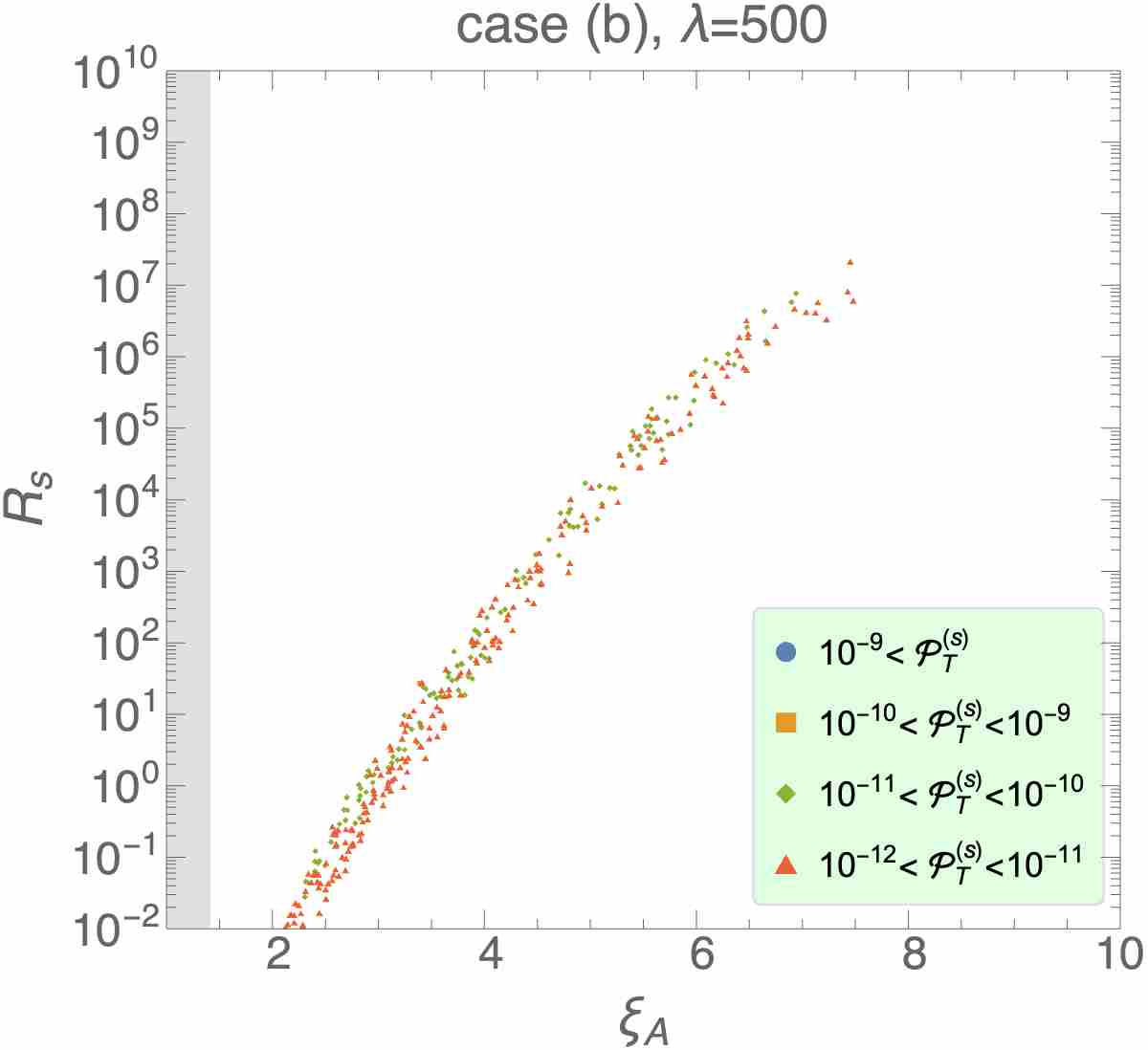}
    \includegraphics[width=7cm]{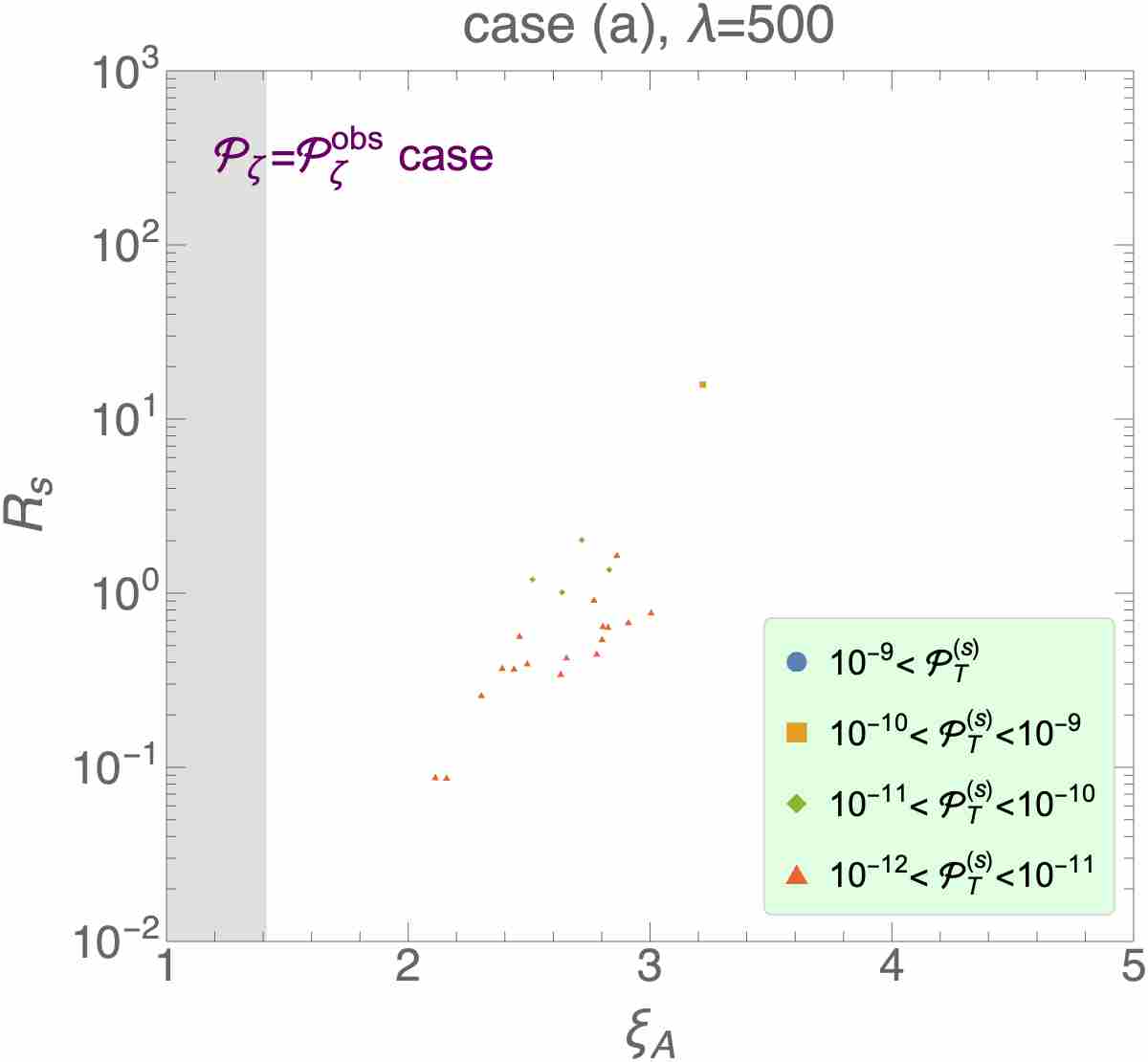}   
    \includegraphics[width=7cm]{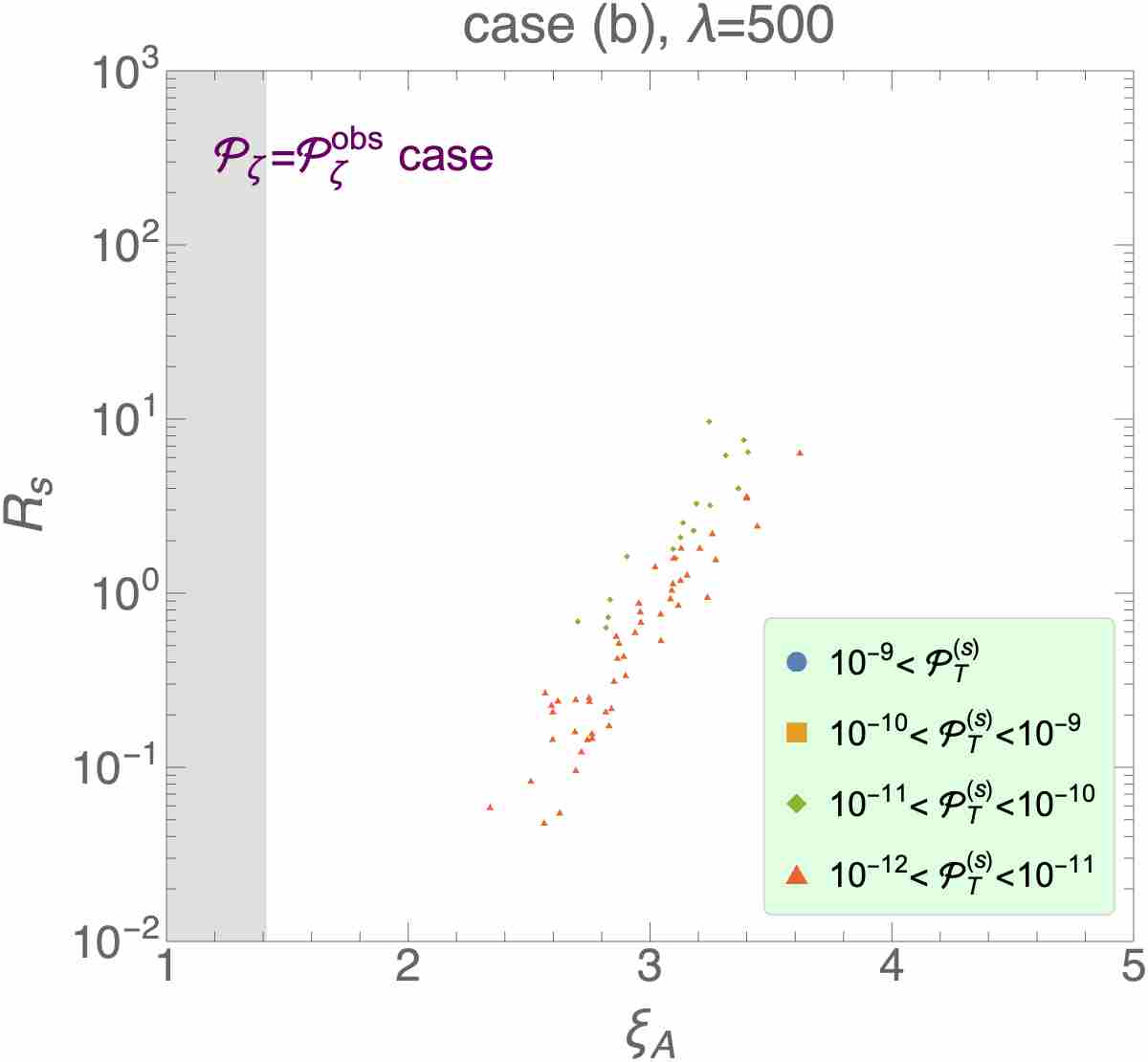}
    \includegraphics[width=7cm]{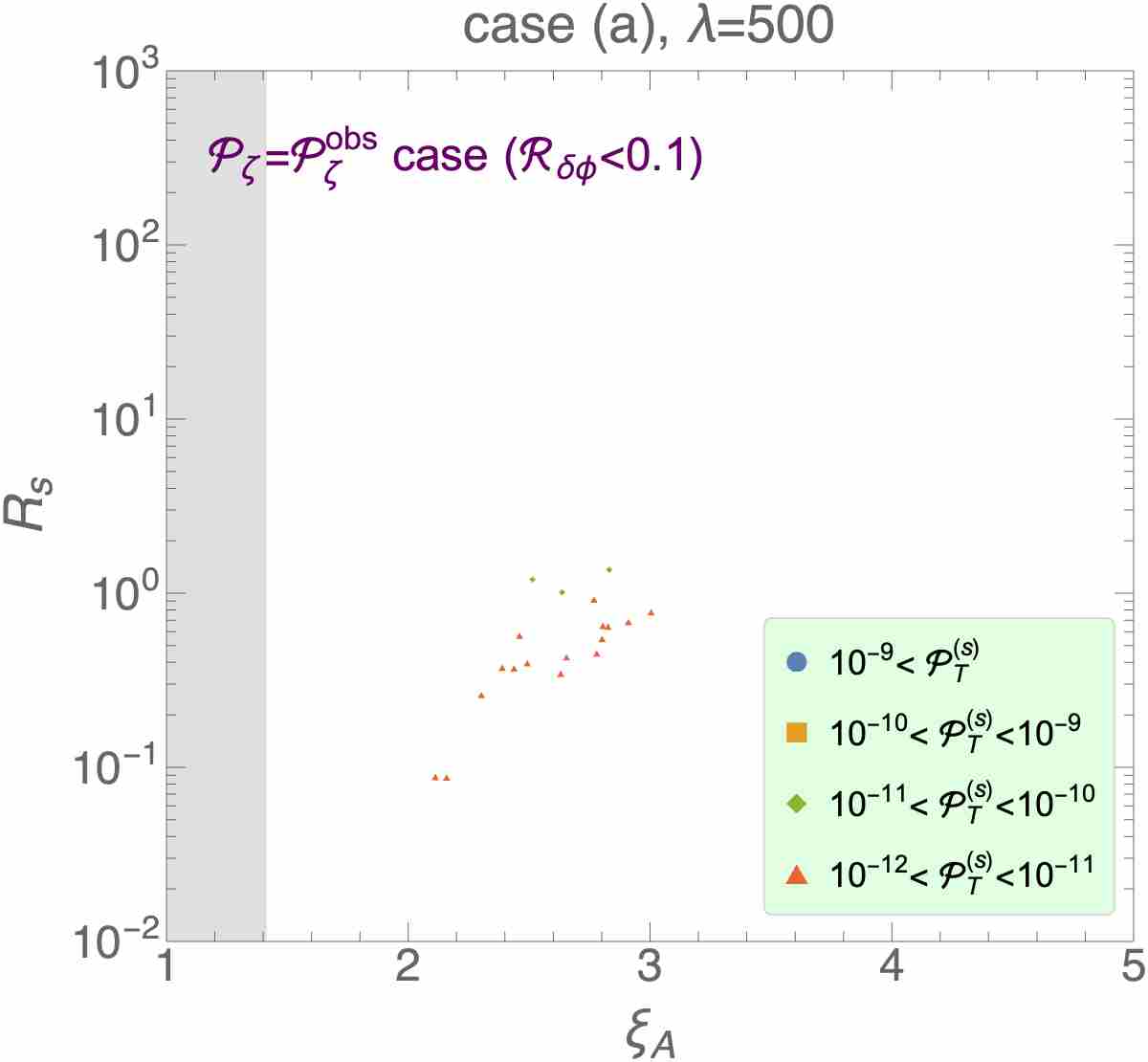}   
    \includegraphics[width=7cm]{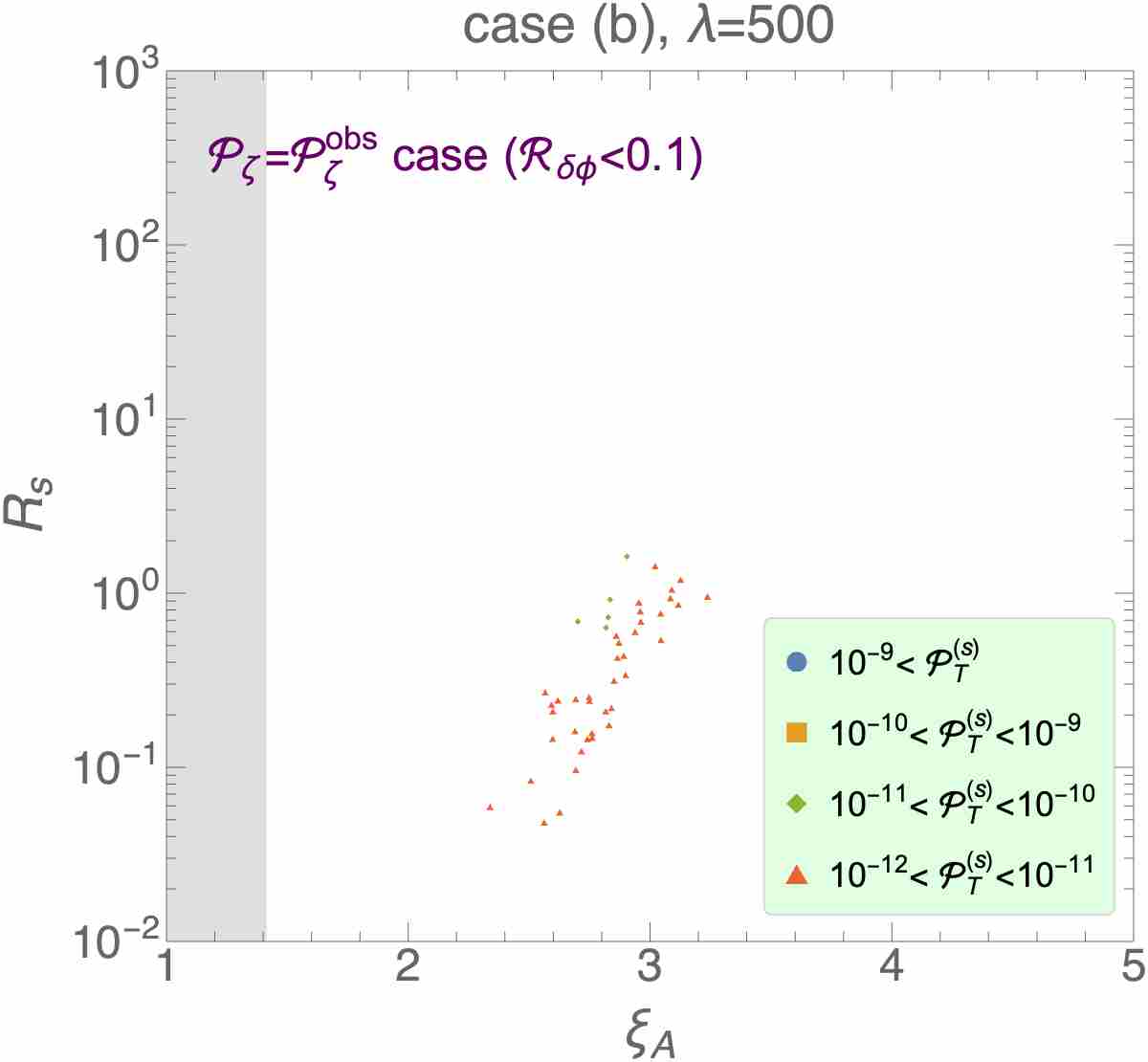}
    \caption{Same as Fig.\,\ref{fig:GW_lam_100_Rs} but for $\lambda=500$. }
  \label{fig:GW_lam_500_Rs}
 \end{center}
\end{figure}

\bibliographystyle{JHEP}
\bibliography{refs.bib}

\end{document}